\documentclass[12pt]{article}
\usepackage[utf8]{inputenc}
\usepackage{booktabs}
\usepackage{eqparbox}
\usepackage[dvipsnames]{xcolor}
\usepackage{soul}

\usepackage{graphicx}
\usepackage{fullpage} 
\usepackage{subcaption} 
\usepackage{braket}
\usepackage{amsmath,amssymb}
\usepackage{bbold}
\usepackage[bottom]{footmisc}
\usepackage{tikz-cd}
\usepackage{dsfont}
\usepackage{amsfonts}
\usepackage{amsthm}
\usepackage{amssymb,graphicx}
\usepackage{hyphenat}
\usepackage{csquotes}
\usepackage{hyperref}
\newcounter{dummy}
\numberwithin{equation}{section}
\hypersetup{
    colorlinks=true,
    linkcolor=blue,
    filecolor=blue,      
    urlcolor=blue,
    citecolor=blue,
    linktoc=page,
}

\usepackage[font={small}]{caption}
\usepackage{float}
\usepackage[parfill]{parskip}
\usepackage{multirow}
\usepackage{braket}
\newcommand{\e}[1]{é}

\newcommand{\edit}[1]{\textcolor{red}{#1}}
\newcommand{\LL}{\mathcal{L}}
\newcommand{\ls}{\left[}
\newcommand{\rs}{\right]}

\begin{document}

\begin{titlepage}
\thispagestyle{empty}
\begin{centering}
	{\scshape\Huge Fractons: gauging spin models and tensor gauge theory \par}
	\vspace*{.25cm}
	{\mdseries\Large Jason Bennett\par}
	\vspace*{.25cm}
	\end{centering}

The objective of the present work — a literature review on both gapped and gapless fractonic theories — is to pedagogically fill in the gaps between the research on fractons, and an undergraduate physics education (particularly quantum mechanics and statistical mechanics). Some familiarity with classical field theory is assumed. We will begin this work by reviewing the gauging of the Ising model to obtain the toric code. Then, following the chronological order for developments in theories of fractons, we will establish the gauged spin model picture of gapped fractonic theories. Next (after establishing the preliminaries of lattice gauge theory) we will cover the developments on the tensor gauge theory front in describing gapless fractonic theories. We then explain how conservation of dipole moment is key in the development of field-theoretic descriptions of fracton models, and conclude by providing future plans for work on gauging such field theories. Finally, we point out two more future plans: one being an established theoretical advance which in fact inspired this work, and the other being exciting experimental advances in ultracold atomic physics which could lead to the physical realization of fracton physics. 
%Aspects of this work that are (to our knowledge, after paying due diligence to the references of all works cited herein) novel  — in the sense that the main conclusions are stated without proof elsewhere — include the following: 
Highlights of this work include: a derivation of the gauge field dynamics (``field strength" or ``curvature") term in the gauging of a spin system to obtain the toric code, a reduction of the 3 terms of the Kogut-Susskind lattice quantum electrodynamics Hamiltonian to the 2 terms of the toric code Hamiltonian, a proof of the correspondence between the invariance of a gauge field and a particular Gauss’s law constraint in a tensor gauge theory context, a demonstration that conservation of dipole moment (in the gapless tensor gauge theory models of fractons) accounts for the same immobility/``fractalization” phenomenology that was shown earlier for the case of gauged spin model picture of gapped fracton theories, and proof of the necessity for polynomial shift symmetries and higher-order spatial derivatives in a Lagrangian in order to have a field theory with dipole moment conservation.

\begin{centering}

\vspace*{.25cm}
{\Large Master's Thesis \par}
{\large Department of Physics and Astronomy, Stony Brook University \par}
\vspace*{.4cm}
{\large Supervised by: Professor Tzu-Chieh Wei, Hiroki Sukeno, and Li Yabo \par}
{\large Approved by: Professors Tzu-Chieh Wei, Jennifer Cano, and Dominik Schneble \par}
\vspace*{.4cm}
{\large May 13, 2022}

\end{centering}	

\end{titlepage}

\iffalse
\begin{titlepage}
\thispagestyle{empty}
\begin{centering}
	\vspace*{1cm}
	{\scshape\Huge Fractons: gauging spin models and tensor gauge theory \par}
	\vspace*{1cm}
	{\mdseries\Large Jason Bennett\par}
	\vspace*{1cm}
	{\Large\mdseries Master's Thesis \par}
	
\vspace*{1cm}	

{\large Department of Physics and Astronomy \par
Stony Brook University \par
May 13, 2022 \par}
\vspace*{3cm}
\end{centering}

\hspace{7cm}
{\large Approved by: \\ [1ex] \par}
\hspace{7cm}
\begin{tabular}{l}
\makebox[3.5in]{\hrulefill} \\
Professor Tzu-Chieh Wei, Advisor \\ [5ex]
\makebox[3.5in]{\hrulefill} \\
Professor Jennifer Cano, Committee chair \\ [5ex]
\makebox[3.5in]{\hrulefill} \\
Professor Dominik Schneble, Committee member
\end{tabular}

\end{titlepage}
\fi
\pagebreak
\begin{titlepage}
\thispagestyle{empty}
\tableofcontents
\thispagestyle{empty}
%\JL{Johannes's comments}

%CS{Ceyda's comments}

%\edit{Things I need to edit in the future}
\end{titlepage}
\pagebreak

\section*{}
\addcontentsline{toc}{section}{List of figures}
\listoffigures

\iffalse
\begin{titlepage}
\thispagestyle{empty}
\listoffigures
\addcontentsline{toc}{section}{List of figures}
\thispagestyle{empty}
\end{titlepage}
\fi

\pagebreak

\section*{Acknowledgements}
\addcontentsline{toc}{section}{Acknowledgements}

I would like to thank my advisor, Professor Tzu-Chieh Wei for supervising my work on this topic through an independent study and ultimately through this thesis work. Taking all five of the graduate physics program's core courses in my first year was quite arduous, and I am incredibly grateful to Professor Wei for giving me the chance to get excited about research again through an independent study in the second semester of that first year. After taking Professor Wei's course on quantum information science and experiencing his stellar pedagogy, I knew he would make the perfect advisor for my path towards a Master's thesis — exploring the further directions I had envisioned after the independent study.

Alternating between meeting with Professor Wei and meeting with Professor Wei's PhD students, Li Yabo and Hiroki Sukeno, was a huge blessing. While meetings with all four of us were useful as a progress check to update them on my efforts and results, taking time to ask Yabo and Hiroki detailed questions was incredible fruitful as they were able to provide great insight into next routes to take as well as see paths through a particular derivation that I'd been stuck on for days. I'm very grateful for their help. On that note, I also want to thank Kevin Grosvenor, currently a postdoc at Universiteit Leiden in the Netherlands. I met Kevin when he gave a brilliantly pedagogic talk on holographic entanglement entropy at Rijksuniversiteit Groningen the year I was there. Having that memory of him, and also due to: the fact that his paper (which a priori had nothing do with fractons) was used by another author in what was ultimately my inspiration for working on fractons (see section \ref{inspriation} for the full story), and the fact he has been working more directly on fractons recently — I was motivated to reach out to Kevin for help in the Noether currents approach to dipole conservation in section \ref{kevhelp}. Kevin, Yabo, and Hiroki's help in that part of thesis thesis was pivotal.

My committee for this Master's thesis consists of two of brilliant researchers and teachers: Professor Jen Cano, and Professor Dominik Schneble. Professor Cano's condensed matter class was categorically my favorite breath course because of her incredible lecturing style. Add in working together and studying with friends Megan McDuffie and Quinn White, and this class full of novel material quickly became the most fun class I'd had at Stony Brook. Given my first exposure to research coming as early as sophomore year in college, it is not surprising that my theoretical understanding of what I was working on was swept under the rug a tad in favor of learning computational methods to simulate the ultracold atomic system. Taking Professor Schneble's ultracold atomic physics class, which was theoretically the most advanced course I'd taken at Stony Brook, opened up my eyes to the wild possibilities for depth in atomic physics. The opportunity to write as well as look into topics of our own personal interest was also a wonderful aspect to the course. I am very grateful to them both for sitting on this committee.

Coming to a huge department within a huge university was jarring given my small college origins. The older graduate students: Samet Demircan, Artemis Giannakopoulou, Henry Klest, and Farid Salazar were all tremendously caring and helpful as questions and/or struggles arose in the program. While graduate school is very hands-off, there is help around the corner should you ask — Professor Matt Dawber, Professor Jac Verbaarschot Professor Derek Teaney, and Donald Sheehan were all there to answer questions and provide tips when needed.

As someone with an obsession for details, I was never one to do homework with friends as I would feel bad dragging out processes that other did seemingly trivially. Going it alone is simply not possible in a program like this. More than anyone else, I have Greg Suczewski to thank for constant camaraderie in all aspects of our first year together. Moreover, I'm thankful for the other 1st/2nd year classmates and friends that created an environment of helpful collaboration and support: Yaman Sanghavi, Bharath Thotakura, Chase Wallace, Edoardo Buonocore, Aswin Parayil Mana, Will Bidle, and Sabina Sagynbayeva.

Last but not least: I'm thankful to significant others and family for making all of this possible. Leaving the heaven of the Netherlands after the first few months of the covid era to pursue the program at Stony Brook was a hard decision, and moreover was one I would have failed at without your love and support over the first year here. With no end to covid lockdowns in sight, you continued to be the best friend I could have imagined even without a reunion on the horizon. I couldn't have done it without you Charley, thank you. And thank you, Gess. To have met you during such a pivotal transition point in my life has been such a boon. Your gregariousness, humor, and thoughtfulness have turned the word anxious into exciting. Thank you for unwavering support, unending fun, and perpetually flourishing love. I can't wait for what's next for us. Thank you — mom, dad, and Sean — your words of support, physical visits, care packages, and all the rest you've done for me these last two years has strengthened our relationships which I hadn't thought could get stronger.

\pagebreak

\section*{Outline}
\addcontentsline{toc}{section}{Outline}
\label{abstract}

We will begin this work by reviewing the gauging of the Ising model to obtain the toric code in Section \ref{spin1}. Then, following the chronological order for developments in theories of fractons, we will in Section \ref{spin2} establish the gauged spin model picture of gapped fractonic theories.

Next (after establishing the preliminaries of lattice gauge theory in Section \ref{LGT}) we will cover the developments on the tensor gauge theory front in describing gapless fractonic theories in Section \ref{TGT}. We then explain how conservation of dipole moment is key in the development of field-theoretic descriptions of fracton models in Section \ref{fft} and conclude by providing future directions in gauging such field theories in Section \ref{gaugetheory}.

Finally, we point out two more future directions in Sections \ref{inspriation} and \ref{rydberg}, the former being an established theoretical advance which in fact inspired this work, and the latter being an exciting experimental advance in ultracold atomic physics that could lead to the physical realization of fracton physics.

Aspects of this work that are (to our knowledge, after paying due diligence to the references of all works cited herein) novel  — in the sense that the main conclusions are stated without proof elsewhere — include the following:
\begin{enumerate}
\item the derivation in Section \ref{novel1} of the gauge field dynamics term in the gauging of a spin system to obtain the toric code,
\item the reduction of the 3 terms of the Kogut-Susskind lattice QED Hamiltonian to the 2 terms of the toric code Hamiltonian in Section \ref{return},
\item showing, in Sections \ref{gauss} and \ref{novel3}, that the invariance of a gauge field implying a corresponding Gauss’s law constraint in a tensor gauge theory context,
\item showing, in Sections \ref{novel4a} and \ref{charge}, that the conservation of dipole moment (in the gapless tensor gauge theory models of fractons) accounts for the immobility/“fractalization” phenomenology of fractons, as exemplified in the gauged spin model picture of gapped fracton theories, and 
\item showing that polynomial shift symmetries and higher-order spatial derivatives in the Lagrangian are necessary ingredients in order to have a field theory with dipole moment conservation in Chapter \ref{fft}.
\end{enumerate}

\pagebreak

\section*{Preface}
\addcontentsline{toc}{section}{Preface}

\begin{quote}
Allow me to explain the objective of this thesis in what I think is a pedagogically helpful manner — through the transparency of ``diary entries." The majority of these diary entries were originally either the introduction or outline for the independent study course this project began as. They will cover: 
\begin{enumerate}
\item the inception of the thesis when it was only the topic of an independent study course, 
\item the results of the work of the independent study course, as well as 
\item the questions that were roused during that work that lead to the plan and timeline for this thesis.
\end{enumerate}
A proper outline — what was accomplished in this thesis — appears above in the Outline and serves to supplement the table of contents.
A proper introduction follows immediately after the current preface in Chapter \ref{intro}.
    
\end{quote}

\begin{center}
    \rule{4cm}{0.05cm}
\end{center}

\textsc{Diary entry \# 1: the buzzword brainstorm (March 2021)}

Let me begin with a list of topics that I have encountered in the various introductions to papers on fractons, spin liquids, lattice spin systems, quantum error codes, and symmetry protected topological phases. Either through this project or in the future, I hope to concretely define all of these terms, procedures, and concepts:

\begin{itemize}
    \item ground state degeneracy and its relation to topological phases
    \item symmetry protected topological phases
    \item why fractons are not aptly described by current topological order or quantum field theory paradigms
    \item quasiparticles
    \item the progression of Landua theory $\rightarrow$ Landua-Ginzburg theory $\rightarrow$ topological order
    \item quantum error correcting codes, more specifically, stabilizer codes on lattices à la Kitaev's toric code \cite{kitaev}
    \item polynomial formalism of the Haah code  \cite{haahthesis} \cite{williamson} \cite{zijianthesis} (with particular emphasis on how it relates to the polynomial shift symmetry algebra of \cite{gromov})
    \item gauging the quantum Ising model to obtain the toric code
\end{itemize}

For the sake of time and due to the ``not knowing what to learn first" syndrome of beginning a new field, I will ``shut up and calculate" for some time before circling back here and using the introduction on the next page as description for what route the project has taken.

\begin{center}
    \rule{4cm}{0.05cm}
\end{center}

\begin{quote}
    Now that we have more to cover through this thesis, a new introduction is of course in order, see Chapter \ref{intro}. However the introduction from the independent study still serves as a suitable synopsis of the results of that project. It was indeed the last item above (``\textit{gauging the quantum Ising model to obtain the toric code}") that captured my attention most. I have a history with gauge theory at this point, in particular when ``gauge" is used as a verb, but only in the gravity setting \cite{thesis}. Understanding what gauging entailed in a condensed matter and quantum information setting became my focus.
\end{quote}

\begin{center}
    \rule{4cm}{0.05cm}
\end{center}

\textsc{Diary entry \# 2: seduced by gauge theory yet again (May 2021)}

A current hot topic bringing together the quantum information, condensed matter, and high energy theory communities is the study of systems know as fractons. These mobility-restricted excitations have been shown to touch on the following areas: topological phases, quantum information, gravity, and QFT dualities \cite{fractons} \cite{primarysource}   \cite{seibergshao}.

In the following work, after showing how gauging the transverse field Ising model can be viewed
as Kitaev's toric code, we explore a recent advance in lattice gauge
theory via gauging sub-system symmetries. This advance spawned the study of so-called fractonic excitations, and remains at the forefront of condensed matter, high-energy theory, and quantum information research. Along the way, we will also touch on the motivation for the toric code by viewing it as
QED on the lattice, and finally we will briefly discuss some further
directions for fracton research.

This work is based primarily on understanding the in's and out's of the following lecture series \cite{XieChenTalk}.

\begin{center}
    \rule{4cm}{0.05cm}
\end{center}
\begin{quote}

I gave a talk at the conclusion of the independent study which can viewed here \cite{talk}.

Returning to the independent study project as the subject matter of my Master's thesis was motivated primarily by a desire to understand a new breed of gauge theory that has become pivotal in the literature on fractons — tensor gauge theory. I found that many of the first examples in papers that developed this tensor gauge prescription for a theory of fractons would surround extending old notations of QED on the lattice. Of all the sections of the independent study, this was by far what I struggled with the most. Other than the original Kogut and Susskind papers, I couldn't find anything more pedagogically-minded to explain the Hamiltonian formalism of lattice gauge theory. That is, until finding exactly that in the ultracold atomic community — in papers that aimed to simulate lattice gauge theory in optical lattices. And so the route which my planned extension from an independent study project to a Master's thesis took the following form: I will split my time looking backwards and forwards. Looking backwards — I will use these newfound AMO papers to inform my understanding of the basics of lattice gauge theory and notions of QED on the lattice in the Hamiltonian picture of Kogut and Susskind. Then I will understand what the tensor gauge theory examples intend to reproduce in higher dimensions, and so looking forward — I will tackle that theoretical picture of fractons.
\end{quote}

\begin{center}
    \rule{4cm}{0.05cm}
\end{center}

\textsc{Diary entry \# 3: The gameplan (Jan 2022)}

Task 1: Update introduction with interest in fractons and other subsystem codes from a quantum info perspective
\newline
Timeline: December
\newline
Citations: \cite{terhal}

Task 2: Lattice gauge theory 
\newline
Timeline: January and February
\newline
Citations: \cite{kogutsusskind} \cite{kogut}  \cite{QSL} \cite{oleg} \cite{lgt1} \cite{lgt2} \cite{lgt3} \cite{lgt4}

Task 3: Tensor gauge theory
\newline
Timeline: March and April
\newline
Citations: \cite{primarysource} \cite{pretko} \cite{genEM} \cite{pretEM}

%Task 5: Update future directions with some new understandings of Gromov’s algebra formalism
%\newline
%Timeline: April
%\newline
%Citations: \cite{newkev}

Task 4: Polish thesis and create LaTex Beamer presentation for defense
\newline
Timeline: May
\newline
Successfully defended thesis needs to be sent to Professor Dawber by May 18
\begin{enumerate}
\item Send thesis to committee Friday April 29th (with Overleaf link to be privy to any finishing touches)
\item Defend thesis Friday, May 13th
\end{enumerate}

\pagebreak

\section{Introduction}
\label{intro}

Although Feynman's vision of a quantum computer (to him — a quantum device capable of faithfully simulating quantum systems of interest) came as early as 1982 \cite{feynman}, it was not until the mid 1990's that the second quantum revolution (in contrast to the first, the inception of quantum mechanics in the early 1900's) began due to the advent of quantum algorithms and quantum error correction schemes for quantum computing \cite{algor} \cite{shor} \cite{DiVincenzo} \cite{bennett}.

Against this background, Kitaev ushered in the beginning of implementing topology in quantum error correction \cite{kitaev} \cite{boundary}. We will return to Kitaev's so-called toric code throughout this work. For more on the role of topology in quantum information science, see \cite{danbrowne}. While from a quantum information  perspective topology could be described as a tool from which new platforms for qubits can be envisioned and crafted, from a condensed matter perspective topology is held in high regard because of its potential for a ``grand unified theory" picture of the phases of matter our world experiences. Beginning with the breakdown of the Landua symmetry breaking theory of phases of matter due to the fractional quantum Hall effect, it was known that a new way to classify materials was needed \cite{laughlin}. Topological order (degeneracy of a material's ground state classified by its topology) arose as a yardstick by which these exotic emergent quasiparticles could be arranged into different phases of matter \cite{wen90} \cite{stringnet} \cite{wen10} \cite{PEPS} \cite{ryu} \cite{wen17}. In fact it is the robustness of topological order which \textit{leads} to quantum information interest in topology — if a ground state has for example 4 distinct realities, i.e. there are 4 ground states, based on possible differing topologies that are protected from local perturbation, you now have 2 qubits (spin up and down) with protection from their environment.
\refstepcounter{dummy}
\label{here2}
The standard prescription for learning about the topological order of a material is investigating how its dynamics change when it is placed in a torus. A torus is 2-dimensional, things got weird when researchers began looking at 3-dimensional spaces. Notably, the classification in terms of topological tensor categories (that arise due to long-range entanglement in the ground state and describe the braiding statistics of the particles — and were used successfully used to enumerate all possible 1-dimensional gapped topological orders \cite{PEPS}) has not yet been able to bear any fruit in 3D \cite{18quanta}.

In addition to lacking a description in terms of previous topological order paradigms, 3D spaces also offer a a practical edge over 2D, since in 2D logical strings from error syndromes can perturb the topological state of the system unless the quantum computer utilizing the code is kept at extremely low temperatures. Haah set off in the 3D space code in search of a stable ground state for quantum memory \cite{18quanta}. What he found has changed what we know about phases of matter, topological order, and even quantum field theory. In an effort to eschew those problems of 2D code spaces, Haah enumerated all possible 3D codes that would fit the error-correction requirements of quantum computing. Of the codes that were absent of string error operators, the one with the most symmetry became known as Haah's code as news of the weird nature of the discovery spread
\cite{haah2011} \cite{haahthesis}  \cite{haah15} \cite{xcube}. \footnote{Alongside this novel phase of matter Haah brought to light, advancements continued in the quantum information community that pulled from the use of topology in condensed matter physics \cite{topquanmem} \cite{nayak}. Raussendorf and Fowler's overlap in tenure at the University of Waterloo and Perimeter Institute for Theoretical Physics (in Waterloo) would mark the start of practical applications in quantum computing through the surface code — generalizations of Kitaev's topological quantum error correcting codes \cite{rauss} \cite{fowler}. Fowler and Martinis's seminal work at the University of Santa Barbara \cite{surface} \cite{first} led to the UCSB team being hired by Google in their task of reaching the era of quantum supremacy — when a quantum computer can reliably outperform the world's strongest classical supercomputer  \cite{firstcollab}. This was achieved to much critical and public acclaim after years of collaboration \cite{quantsupr}. First applications of the quantum supreme processor include cryptography, quantum optimization, and quantum chemistry \cite{aaronson} \cite{use1} \cite{use2}. }

The weirdness of Haah's code stems from subsystem symmetry\footnote{IBM's competing surface code processor uses hexagonal honeycomb lattice surface codes, with Fowler leading Google's own work on subsystem honeycomb lattices, and Haah leading Microsoft's honeycomb code work \cite{honeycomb1} \cite{honeycomb2} \cite{honeycomb3}. That is to say, quantum supremacy is not the endgame, subsystem codes like honeycombs or fractons are at the frontier of the field \cite{fracapp1} \cite{fracapp2}.}, a concept we will dive deeper into below in Section \ref{spin2}. The consequence of enforcing various symmetries across a system is immobile particles — immobile insofar that the only way for them to move around is either by letting up on a subset of the symmetries (for instance moving along a plane or line) or in the even stranger case, by creating more particles \cite{fractons} \cite{primarysource}. The latter method of ``mobility," has led to these emergent excitations being labeled fractons (in accordance with the phonon, anyon, axion, exciton, etc. tradition) due to the fractal nature of their propagation. 
\refstepcounter{dummy}
\label{here3}
This fractal behavior leads to the two most rigorous descriptions of condensed matter systems, topological order and quantum field theory, failing to account for these systems' dynamics. Because of the exponential growth of the number of ground states on this fractal lattice, the notion of topological invariant ground states goes out the window — for exactly what differentiates topology from geometry is that the topology of a 100 meter doughnut is the same as the topology of a 10 centimeter coffee cup. This M.C. Escher fractal landscape also eludes a quantum field theory description based on a continuum approximation of a lattice, since that approximation will never hold. Significant effort is being funneled from high-energy theory and condensed matter theory into resolving this quantum field theory dilemma, leading even to using supersymmetric quiver gauge theory and string theory to study fractons \cite{qft} \cite{seibergshao} \cite{shao2} \cite{chamon2} \cite{quiver} \cite{branes} \cite{distler}.

It terms out that, when Haah and his collaborators made more rigorous the framework of his code in 2015, systems displaying these behaviors were observed in a work concerning exactly solvable spin models a decade earlier by Chamon \cite{chamon}. These models, the toric code for example, are called gapped — their ground state is separated in energy from excited states. Gapped fractons, like Haah's original code, as well as he and his collaborator's ``X-cube" model, are suitably described by the spin model picture of lattice gauge theory and quantum codes \cite{xcube}.
The models that perplex high-energy theorists and gravity researchers alike are gapless, where a particle picture is not possible, and revolutionary work in tensor gauge theory was needed before any description of these phases was fathomable \cite{pretko} \cite{genEM} \cite{pretEM}. The main insight of this model of fracton physics is the effect of upgrading the gauge connection in a gauge theory. When the gauge connection is a vector, as is the $A_\mu$ vector potential of the standard treatment of electrodynamics, the conservation law that arises tells us that total charge is conserved. In this case, isolated positive or negative charges are free to roam around space. It turns out that when the gauge connection is tensor, $A_{\mu\nu}$ for instance, higher order conservation laws hold — notably a conservation of dipole moment. This leads to dipoles being able to roam around, but immobilizes single charges, hence a picture of the immobility of fractons emerges. Moreover, although it is beyond the scope of this work, it has been shown that these tensor gauge theory models of gapless fracton phases can be Higgsed down to resemble the gapped models as well \cite{higgs1} \cite{higgs2}.

\pagebreak

\section{Gauging spin models: toric code}
\label{spin1}

\subsection{Intro to the toric code}
We begin with the coverage of Kitaev's toric code \cite{kitaev} in the following lecture series \cite{XieChenTalk}.

The model has a qubit, or rather a spin-$\frac{1}{2}$ degree of freedom, on each link between vertices on the lattice as follows.

\begin{figure}[H]
\centering
\includegraphics[scale=0.5]{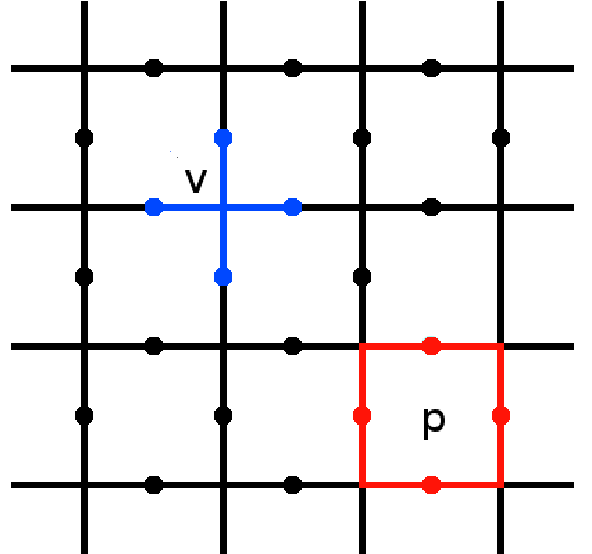}
\captionsetup{format=hang}
\caption[Physical set up for the toric code]{The ``physical" set up for the toric code: qubits placed on the edges of a lattice system. A particular qubit would be ``in" the vertex v or the plaquette p if it is in the respective highlighted region.}
\end{figure}

The Hamiltonian for the model is given as 

\begin{equation}
\label{toric}
    H = - \sum_v A_v - \sum_p B_p,
\end{equation}

where the vertex operator $A_v$ is a tensor product of Pauli x matrices
\begin{equation}
    A_v = \prod_{i \in v} \tau_i{}^x,
\end{equation}

and the plaquette operator $B_p$ is a tensor product of Pauli z matrices

\begin{equation}
    B_p = \prod_{i \in p} \tau_i{}^z.
\end{equation}

Investigating the ground state situation is the first step, since ground state degeneracy is linked to topological order.

What is the ground state? Instead of looking at how we can minimize the total energy (full Hamiltonian) we can aim to minimize the energy due to just each term in the Hamiltonian (vertex and plaquette operators), since the terms commute with one another.

The (energy) eigenvalues of the vertex and plaquette operators are clearly $\pm 1$ as they are composed of Pauli matrices. Now because of the minus signs in the Hamiltonian, it follows that the ground state (lowest energy state) is one with eigenvalues $+1$ so that the energy is negative.

Thus we will call states with $+1$ eigenvalue unoccupied, and states with $-1$ eigenvalues occupied since they increase the energy. So if we want to the vertex operator to (in total) have a $+1$ eigenvalue, the Pauli $\tau^x$ operators in the product can either all four being $+1$ (unoccupied), all four being $-1$ (occupied), or two being each $+1$ and $-1$ represented respectively by a,b, and c in the figure below.

\begin{figure}[H]
\centering
\includegraphics[scale=0.8]{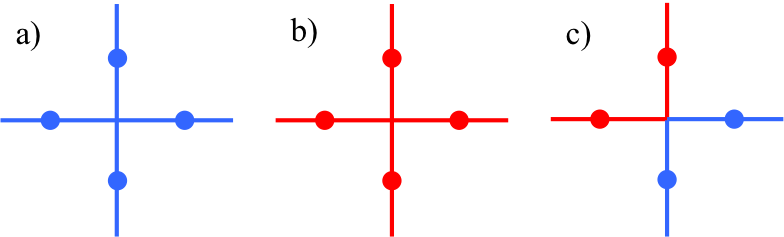}
\captionsetup{format=hang}
\caption[Spin orientations that lead to ground state of vertex operator]{Possible orientations of unoccupied (blue) and occupied (red) sites about a vertex that yield a total $+1$ eigenvalue for the vertex operator.}
\end{figure}

The key take away from this analysis is that none of these configurations involve (un)occupation ending at a vertex. And so we can say that the energetically favored configurations of (un)occupation for the vertex operator involve closed loops, since those would disallow any string of (un)occupation to end at a vertex.

Moreover, let's take note of what the the plaquette operator does. It either: creates a loop by turning unoccupied sites around a plaquette to occupied, flips a loop by turning occupied sites into unoccupied, moves a loop, or enlarges a loop. And so it is clear that loops are energetically favorable to both the vertex and plaquette operators, and we can use this information to note what the ground state wavefunction looks like: simply all possible loop configurations all the way from no loops, to multiple loops of differing sizes/overlap patterns.

The model becomes more complex when we place the lattice on a torus (by identifying both the top and bottom and the left and right sides). The effect of this is that the plaquette operator cannot complete contract some loops down to nothing as it could in the previous set up. There exist non-trivial loops around the torus that cannot be contracted to nothing because of the genus of the torus.

\begin{figure}[H]
\centering
\includegraphics[height=6cm,width=17cm]{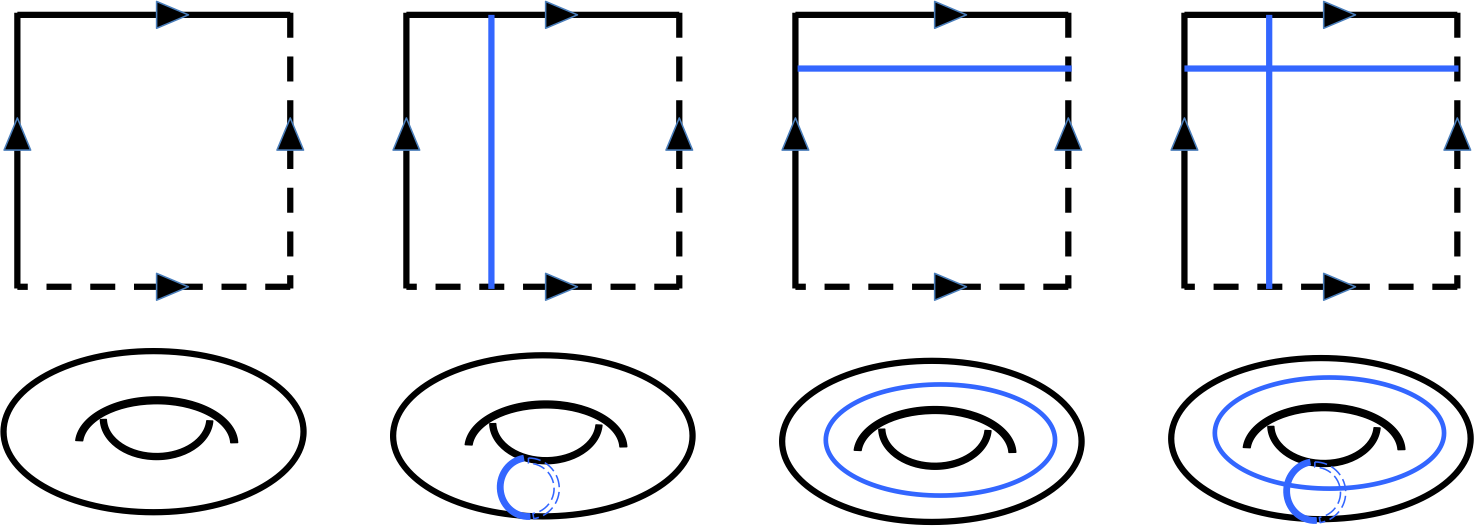}
\captionsetup{format=hang}
\caption[Topology of loop operators around a torus]{The four distinct setups for the model on a torus. Loops around the torus cannot be contracted with the plaquette operator because of the non-trivial topology of the torus. There are not more than four because two radial or transverse loops would enable the plaquette operator to be continually applied until the two loops were eliminated.}
\end{figure}

This leads to fours distinct ``universes" for the ground state wavefunction, each configuration has its own groundstate wavefunction (still constructed in the same way as before, all possible loop combinations). The question of ground state degeneracy (GSD) can seemingly now be answered, there is four-fold degeneracy corresponding to the four distinct setups above, but is defined more technically as follows \cite{QSL}.

\begin{enumerate}
    \item $\prod_v A_v=\prod_p B_p=1$ since every $\sigma_i^{x/z}$ would appear twice and we know that $(\sigma_i^{x/z})^2=\mathbb{1}$
    \item for a lattice with N sites, there are N-1 independent choices of eigenvalues for the vertex and plaquette operator,
    \item there are $2^{2N-2}$ specific choices for eigenvalues of both operators,
    \item there are 2 bonds per site and so there are $2^{2N}$ spin states,
    \item GSD $= \frac{\text{ spin states}}{\text{specifications for eigenstate}}=\frac{2^{2N}}{2^{2N-2}}= \frac{4^N 4}{4^N}=4$ 
\end{enumerate}

This can be extended to lattices on higher genus $g$ surfaces according to GSD $= 4^g$ \cite{haahthesis}.

This GSD is topological, i.e. the degeneracy is not broken by perturbing the Hamiltonian. To see why, consider the following argument.

Let $|\psi_1\rangle$ and $|\psi_2\rangle$ be two degenerate states such that 

\begin{eqnarray}
H |\psi_1\rangle &=& E |\psi_1\rangle, \nonumber \\
H |\psi_2\rangle &=& E |\psi_2\rangle.
\end{eqnarray}

If we perturb the Hamiltonian by applying a $\tau^x$ at each site with a small perturbing parameter $\epsilon$,

\begin{equation}
    H = H_0 +\epsilon \sum \tau^x,
\end{equation}

most of the terms one would encounter in the standard degenerate perturbation theory equations (see section 5.2 of \cite{sakurai}) would be zero since 

\begin{equation}
    \langle \psi_1 | \tau^x |\psi_2\rangle =0.
\end{equation}

However, a string of $\tau^x$ operators across the whole system would map the state to each other

\begin{equation}
    \langle \psi_1 | \prod \tau^x |\psi_2\rangle = 1.
\end{equation}

Thus the term in the perturbation would read

\begin{equation}
    \langle \psi_1 | \epsilon^L \prod \tau^x |\psi_2\rangle \propto \epsilon^L,
\end{equation}

where L is the length of the system.
If we then consider the thermodynamic limit (the system is large, i.e. $L\rightarrow \infty$) then the $\epsilon^L$ is exponentially small and can be ignore. Thus no perturbation can break the GSD, and so it is topological. 

Now we look at excitations, or equivalently violations of vertex and plaquette terms leading to quasiparticles.

We call the the violation of the vertex operator $A_v$ a charge excitation, $e$, and it consists of a string of $\tau^x$ operators. Similarly we call the violation of the plaquette operator $B_p$ a magnetic flux excitation, $m$, and it consists of a string of $\tau^z$ operators. The motivation for these names will come from the gauging procedure to follow, as well as the quantum electrodynamics motivation for the form of the toric code Hamiltonian that we will establish in Section \ref{return}.

\pagebreak
\subsection{Gauging the transverse field Ising model to get the toric code}
For this discussion we still use 
\cite{XieChenTalk} as the main source material, but the lecturer's paper
\cite{shirleyslaglechen} and the original source material
\cite{kogut} are also consulted. The general convention for what follows is that $\tau$ stand for gauge fields and $\sigma$ stand for matter fields.

When no external field the Ising model (spin degrees of freedom on the vertices) Hamiltonian reads

\begin{equation}
    H_{\text{Ising}} = -\sum_{<vw>} J_{vw}\sigma_v{}^z \sigma_w{}^z,
\end{equation}

where $v$ and $w$ are two different vertices.

If an external field is present, a term is added and the transverse field Ising (TFI) model reads

\begin{eqnarray}
\label{almost}
    H_{\text{TFI}} &=& -\sum_{<vw>} J_{vw} \sigma_v{}^z \sigma_w{}^z - \sum_v h_v \sigma_v{}^x \nonumber \\
    &=& -\sum_{<vw>} J_{vw} \sigma_v{}^z \sigma_w{}^z - \sum_v  \sigma_v{}^x,
\end{eqnarray}

where we let the external field be uniform, $h_v=1$ $\forall$ $v$.

In the limit $J\ll1$ the transverse field dominates and there is a symmetric phase. However in the limit $J\gg1$ there is a symmetry breaking phase, and moreover there is a GSD — either all spins are up or all spins are down. Thus the GSD is two-fold, and the global symmetry is a $\mathbb{Z}_2$ symmetry, represented by the following unitary operator

\begin{equation}
    U = \prod_i \sigma_i{}^x,
\end{equation}

where the conservation of the value of a measurement of the this operator is guaranteed by the Heisenberg equation of motion (for the time-independent operator $U$) \cite{sakurai}

\begin{equation}
\label{heis}
    \frac{d}{dt} U = \frac{1}{i \hbar} \left[U, H\right]
\end{equation}

The commutator $\left[U, H\right]$ is indeed zero (as we will prove following equation \ref{QIcom} once the proper Pauli operator algebra has been discussed), thus by equation \ref{heis}, the measurement of $U$ is conserved, and by Noether's theorem, corresponds to a symmetry of the theory \cite{noethercite}.\footnote{Noether's theorem will be more formally treated in Section \ref{noether}.}

Gauging this theory (coupling it to a local $\mathbb{Z}_2$ gauge field) amounts to:

\begin{enumerate}
    \item connecting nearest neighbors via gauge fields between the lattice sites, i.e. we place gauge fields $\tau$ on the edges, while the matter fields $\sigma$ are on the vertices,
    \item obeying Gauss's law in the local symmetry — integrating around a vertex by drawing a circle around the 4 gauge fields along the 4 edges connected to a vertex tells us how much charge the enclosed matter field has, and finally
    \item enforcing a no-flux condition for any particular plaquette in the lattice.
\end{enumerate}

Regarding the first step, the 2nd term in $H_{\text{TFI}}$ is symmetric under even the local symmetry, but the 1st term needs to be modified to respect the local symmetry. The 1st term becomes

\begin{equation}
    - \sum_{<vw>} J_{vw} \sigma_v{}^z \tau_{<vw>}^z \sigma_w{}^z
\end{equation}

Step 2 involves bringing in a term into the Hamiltonian that accounts for the local symmetry, or Gauss's law — which amounts to an equivalence between integrating through the gauge fields on the edges around a vertex and the charge on the vertex. This is represents mathematically as

\begin{equation}
\label{local2}
   A_v{}'= \sigma_v{}^x \prod_{i \in v} \tau_i{}^x
\end{equation}

At this point, the TFI Hamiltonian, $H_{TFI}$ in equation \ref{almost} has become,

\begin{eqnarray}
\label{local}
 H_{\text{TFI}} \rightarrow   H_{\text{local}} &=& -\sum_{<vw>} J_{vw} \sigma_v{}^z \tau_{<vw>}{}^z \sigma_w{}^z - \sum_v  \sigma_v{}^x - \sum_v A_v{}'  \nonumber\\
    &=& -\sum_{<vw>} J_{vw} \sigma_v{}^z \tau_{<vw>}{}^z \sigma_w{}^z - \sum_v  \sigma_v{}^x - \sum_v \sigma_v{}^x \prod_{i \in v} \tau_i{}^x 
\end{eqnarray}

These modifications make $H_{\text{TFI}}$ local, but the gauge fields themselves are contributing unwanted degeneracy to the theory because there is no term telling the gauge fields what to do by accounting for their dynamics. 

\pagebreak

\subsubsection{Enforcing a no-flux condition}
\label{novel1}
Step 3 is significant in that we made a mistake in the original derivation of a term that, as mentioned in the previous, tells the gauge fields what to do by accounting for their dynamics. 

\begin{quote}
    Before explaining how to properly think of this derivation, for the sake of posterity, we will include the incorrect derivation here before returning (with this indented quote LaTeX environment) to discuss the error in logic. \footnote{We thank Professor Tzu-Chieh Wei for pointing this error out during the independent study phase of the this project during a talk \cite{talk} (around the 12:55 mark) \cite{tzu}.}
\end{quote}

We do so by introduced a so-called ``no-flux" condition by placing $\tau^z$'s at all the edges. This is represents mathematically as

\begin{eqnarray}
\label{error}
  \prod_{i \in p} \sigma_i{}^z \prod_{i \in p} \tau_i{}^z &=& (\sigma^z)^4 \prod_{i \in p} \tau_i{}^z \nonumber\\
   &=& (\sigma^z)^2 (\sigma^z)^2 \prod_{i \in p} \tau_i{}^z \nonumber\\
    &=& (1) (1) \prod_{i \in p} \tau_i{}^z \nonumber\\
    &=& \prod_{i \in p} \tau_i{}^z \nonumber\\
    &=&  B_p 
\end{eqnarray}

we can see that this is the same plaquette operator $B_p$ from the toric code. (We will shift to the old $A_v$ from the toric code in a few pages. As it stands, it has a matter field $\sigma$ in it and is not the same $A_v$ from $H_{\text{toric code}}$.)

\begin{figure}[H]
\centering
\includegraphics[scale=0.7]{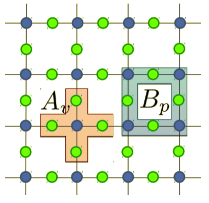}
\captionsetup{format=hang}
\caption[Vertex and plaquette operators of the toric code]{The $A_v{}'$ local symmetry/Gauss's law term and the $B_p$ no-flux plaquette term. Green dots on the edges are gauge fields $\tau$ and blue dots on the vertices are matter fields $\sigma$. Image from \cite{shirleyslaglechen}.}
\label{err}
\end{figure}

$A_v{}'$ and $B_p$ can be represented graphically as follows in Figure \ref{err}

Returning to our indented quote LaTeX environment to provide retrospective insight on the mistake above:

\begin{quote}
    The error of the above equation \ref{error} is due in part to Figure \ref{err} above and its ambiguity about BOTH matter and gauge fields residing in the highlighted regions of BOTH the $A_v{}'$ and $B_p$ terms.
    
    Nonchalantly following the precedence of the $A_v{}'$ term's derivation, where matter field and gauge field are both included: we erroneously did the same for the $B_p$ term.
    
    Armed with the figure above (from the reference \cite{shirleyslaglechen}) as well as (what we thought was the \textit{final} form of) the $B_p$ term in the source's caption of their figure (figure 2 in \cite{shirleyslaglechen}) — which seemed too good to be true, since the $A_v{}'$ term needed the ``integrating out the matter fields" procedure to truly be the $A_v$ of the toric code, whereas the $B_p$ was apparently already in it's final form? — we mistakenly ascribed the same matter field/gauge field presence to the $B_p$ term that we did the $A_v{}'$ term.
    
    The first way\footnote{The realization is due to conversation with Hiroki Sukeno \cite{Hiroki}.} to correct this mistake is to note that, by introducing gauge fields, we are privy to the procedures of gauge theory from a field theory perspective. Consulting pages 28 and 29, as well as Section 3.2.1 in our previous work \cite{thesis}, we can see one explanation for why the $B_p$ ought never to have matter fields in it, even before any integrating those fields out.
    
    \begin{displayquote}
the above Lagrangian ``...is invariant under local gauge transformations, but we have been obliged to introduce three [for SU(2)] new vectors fields $A_\mu{}^a$, and they will require their own \textit{free} Lagrangian..." — Griffiths page 364 \cite{griffithsParticles}
\end{displayquote}

\begin{displayquote}

``To complete the construction of a locally invariant Lagrangian, we must find a kinetic energy term for the field $A_\mu$: a locally invariant term that depends on $A_\mu$ and its derivatives, but not on $\psi$." — Peskin and Schroeder page 483 \cite{peskinschroeder}

... ``Using the covariant derivative, we can build the most general gauge invariant Lagrangians involving $\psi_i$. But to write a complete Lagrangian, we must also find gauge-invariant terms that depend only on $A_\mu{}^a$. To do this, we construct the analogue of the electromagnetic field tensor." — Peskin and Schroeder page 488
\end{displayquote}

\begin{displayquote}
``We can now immediately write a gauge invariant Lagrangian, namely [the above Lagrangian]
but the gauge potential $A_\mu$ does not yet have dynamics of its own. In the familiar example
of U(1) gauge invariance, we have written the coupling of the electromagnetic potential
$A_\mu$ to the matter field $\phi$, but we have yet to write the Maxwell term $-\frac{1}{4}F_{\mu\nu}F^{\mu\nu}$ in the Lagrangian. Our first task is to construct a field strength $F_{\mu\nu}$ out of $A_\mu$." — Zee page 255 \cite{zeeQM}
\end{displayquote}
    
    For instance, from Section 3.3 in \cite{thesis}, the field strength/curvature reads
    
    \begin{equation}
        F_{\mu\nu}{}^a = \partial_\mu A_\nu{}^a - \partial_\nu A_\mu{}^a +gf_{bc}{}^aA_\mu{}^bA_\nu{}^c
    \end{equation}
   
    and this object is used to construct the term in the Lagrangian accounting for the dynamics of the gauge field itself (what we are looking to do here in the $B_p$ term). Notice the field strength/gauge curvature above only depends on that theory's gauge field (and it's derivatives), but not the matter fields (such as the $\psi$ mentioned in the Peskin and Schroeder quote above).
    
    A pertinent example of this procedure is found in the QED section of Wikipedia's gauge theory article \cite{wikigauge} (also cited in \cite{thesis}) where the interaction term (our $A_v$) is formed from a minimal coupling of the current to the gauge field, but the ``gauge field" term in the Lagrangian (our $B_p$) is constructed from the field strength (only depending on the gauge field and its derivatives).
    
    \pagebreak
    
    The second way\footnote{The realization is due to conversation with Hiroki Sukeno \cite{Hiroki} and Kevin Slagle \cite{slagle}.} to correct this mistake is to note a very important property of the operators in the toric code we have totally neglected in our first study of this topic.\footnote{\label{foothere}And will return to haunt us in another mistake that was found during the ``polishing" of the slides for the defense of this thesis in Section \ref{return}.} The $A_v$ and $B_p$ terms commute $\forall$ $v$ and $p$. 
    \refstepcounter{dummy}
\label{where}

    This is stated in the sentence immediately following the defining of the $A_v$ and $B_p$ terms in equation 1 of the original source \cite{kitaev}, but is worked out fully in equations 2, and 3 of \cite{surface}. The commutation is trivially satisfied when there is no overlap between the $A_v$ and $B_p$ edges, but even in the non-trivial case, there are only 2 shared edges, and the commutation follows directly from the commutation of the Pauli gates that describe the qubits on the edges. 
    
    For completeness, we will write out the two pertinent equations from \cite{surface}.
    
    In equation 1, the authors of \cite{surface} outline some necessary properties of Pauli operators (see \cite{qiskit} for more on this nomenclature)
    
    \begin{eqnarray}
X \equiv \sigma_x / \tau_x &=& \begin{pmatrix}
0 & 1 \\
1 & 0
\end{pmatrix} \nonumber\\
Z \equiv \sigma_z / \tau_z &=& \begin{pmatrix}
1 & 0 \\
0 & -1
\end{pmatrix}\nonumber\\
\left[X_a,X_a\right] &=& 0 = \left[Z_a,Z_a\right] \label{foraside} \\
\{ X_a,Z_a \} &=& 0 \rightarrow X_a Z_a = -Z_a X_a \label{xzanti} \\
\left[X_a,Z_b\right] &=& 0 \rightarrow X_a Z_b = Z_b X_a \label{xzcom}
    \end{eqnarray}
    
    where $a$ is one qubit and $b$ another.
    
    In equation 2, the authors of \cite{surface} are considering a two-qubit system (labeled by $a$ and $b$ as above) being measured with two-qubit operators $X_a X_b$ and $Z_a Z_b$. 
    
    \begin{eqnarray}
    \label{QIcom}
        \left[ X_a X_b, Z_a Z_b \right] &=& X_a X_b Z_a Z_b - Z_a Z_b X_a X_b \nonumber \\ 
        &=& X_a Z_a X_b  Z_b - Z_a X_a Z_b  X_b \nonumber \\
        &=& (X_a Z_a) (X_b  Z_b) - Z_a X_a Z_b X_b \nonumber \\
        &=& (-Z_a X_a) (-Z_b  X_b) - Z_a X_a Z_b X_b \nonumber \\
         &=& Z_a X_a Z_b  X_b - Z_a X_a Z_b X_b \nonumber \\
         &=& 0
    \end{eqnarray}
    
    where we used equation \ref{xzcom} twice in the 2nd line, and equation \ref{xzanti} in the 4th line.\footnote{As promised in equation \ref{heis}, we are now privy to the equation \ref{QIcom} necessary to prove that $\left[U,H\right]$. The external field term in equation \ref{almost} for the transverse field Ising Hamiltonian ($H_{\text{TFI}}$) commutes with $U$ since both are the same type of Pauli operators. By equation \ref{foraside}, we know Pauli operators of the same type ($X$/$Z$) commute. The Ising coupled term of $H_{\text{TFI}}$ \textit{also} commutes with $U$ precisely as shown in equation \ref{QIcom}. For every choice of vertices for the two Ising coupled Pauli operators, the corresponding two Pauli operators from $U$ come together to form the exact commutator of \ref{QIcom}.}
    
    In equation 3 the authors of \cite{surface} consider the non-trivial case of 2 edges overlapping in the $A_v$ operator ($X_a X_b X_c X_d$) and the $B_p$ operator ($Z_a Z_b Z_e Z_f$)
    
    \begin{eqnarray}
        \left[ X_a X_b X_c X_d, Z_a Z_b Z_e Z_f \right] &=& X_a X_b X_c X_d Z_a Z_b Z_e Z_f - Z_a Z_b Z_e Z_f X_a X_b X_c X_d \nonumber \\
        &=& X_a X_b Z_a Z_b X_c X_d Z_e Z_f - Z_a Z_b X_a X_b Z_e Z_f  X_c X_d \nonumber \\
        &=& X_a Z_a X_b Z_b X_c X_d Z_e Z_f - Z_a X_a Z_b X_b X_c X_d Z_e Z_f \nonumber \\
        &=& X_a Z_a X_b Z_b X_c X_d Z_e Z_f - (-X_a Z_a) (-X_b Z_b) X_c X_d Z_e Z_f \nonumber \\
        &=& X_a Z_a X_b Z_b X_c X_d Z_e Z_f - X_a Z_a X_b Z_b X_c X_d Z_e Z_f \nonumber \\
        &=& 0
    \end{eqnarray}
    
    where we used equation \ref{xzcom} in the 2nd and 3rd lines;
    \newline
    and used equation \ref{xzanti} in the 4th line.

    And SO, if we included all 4 of the matter fields in the $B_p$ term as we erroneously did above, the gauge fields on the edges would still commute (based on the above calculation), however the matter fields on the shared corner would be left over and lead to a non-zero commutator between the two terms.
    
\begin{figure}[H]
\centering
\includegraphics[scale=1]{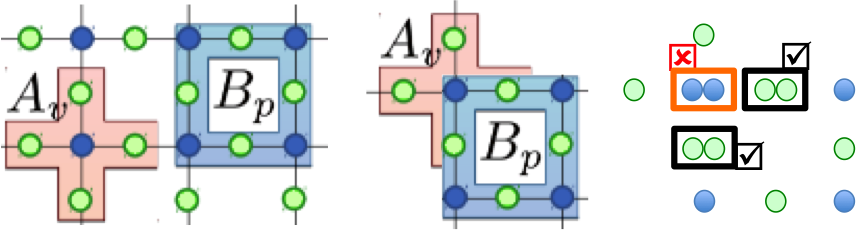}
\captionsetup{format=hang}
\caption[Non-commuting erroneous $B_p$ term]{If the $B_p$ term had the (blue) matter fields on all corners, than the shared corner would leave a non-zero commutator between the $A_v$ and $B_p$ terms in our erroneous prior construction. The (green) gauge fields on the 2 shared edges however would commute just as in the standard toric code. Leftmost figure is from \cite{shirleyslaglechen}.}
\label{conv}
\end{figure}
    
    This concludes our remedying of the $B_p$ term for the gauging procedure.

\end{quote}

So our full Hamiltonian (summing over all the vertices) that takes into account local symmetry (as we covered through arriving at $H_{\text{local}}$ in equation \ref{local}), AND accounts for the dynamics of the introduced gauge field (as we justified with the discussion of including the $B_p$ term above) now reads as follows,

\begin{eqnarray}
H_{\text{local}} \rightarrow  H_{\text{gauge}} &=& -\sum_{<vw>} J_{vw} \sigma_v{}^z \tau_{<vw>}{}^z \sigma_w{}^z - \sum_v  \sigma_v{}^x - \sum_v A_v{}' - \sum_p B_p  \\
    &=& -\sum_{<vw>} J_{vw} \sigma_v{}^z \tau_{<vw>}{}^z \sigma_w{}^z - \sum_v  \sigma_v{}^x - \sum_v \sigma_v{}^x \prod_{i \in v} \tau_i{}^x - \sum_p \prod_{i \in p} \tau_i{}^z  \nonumber
\end{eqnarray}

Before taking a limit to recover the toric code, a connection to the gauge theory of particle physics and such is feasible at this point.

The swapping of $\sigma_i{}^x \sigma_j{}^x$ for $\sigma_i{}^x \tau_{<ij>}^x \sigma_j{}^x$ in our gauging procedure is similar to swapping a normal derivative for a covariant derivative in gauge theory (see equations 2.4 - 2.8 in \cite{thesis}) and the inclusion of the plaquette operator $B_p$ to account for the dynamics of the gauge field is similar to the inclusion of the field strength for the same purposes (see pages 28 - 40 in \cite{thesis}). The inclusion of the Gauss's law is like obeying the geometry of your background.

The identification of this gauged TFI with the toric code comes from 2 final steps:

\begin{enumerate}
    \item taking the following limit for the Ising coupling: $J_{vw}=0$. This is physically the symmetric paramagnetic phase, i.e. there is no permanent magnetization due to non-interacting spins, and 
    \item ``integrate out" the matter fields, i.e. looking for when the energy of the matter fields $\sum_v \sigma_v {}^x$ is minimized (the matter fields $\sigma^z$ are taken out of the equation by the first step above). Minimized in this context means $\sigma_v {}^x \rightarrow 1$. This procedure is more standard in quantum field theory when looking into effective field theory limits for low energy \cite{lobus}.
\end{enumerate}

As a result of this 1st step, the $\sum_{<vw>} J_{vw} \sigma_v{}^z \tau_{<vw>}{}^z \sigma_w{}^z$ term goes to zero.

The 2nd step implies that the second term in the Hamiltonian becomes (noting that the lattice is finite, so the sum $\sum_v 1$ over the vertices converges to a constant $C$, which depends on the size of the lattice)

\begin{equation} 
\label{ref1}
 \sum_v  \sigma_v{}^x \rightarrow  \sum_v 1  \rightarrow C
\end{equation}

and since adding a constant to a Hamiltonian does nothing to the dynamics of the system, we can safely ignore the resulting $C$.

The 2nd step also implies that the third term in the Hamiltonian becomes

\begin{equation}
\label{ref2}
A_v{}' =  \sum_v \sigma_v{}^x \prod_{i \in v} \tau_i{}^x \rightarrow  \sum_v (1) \prod_{i \in v} \tau_i{}^x  = \sum_v A_v.
\end{equation}

where we realize that this results in none other than the vertex operator $A_v$ from our toric code — 4 $\mathbb{Z}_2$ objects on the edges around a vertex. 

And so we have reduced the gauged symmetric phase of TFI to the toric code

\begin{eqnarray}
    H_{\text{gauged TFI}} 
    &=& -\sum_{<vw>} J_{vw} \sigma_v{}^z \tau_{<vw>}{}^z \sigma_w{}^z - \sum_v  \sigma_v{}^x - \sum_v \sigma_v{}^x \prod_{i \in v} \tau_i{}^x - \sum_p \prod_{i \in p} \tau_i{}^z  \nonumber \\
    &=& - 0 - C - \sum_v  (1) \prod_{i \in v} \tau_i{}^x - \sum_p \prod_{i \in p} \tau_i{}^z  \nonumber \\
    &\approx& - \sum_v  A_v - \sum_p B_p  \nonumber \\
    &=& H_{\text{toric code}},
\end{eqnarray}

where we use $\approx$ to signify ``up to a constant" due to the $-C$ being neglected.

\pagebreak

\subsection{Interlude: gauge theory comments}

Some comments to add to the previous discussion on gauging:

\begin{itemize}
    \item introducing a gauge field doesn't change the dynamics of the matter field at least in the low-energy limit, i.e. there is no non-trivial flux.
    \item the charge of the gauge field comes from the symmetry charge of the system. For instance with a $\mathbb{Z}_2$ symmetry, the symmetry charge would read $0$ or $1$ and so the gauge charge would have the bosonic statistics $e \times e = \mathbb{1}$, since the $e$ particles are bosons. This occurs in the same way that the Aharanov-Bohm effect's phase between gauge charge and gauge flux is determined by the original global symmetry group.
    \item a flux excitation is one added in by hand. As two examples, for the symmetric phase of the transverse field Ising model of the last section, the flux excitation $m$ would be bosons, whereas if we were to look at a symmetry protected topological phase with a $\mathbb{Z}_2$ symmetry, the flux excitation $m$ would be anyons. In this way, coupling your new theory to a gauge field and looking at the $m$ excitation allows you the garner new information about your model.
    \item the Higgs mechanism is the reduction in gauge group as a result of spontaneous symmetry breaking, for example in the Standard Model — the electroweak SU(2) $\times$ U(1) breaks down to U(1), and in superconductivity — U(1) gauge group breaks down to $\mathbb{Z}_2$.
\end{itemize}

\pagebreak

\subsection{Further directions}

There exists a procedure to gauge at the level of the quantum states as opposed to Hamiltonians as we did here and Lagrangians as we do in particle physics \cite{gaugestates}. Motivated by the language of the above citation and the ubiquity of terms like projected entangled pair states, matrix product operators, tensor networks, density matrix renormalization group, and Affleck–Kennedy–Lieb–Tasaki states in condensed matter and quantum information, the following papers warrant their own sections in future work \cite{Weichselbaum} \cite{vidal} \cite{WilliamsonHaegemanSchuchVerstraete} \cite{wei}

\pagebreak

\section{Gauging spin models: fractons}
\label{spin2}

These theories were originally brought up in the quantum information communities in attempts to build quantum hard drives. Along this path, a future direction for the work is from the quantum error correcting code perspective. The following references are helpful in that regard
\cite{zijianthesis} \cite{terhal} \cite{haah2011} \cite{topquanmem}  \cite{surface} \cite{beginners} \cite{lauhman}.

A characteristic feature of fractonic models is the presence of excitations (such as e and m excitations from earlier) which cannot move, or can only move along a line. 

The most pressing questions about these models from a condensed matter perspective are: what kind of phases do they belong to, and what kind of physical mechanism describes this behavior?

Another characteristic feature of the models is subsystem symmetry, where the entire system doesn't have the same global symmetry, but rather certain parts of the system can have different symmetries. For instance, as shown below, the a 3-dimensional space may be sliced into many x-y planes and each may be characterized by different symmetry generators.\footnote{Only very recently, have subsystem symmetries of non-Abelian structure become to be investigated \cite{NA1} \cite{NA2} \cite {NA3} \cite {NA4} \cite {NA5}. Here we are considering to more solidly establish in the literature Abelian cases.}

\begin{figure}[H]
\centering
\includegraphics[height=10cm,width=12cm]{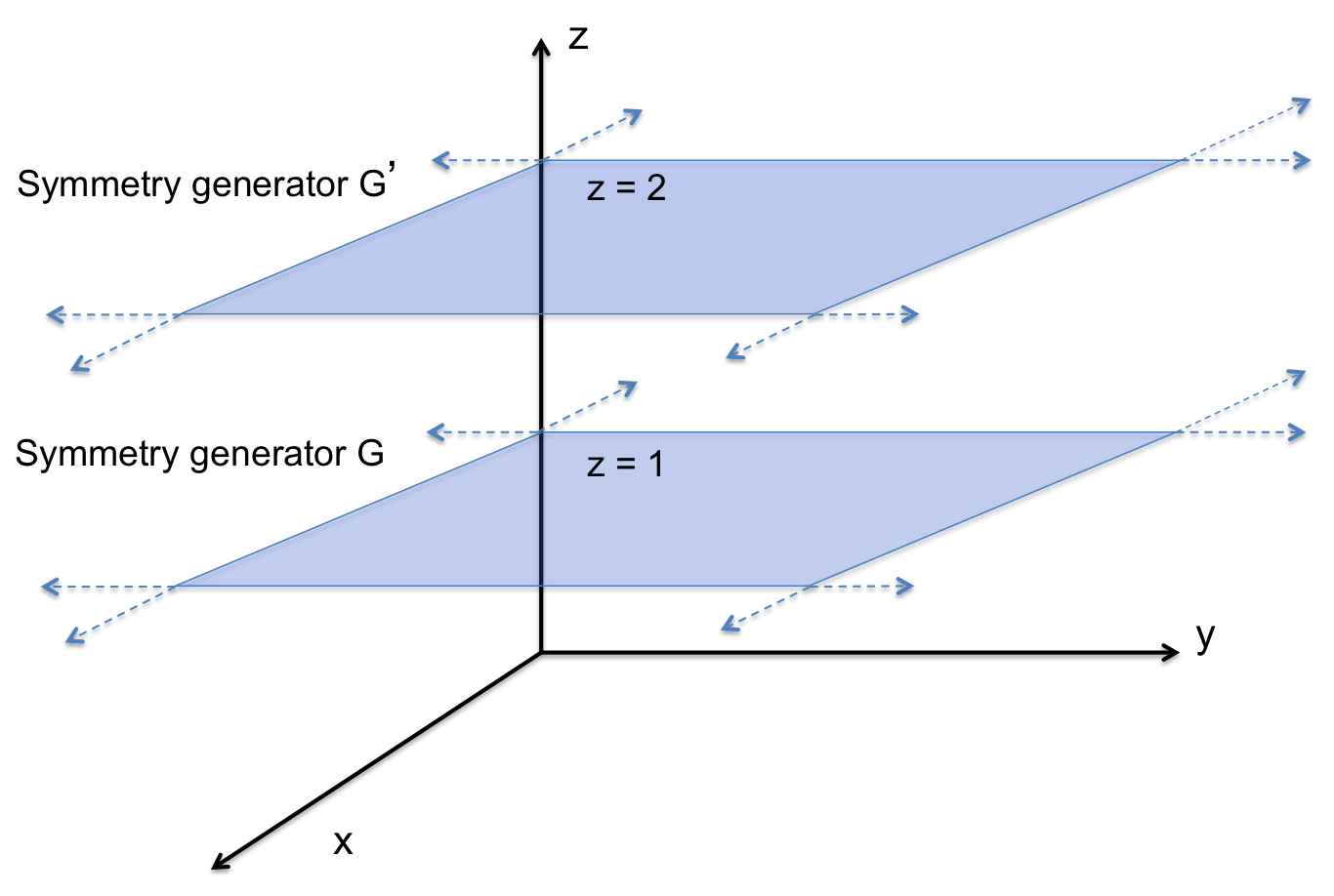}
\captionsetup{format=hang}
\caption[Subsystem symmetries]{Foliation of 3-dimensional space into various xy planes, each potentially with different sub-system symmetries in each plane.}
\end{figure}

\subsection{X-cube model}
\label{novel4a}
The first model of fracton excitations is coined the X-cube model, where qubits are places at every side around a cube. So for example around a single cube would be 12 qubits.

\begin{figure}[H]
\centering
\includegraphics[height=6cm,width=6cm]{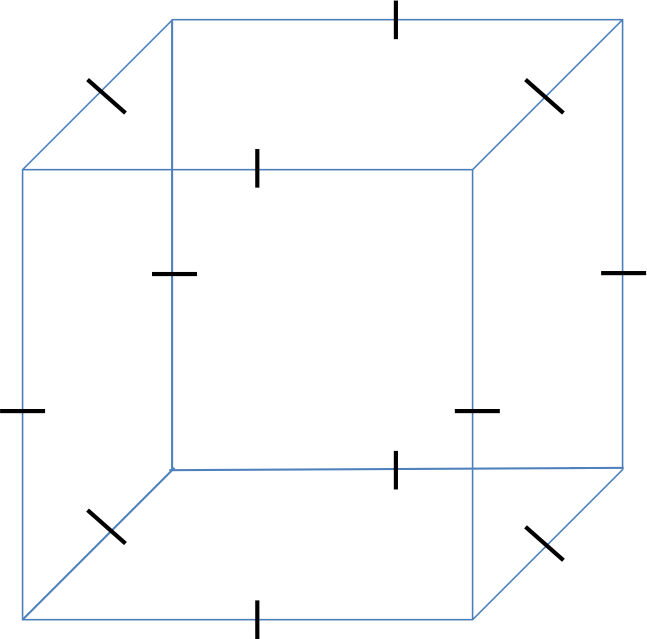}
\captionsetup{format=hang}
\caption[Physical set up for the X-cube model of fractons]{The X-cube model consists of qubits around a cubic lattice.}
\end{figure}

In hindsight, we know these qubits on the edges represent gauge fields.

\begin{figure}[H]
\centering
\includegraphics[height=4cm,width=16cm]{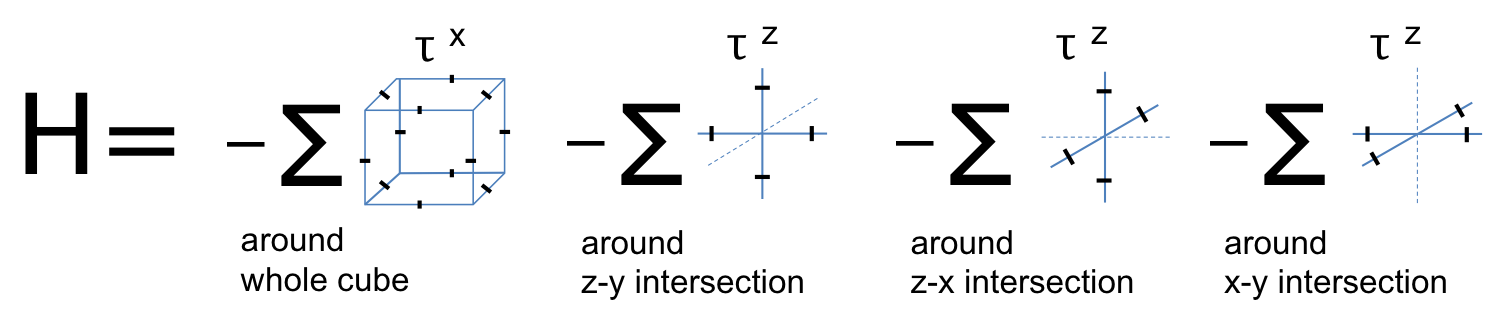}
\captionsetup{format=hang}
\caption[Pictorial Hamiltonian of the X-cube model]{The X-cube model consists of 12 $\tau^x$ around every edge of the cube, and 4 $\tau^z$ around a vertex in the zy, zx, and xy planes.}
\label{pictorial}
\end{figure}

The Hamiltonian for the X-cube can be depicted pictorially as in Figure \ref{pictorial}.

We will simplify this to the following now that the notation is clear,

\begin{eqnarray}
\label{AcBv}
    H &=& -\sum_{\text{cubes}} \prod_{\text{cube}} \tau^x -\sum_{\text{vertices}} \prod_{{\color{blue}z-y}} \tau^z  -\sum_{\text{vertices}} \prod_{{\color{green}z-x}} \tau^z -\sum_{\text{vertices}} \prod_{{\color{red}x-y}} \tau^z \nonumber \\
    &=& - \sum_c A_c - \sum_v B_v{}^x - \sum_v B_v{}^y - \sum_v B_v{}^z \nonumber \\
    &=& - \sum_c A_c - \sum_{v,\mu} B_v{}^\mu
\end{eqnarray}

where the colors will become clear in the next figure, and where $\mu = \{ x,y,z\}$. 

The definitions $A_c$ and $B_v$ are meant to illicit juxtaposition with the toric code's $A_v$ and $B_p$ and will become clearer in the next section.

As in the toric code example where we looked at Gauss's law on the lattice via excitations, we can see excitations from various terms in the Hamiltonian by placing a $\tau^x$ operator at a particular link on the lattice. Without loss of generality, we choose a link along the y-axis of the lattice. Between the 4 terms in the Hamiltonian, this operator will overlap with 3 of them: the $\tau^x$ on that corresponding side of the cube, the blue $\tau^z$ at the z-y intersection, and the red $\tau^z$ at the x-y intersection. Firstly, the cube term overlap will not cause an excitation since those two $\tau^x$ will commute, but there are two excitations from the $\tau^z$, and so those two excitations (represented by two red and blue dots in the figure below) can be moved along a particular direction along an axis of the lattice.

\begin{figure}[H]
\centering
\includegraphics[height=3cm,width=12cm]{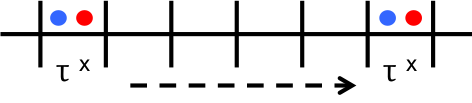}
\captionsetup{format=hang}
\caption[Lineon motion of fractons]{A $\tau^x$ operator placed on the y-axis will create 2 excitations due to the interaction with the $\tau^z{}_{{\color{blue}z-y}}$ and the $\tau^z{}_{{\color{red}x-y}}$ terms in the Hamiltonian.}
\label{lineons}
\end{figure}

Generalizing to all three axes, one can see that the only option for mobility is along a straight line, and moreover in pairs of excitations. The twin excitations traveling along the y-axis are blue and red, those along the x-axis are red and green, and those along the z-axis are blue and green. These excitations along one of the x, y, z-axes are called lineons. 

\begin{figure}[H]
\centering
\includegraphics[scale=0.4]{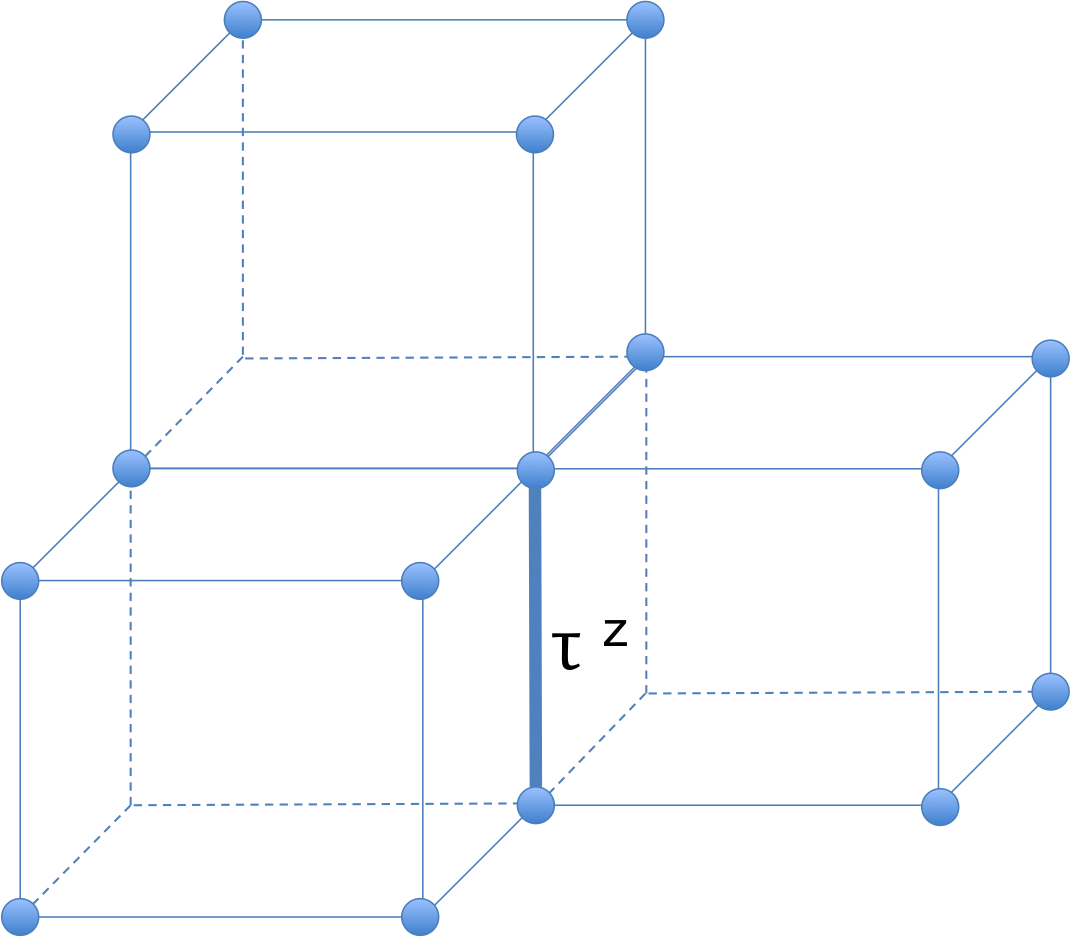}
\captionsetup{format=hang}
\caption[3D action of bit flip operator in X-cube model]{A $\tau^z$ operator placed on the z-axis will interact with the $\sum \tau^x{}_{\text{cube}}$ term in the X-cube Hamiltonian.}
\label{fig10}
\end{figure}

Fractons come when we consider excitations due to $\tau^z$. Applying a $\tau^z$ along a particular edge, as seen in Figure \ref{fig10}, leads to excitations along other edges due to the $\tau^z$ interacting with the $\tau^x$ from the $\sum \tau^x{}_{\text{cube}}$ term in the X-cube Hamiltonian. 

It is most helpful to analyze this behavior from the top-down perspective.

\begin{figure}[H]
\centering
\includegraphics[scale=0.6]{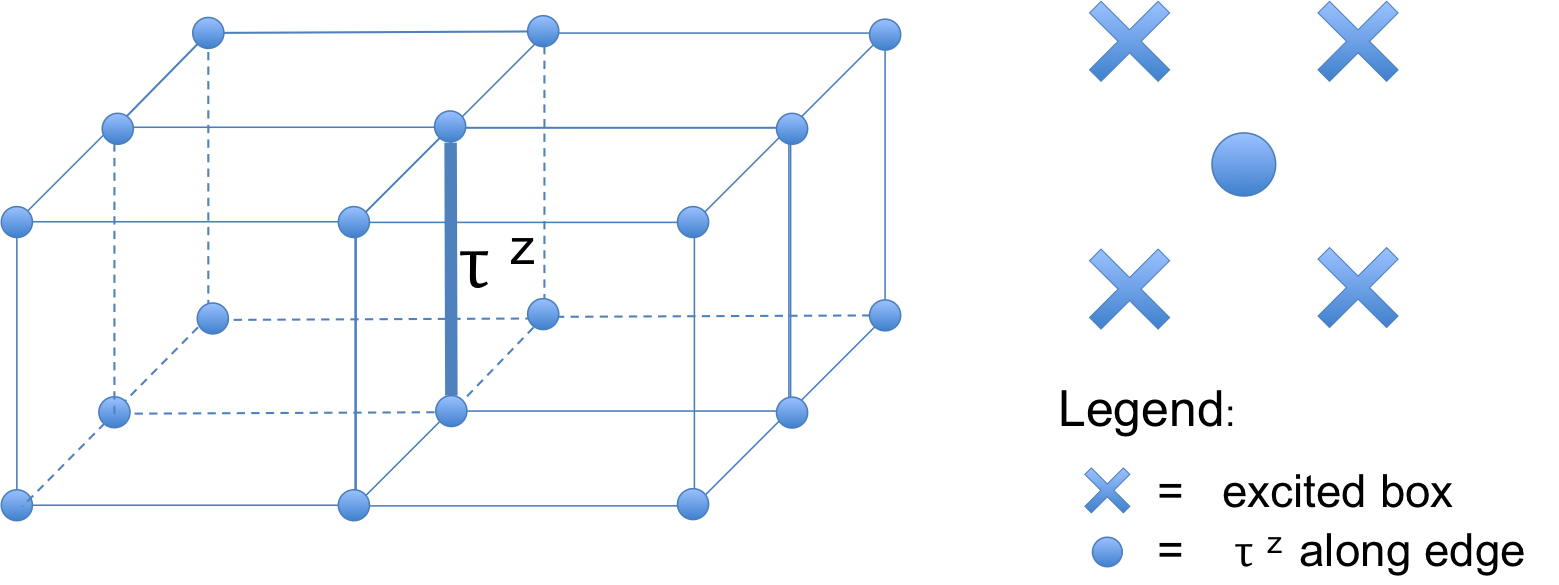}
\captionsetup{format=hang}
\caption[Top-down action of bit flip operator in X-cube model]{A 3D and top-down depiction of $\tau^z$ operator interacting with the $\sum \tau^x{}_{\text{cube}}$ term in the X-cube Hamiltonian.}
\end{figure}

\begin{figure}[H]
\centering
\includegraphics[scale=0.5]{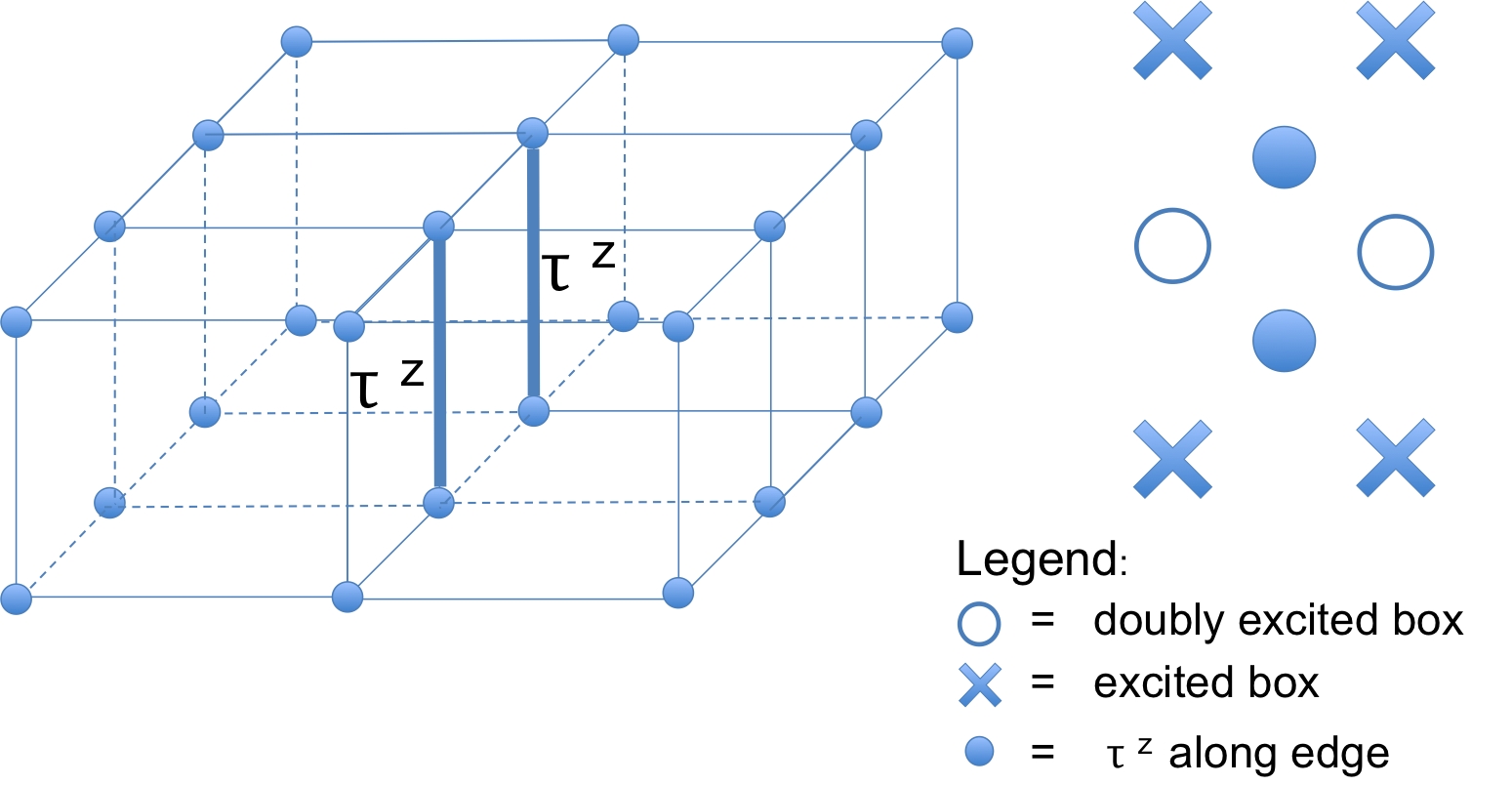}
\captionsetup{format=hang}
\caption[3D and top-down action of two bit flip operators in the X-cube model]{A 3D and top-down depiction of 2 $\tau^z$ operators interacting with the $\sum \tau^x{}_{\text{cube}}$ term in the X-cube Hamiltonian to produce excitations along the other edges of the cube that shared an edge with the perturbing $\tau^z$. Since more than 1 $\tau^z$ acts on 2 of the 6 cubes in this case, the qubit $\tau^z$ flips twice to return the same state as before for 2 of the 6 cubes.}
\label{fig12}
\end{figure}

More pronounced features of these fracton excitations can be seen when $\tau^z$ is applied along multiple edges, such as in Figure \ref{fig12}.

\begin{figure}[H]
\centering
\includegraphics[scale=0.5]{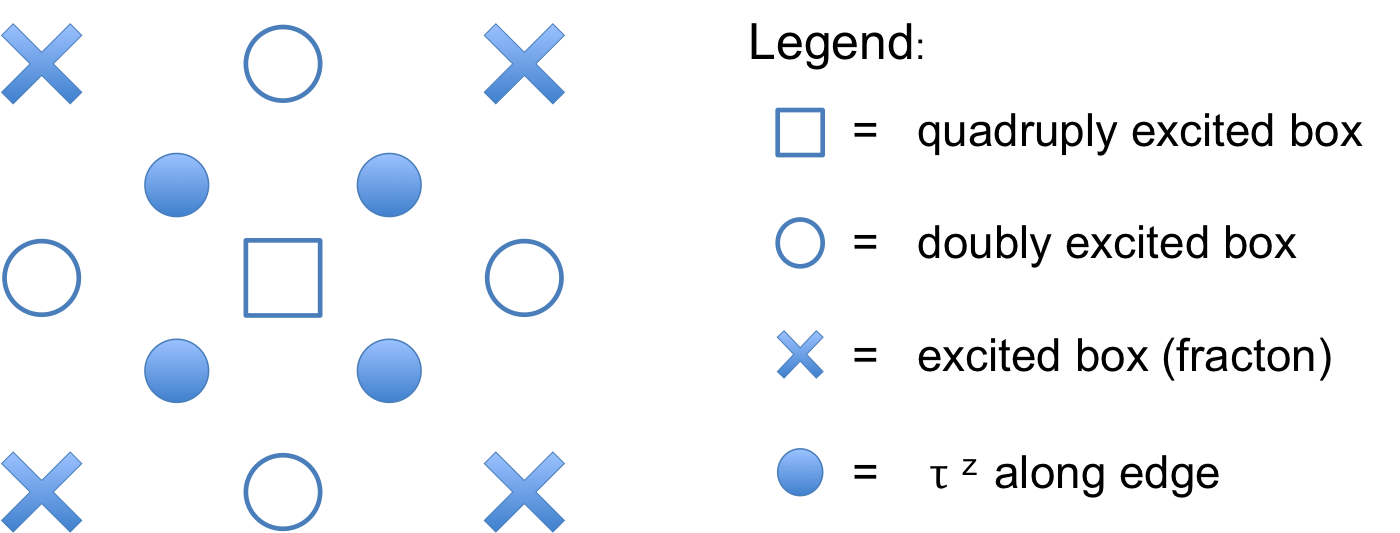}
\captionsetup{format=hang}
\caption[Top-down action of four bit flip operators in the X-cube model]{A top-down depiction of four $\tau^z$ operators interacting with the $\sum \tau^x{}_{\text{cube}}$ term in the X-cube Hamiltonian to produce excitations along the other edges of the cube that shared an edge with the perturbing $\tau^z$. Since more than 1 $\tau^z$ acts on 4 of the 9 cubes in this case, the qubit $\tau^z$ flips twice to return the same state as before, these are represented by un-flipped circle symbols. Moreover, 1 of the 9 cubes is flipped 4 times (remains in original state like in doubly-flipped states), this is represented by a square symbol.}
\label{fig13}
\end{figure}

Continuing with $\tau^z$ along multiple edges but only looking at the top down view for brevity, we can see four $\tau^z$'s act in Figure \ref{fig13}.

\begin{figure}[H]
\centering
\includegraphics[scale=0.4]{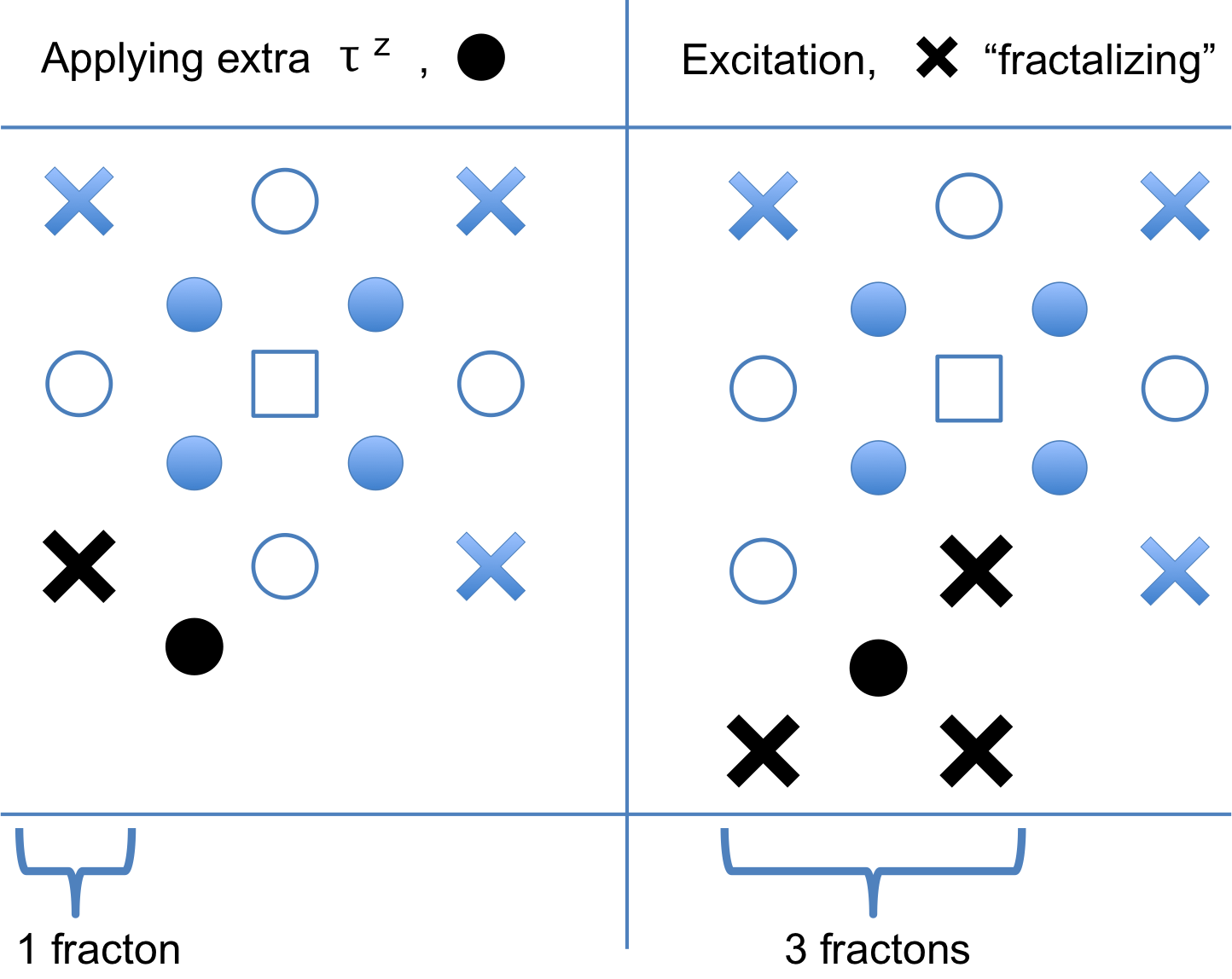}
\captionsetup{format=hang}
\caption[Fractalization resulting from attempting to move a fracton]{When we try to move a fracton (the black X on the left) by applying a $\tau^z$ (the black circle on the left), we simply fractalize it by making 3 copies on the fracton.}
\label{single}
\end{figure}

Now that we have some separation between the excitations (marked as X's in the top-down point of view in Figure \ref{fig13}) we can identify them as we did with the lineons above. These singular X excitations are called fractons. As opposed to the lineons, which can move along one of the axes, fractons cannot move. We can see this by applying another $\tau^z$ in the diagram above and noting that instead of moving, they ``fractalize" into 3 copies of themselves, as shown in Figure \ref{single}.

\pagebreak

\subsection{Gauging subsystem symmetries to get the X-cube model}

The starting point, before we introduce gauge fields anywhere in the model, is to place matter fields $\sigma^x$ at all the lattice points

\begin{equation}
    H_{\text{matter}}= -\sum_v \sigma_v{}^x.
\end{equation} 

Note the similarity with the second term in $H_{\text{TFI}}$ from equation \ref{almost}.

The system will have subsystem symmetries based on planes. The xy, yz, and zx planes will have their own subsystem symmetry, and these will be denoted by

\begin{eqnarray}
U_z &=& \prod_{v \in \text{x-y plane}} \sigma_v{}^x \nonumber \\
U_x &=& \prod_{v \in \text{y-z plane}} \sigma_v{}^x \nonumber \\
U_y &=& \prod_{v \in \text{z-x plane}} \sigma_v{}^x.
\end{eqnarray}

Notice that a symmetry charge has a particular subsystem/planar symmetry, when it is confined to move on that plane. As more subsystem symmetries are enforced, more restrictions on the motion occur. For instance, if 2 of the 3 subsystem symmetries are respected, then the charge is confined to the line of intersect between the planes. This is known as a lineon. And if all 3 symmetries are respected, we end up with an immobile symmetry charges like a fracton. 

\begin{figure}[H]
\centering
\includegraphics[scale=0.4]{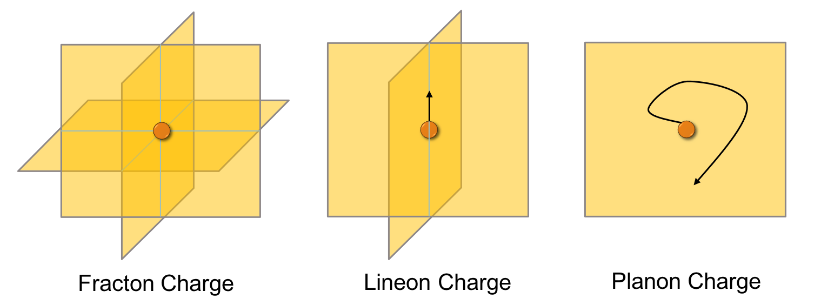}
\captionsetup{format=hang}
\caption[Motion of fractons, lineons, planons via enforcing subsystem symmetries]{Image from \cite{shirleyslaglechen} depicting different movement for particles given different sub-symmetries enforced. When we enforce all 3 (planar) subsystem symmetries on the matter field, the symmetry charge is immobile like the fractons of the X-cube model.}
\end{figure}

\begin{figure}[H]
\centering
\includegraphics[scale=0.6]{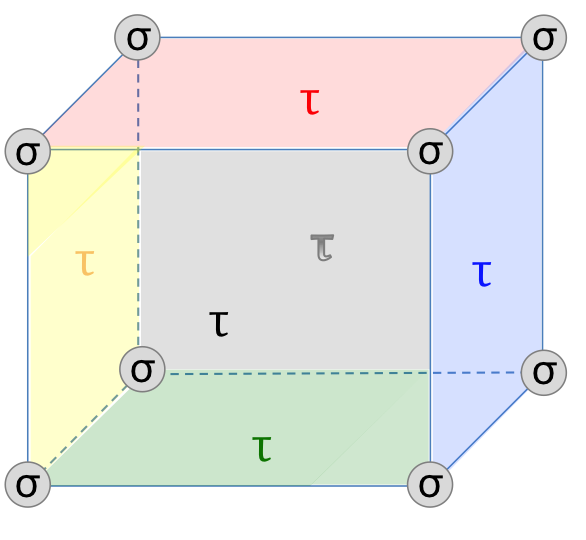}
\captionsetup{format=hang}
\caption[Placement of gauge fields due to enforcing subsystem symmetries]{When we enforce all 3 (planar) subsystem symmetries on the matter field, the gauge fields must reside equally along all axes — at the center of the plaquettes.}
\label{fig15}
\end{figure}

Moreover, to respect all the subsystem symmetries, the only possible locations for the gauge fields are the centers of the plaquettes. So representing the matter fields like $\sigma$'s and the gauge fields by $\tau$'s we have generically, as shown in Figure \ref{fig15},

\begin{figure}[H]
\centering
\includegraphics[scale=0.8]{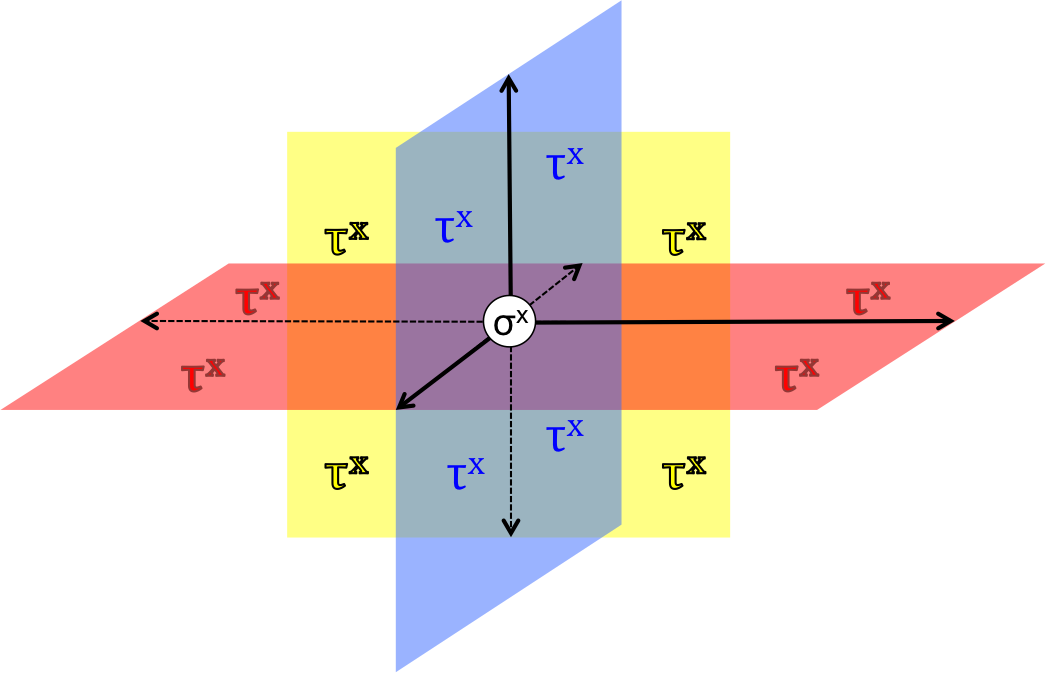}
\captionsetup{format=hang}
\caption[Gauss's law term on the vertices in the X-cube model]{Local symmetry/Gauss's law term in gauge Hamiltonian.}
\label{fig17}
\end{figure}

If we view the fracton confined by the 3 subsystem symmetries as a matter field on the lattice site, then the local gauge symmetry term in our gauged Hamiltonian can be written as the matter field, with the gauge fields around it. Because the matter field joins 4 cubes in 3 planes, there are 12 gauge fields surrounding one matter field as shown in Figure \ref{fig17},

represented mathematically by 

\begin{eqnarray}
    H_{\text{gauge}} &=& H_{\text{matter}} + H_{\text{local}} + H_{\text{flux}} \nonumber \\ 
    &=& -\sum_v \sigma_v{}^x - \sum_v \sigma_v{}^x \prod_{12} \tau^x + H_{\text{flux}}.
\end{eqnarray}

Note the similarity to the procedure for the toric code in equations \ref{local2} and \ref{local}.

The flux term in the Hamiltonian will be represented by influencing a no-flux condition as with the toric code, but with respect to all the subsystem symmetries. We will have 3 terms, where there is no flux exiting in the $\pm$ x, $\pm$ y, and $\pm$z directions. 

\begin{figure}[H]
\centering
\includegraphics[scale=0.8]{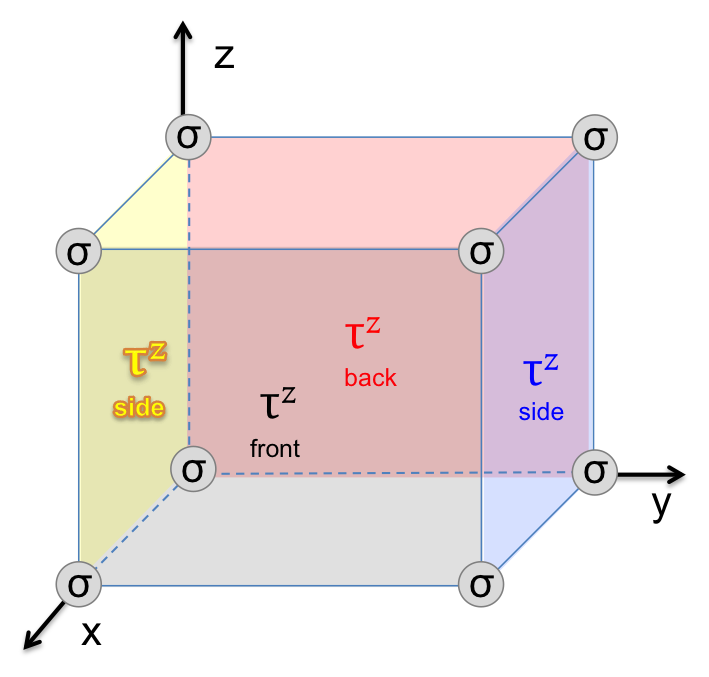}
\captionsetup{format=hang}
\caption[No flux (in a particular direction, z) term for the X-cube Hamiltonian]{No-flux gauge field term in the z-direction, where gauge fields are along the sides, front, and back, which have normal vectors in the x and y directions as opposed to z, so there is no z-flux.}
\label{fig18}
\end{figure}

We will call the no-flux term for the $\pm$ z direction 

\begin{equation}
    H_{\text{z- no flux}}= -\sum_p  \prod_{sides,front,back} \tau^z,
\end{equation}

where the sides, front, and back of the cube are denoted as follows in Figure \ref{fig18} (and the sum over p is short for sum over the plaquettes),

\begin{figure}[H]
\centering
\includegraphics[scale=0.7]{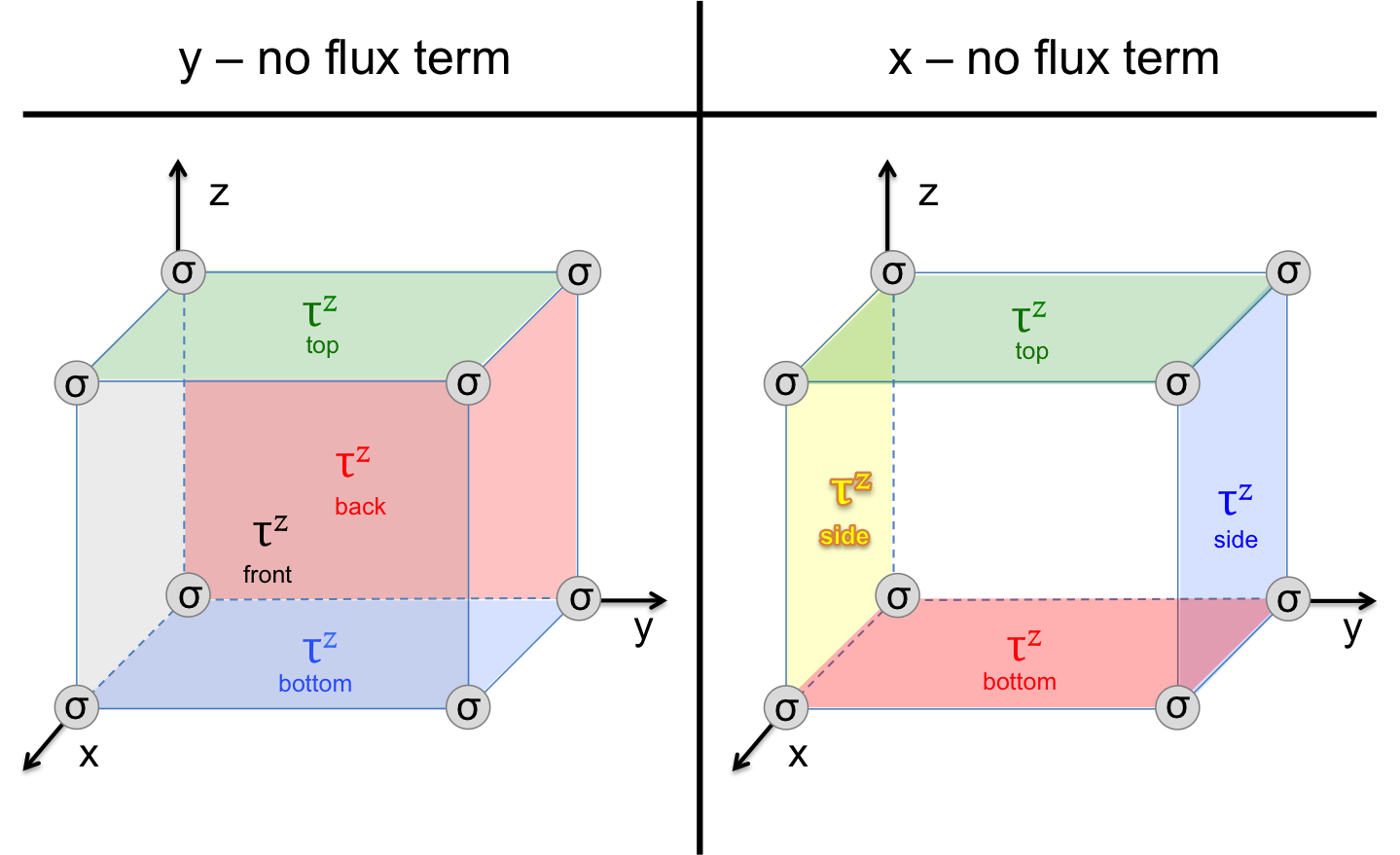}
\captionsetup{format=hang}
\caption[No flux (in x and y directions) term for the X-cube Hamiltonian]{No-flux terms in the y- and x- directions.}
\label{19}
\end{figure}

In the same way for the y and x no-flux terms, we have the following Hamiltonian terms:

\begin{eqnarray}
    H_{\text{y- no flux}} &=& -\sum_p  \prod_{top,bottom,front,back} \tau^z \nonumber \\
     H_{\text{x- no flux}} &=& -\sum_p  \prod_{top,bottom,sides} \tau^z,
\end{eqnarray}

which are represented graphically in Figure \ref{19}.

All together, the gauge Hamiltonian reads

\begin{eqnarray}
    H_{\text{gauge}} &=& H_{\text{matter}} + H_{\text{local}} + H_{\text{flux}}  \\ 
    &=& -\sum_v \sigma_v{}^x - \sum_v \sigma_v{}^x \prod_{12} \tau^x +  H_{\text{x- no flux}} + H_{\text{y- no flux}} +H_{\text{z- no flux}} \nonumber \\
    &=& -\sum_v \sigma_v{}^x - \sum_v \sigma_v{}^x \prod_{12} \tau^x  -\sum_p  \prod_{t , bo , s } \tau^z  -\sum_p  \prod_{t , bo , f, ba } \tau^z  -\sum_p  \prod_{s ,f ,ba } \tau^z, \nonumber
\end{eqnarray}

where ``s, f, ba, t, and bo" stand for ``sides, front, back, top, and bottom" respectively.

And now, just as we did in equations \ref{ref1} and \ref{ref2} for the TFI to toric code identification, we ``integrate out" the matter fields, i.e. looking for when the energy of the matter fields $\sum_v \sigma_v {}^x$ is minimized, i.e. we set $\sigma_v{}^x$ to $1$. The sentences below equation 6 of \cite{shirleyslaglechen} and above equation 16 of \cite{xcube} identify this ``integrating out" procedure with the zero-temperature phase of the Hamiltonian. 

The results of this are the same as with the toric code, 

\begin{equation} 
 \sum_v  \sigma_v{}^x \rightarrow  \sum_v 1 \rightarrow C
\end{equation}

and 

\begin{equation}
 \sum_v \sigma_v{}^x \prod_{12} \tau^x \rightarrow  \sum_v \prod_{12} \tau^x 
\end{equation}

Before making the identification with the X-cube model, a \textit {novel aspect} to this fractonic case that didn't appear in our work on gauging the toric code is shifting from a `gauge field on the plaquettes' to a `gauge fields on the edges' perspective by way of a dual lattice.

\begin{figure}[H]
\centering
\includegraphics[scale=0.7]{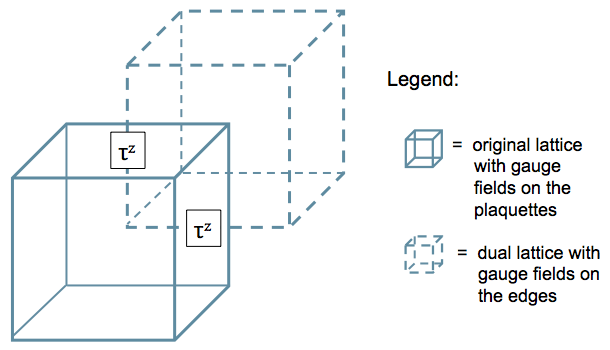}
\captionsetup{format=hang}
\caption[Dual lattice perspective of matter and gauge fields in the X-cube model]{Changing perspectives to a dual lattice, where now all the gauge field move from the plaquettes to the edges.}
\end{figure}

As a result, the 1st term in the remaining Hamiltonian is now simply gauge fields around the 12 edges of a particular cubic cell, just like the term in our original equation \ref{AcBv}.

\begin{equation}
\sum_{\text{vertex}} \prod_{12} \tau^x \rightarrow \sum_{\text{cube}} \prod_{12} \tau^x = \sum_{c} A_c
\end{equation}

\begin{figure}[H]
\centering
\includegraphics[scale=1.1]{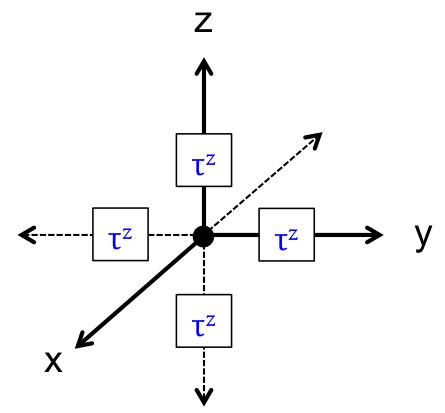}
\captionsetup{format=hang}
\caption[No flux term in the dual lattice perspective]{The no flux term in the x-direction in the dual lattice perspective. Consists of the 4 gauge fields on the edges attach to a particular vertex that are perpendicular to the no flux direction.}
\label{20}
\end{figure}

Similarly for the no-flux terms, for instance the no-flux in the x-direction term would become 

\begin{equation}
    \sum_p  \prod_{\text{top ,bottom ,sides }} \tau^z \rightarrow \sum_v \prod_{i \in \text{y-z plane}} \tau_v{}^z = \sum_v B_v{}^x
\end{equation}

from equation \ref{AcBv}. This no flux in the x-direction term can be visualized in Figure \ref{20} as follows

And thus, our final gauged Hamiltonian is precisely the X-cube model as shown in equation 1 and figure 1 of the review article \cite{fractons} based on the original work \cite{xcube},

\begin{eqnarray}
    H_{\text{gauge}} &=& H_{\text{matter}} + H_{\text{local}} + H_{\text{flux}} \nonumber \\ 
    &=& -\sum_v \sigma_v{}^x - \sum_v \sigma_v{}^x \prod_{12} \tau^x  -\sum_p  \prod_{t , bo , s } \tau^z  -\sum_p  \prod_{t , bo , f, ba } \tau^z  -\sum_p  \prod_{s ,f ,ba } \tau^z \nonumber \\
    &=& -C - \sum_v (1) \prod_{12} \tau^x  -\sum_p  \prod_{t , bo , s } \tau^z  -\sum_p  \prod_{t , bo , f, ba } \tau^z  -\sum_p  \prod_{s ,f ,ba } \tau^z \nonumber \\
    &\approx&  - \sum_v \prod_{12} \tau^x  -\sum_p  \prod_{t , bo , s } \tau^z  -\sum_p  \prod_{t , bo , f, ba } \tau^z  -\sum_p  \prod_{s ,f ,ba } \tau^z \nonumber \\
    &=& - \sum_c A_c - \sum_v B_v{}^x - \sum_v B_v{}^y - \sum_v B_v{}^z \nonumber \\
    H_{\text{X-cube}}&=& - \sum_c A_c - \sum_{v,\mu} B_v{}^\mu,
\end{eqnarray}

where again, the $\approx$ signifies ``up to a constant" due to the $-C$ being neglected.

This matches our equation \ref{AcBv}.

\begin{figure}[H]
\centering
\includegraphics[height=4cm,width=16cm]{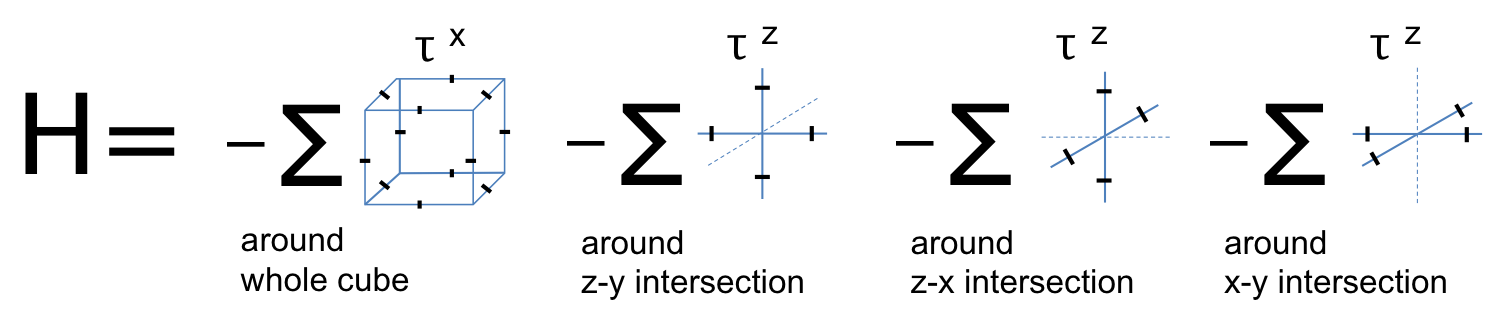}
\captionsetup{format=hang}
\end{figure}

\pagebreak

\section{Lattice gauge theory}
\label{LGT}

Gauge theory can be broadly described as using mathematical symmetry to explain the dynamics of physical systems. By the time Yang and Mills set out on their quest to study more advanced symmetries in the 1950's, it was known that Maxwell's theory of electromagnetism from nearly a century earlier was a gauge theory — the mathematical symmetry underlying it being the U(1) group \cite{ym}. As funding for particle colliders grew and Yang-Mills theory (the mathematical symmetry underlying their original theory being the SU(2) group) was extended, gauge theory became the foundation of one of the most rigorously tested theories of modern physics — the Standard Model (the mathematical symmetry underlying the model being the $SU(3) \times SU(2) \times U(1)$ group). The generic prescription for gauge theory in this context  goes as follows:

\begin{enumerate}
    \item a global symmetry is gauged, i.e. made into a local (gauge) symmetry,
    \item a gauge field (equivalently called a fiber connection or a Lie algebra-valued 1-form) is introduced
    \item a covariant derivative is introduced such that it transforms like the gauge field itself under the gauge symmetry, 
    \item the transformation of the gauge field is determined by requiring that the new derivative is covariant (or in more formal approaches, derived directly from the structure of the particular algebra),
    \item the gauge field's dynamics are accounted for in the Lagrangian by including a new term from the field strength (equivalently called a curvature) which is derived from the commutator of covariant derivatives (or again, directly from the structure of the Lie algebra)
\end{enumerate}

For the reader's own sense of chronology, the following chapter assumes a working knowledge of gauge theory at the level of the following pedagogic review article \cite{thesis}, which outlines the development of gauge theory in the language of Lagrangians, differential geometry, and symmetry algebras. For a more succinct version of that article see the corresponding talk \cite{thesis} and the original source material \cite{bergshoeff} \cite{Roelthesis}. For a more rigorous look at the gauge theory formalism developed there, see \cite{SUGRA}. It pays to point to that background since historically gauge theory was introduced in that context, and it cannot be argued that the theory is not simultaneously beautiful (see any Lie algebras and physics textbook for discussion on Murray Gell-Mann's \textit{Eightfold Way}) and remarkable successful (see the countless successes of the Standard Model of particle physics).

However, things are nowhere near complete. Symmetry only gets us so far. While the physicist's go-to tool of perturbation theory works most of the time to extend our knowledge of what is happening in a particular system, it inherently relies on small interactions to expand about. In particular, the theory of quantum chromodynamics — the last essential mathematical symmetry in the Standard Model being the SU(3) group — has regimes that are very strongly interacting and are thus deemed non-perturbative. Other techniques need to be employed to study these regimes. 

By confining quantum field operators to a lattice, Wilson brought about a revolution in quantum field theory by enabling both the high-energy and condensed matter communities to use methods of statistical mechanics to study strongly-interacting systems \cite{wilson1} \cite{wilson2}. Within five years, the computational power of the day was harnessed by Creutz \footnote{Residing just down the road from Stony Brook at Brookhaven National Lab.} and others to use Monte Carlo methods to do explicit calculations with Wilson's lattice gauge theory — verifying key features of quantum chromodynamics \cite{monte}.

The numerical power of Monte-Carlo methods for lattice gauge theories will of course continue as the power of classical computer processors increases\footnote{The Japanese Fugaku supercomputer at Riken took 1st place away from IBM's Summit supercomputer at Oak Ridge on the TOP500 list in 2020 \cite{top500}.}, and there are still innovations occurring in the field that lighten the computational intensity \cite{solve}. However, researchers have shifted their focus in recent years as the quantum information revolution has come to fruition and researchers from all fields of science see its potential that directly or indirectly modeling lattice gauge theories on computers could have. \refstepcounter{dummy}
\label{here}

Notably, Feynman's own envisioned use of a quantum computer is nearing fruition \cite{lyod}. One of the most developed methods of eschewing the need for Monte-Carlo methods has arrived in the form of simulating lattice gauge theories on using ultracold atoms in optical lattices \cite{article1} \cite{rev1} \cite{rev2}. \footnote{Other approaches to simulating lattice gauge theories involve tensor networks and Floquet theory \cite{article2} \cite{floq}. } Moreover, the topologically ordered systems we discussed in the introduction are likewise being tackled via ultracold atom quantum simulation efforts \cite{topryd}. As opposed to the ``digital quantum simulation" that Google's quantum optimization and quantum chemistry demonstrations presented (recall from the introduction \cite{use1} \cite{use2}), these AMO community efforts are known as ``analogue quantum simulation" — using a finely tunable smaller physical system to model one an intractable large one. \footnote{There has been good progress in the last few years in the middle ground between purely digital and analogue quantum simulation using a procedure known as ``variational" quantum simulation that makes use of both an analogue system and the power of quantum computing \cite{VQS1} \cite{VQS2}.} 

To frame lattice gauge theory in a language suitable for modeling with atomic systems, AMO researchers found themselves further developing the Hamiltonian formalism of Wilson's gauge theory known as the Kogut and Susskind picture. In this prescription, the lattice gauging procedure was performed in a Hamiltonian/second-quantized formalism using canonical creation and annihilation operators as opposed to path-integrals and Wick rotations like Wilson's original work \cite{kogutsusskind}.

One of the primary objectives of this thesis is an understanding of the tensor gauge theory picture of fractons mentioned in the introduction. Some of the most basic examples the creators of this theory use to formulate the theory are extensions of quantum electrodynamics on the lattice in the Hamiltonian formalism of Kogut and Susskind. And so here we explore that formalism to get a grip on the basics of lattice gauge theory before continuing onto tensor gauge theory.

\pagebreak

\subsection{Electrodynamics refresher}

The vector field approach to electricity and magnetism is what we all begin with in high school, but the underlying potential formulation soon supplants this picture at the undergraduate level.

In this picture, the gauge field $A_{\mu}$ comprises the scalar and vector potentials $\phi$ and $\vec{A}\equiv A_i$ and gives us all the information about the electric field $E$ and magnetic field $B$ that we need. 

Using the following Maxwell equations

\begin{eqnarray}
 \nabla\cdot B &=& 0 \nonumber\\
 \nabla \times E &=& -\frac{\partial B}{\partial t}
\end{eqnarray}

in conjunction with vector identities such as the divergence of a curl being zero for the B equation, and combining that B result with the existence of a gradient of a scalar for a conservative vector field for the E equation, one finds the consistent potential formulation to be 

\begin{eqnarray}
E &=& - \nabla \phi - \frac{\partial A}{\partial t} \nonumber \\
B &=& \nabla \times A.
\end{eqnarray}

The concept of gauge freedom arises due to the non-uniqueness of the gauge field. Notably, the physical quantities E and B do not change if the gauge field undergo the following transformation for some scalar function $\alpha$

\begin{eqnarray*}
 \phi &\rightarrow& \phi - \frac{\partial \alpha}{\partial t} \\
 A &\rightarrow& A + \nabla \alpha.
\end{eqnarray*}

The ambiguity around the gauge field enables a choice of gauge known as gauge fixing. The Lorenz and Coulomb gauge are the most commonly encountered, but many others exist. Notably, the Weyl temporal gauge (also known as the Hamiltonian gauge for reasons that will become clear momentarily) makes the choice \cite{temp}

\begin{equation}
    A_0 = \phi =  0
\end{equation}

in which case the potential formulation of $E$ simplifies to 

\begin{equation}
\label{contE}
    E =  - \frac{\partial A}{\partial t}  = - \dot{A}
\end{equation}

The magnetic field $B$ is still represented by $\nabla \times A$ and by Stokes' Theorem we have an expression for the magnetic flux through some contour in terms of $A$ as well

\begin{eqnarray}
\label{contflux}
 \Phi &=& \int \int (B=\nabla \times A) \cdot dS \nonumber \\
 &=&\int A \cdot dr
\end{eqnarray}

\pagebreak

\subsection{The Hamiltonian formalism of lattice gauge theory}

While Kogut and Susskind were certainly interested in the interaction of matter with gauge fields to construct lattice gauge theories, here we eschew matter interactions in favor of a pure gauge theory \cite{oleg}. This is motivated by what we want to understand about the tensor gauge theory extensions of lattice gauge theory, and pure gauge is always the easier construct to get one's hand dirty with first. We will return later on to place matter fields on the lattice sites and investigate the dynamics of both their hopping and their interaction with gauge fields.

As will reappear consistently throughout this work, the gauge fields are located on the edges between lattice sites.

\begin{figure}[H]
\centering
\includegraphics[scale=0.5]{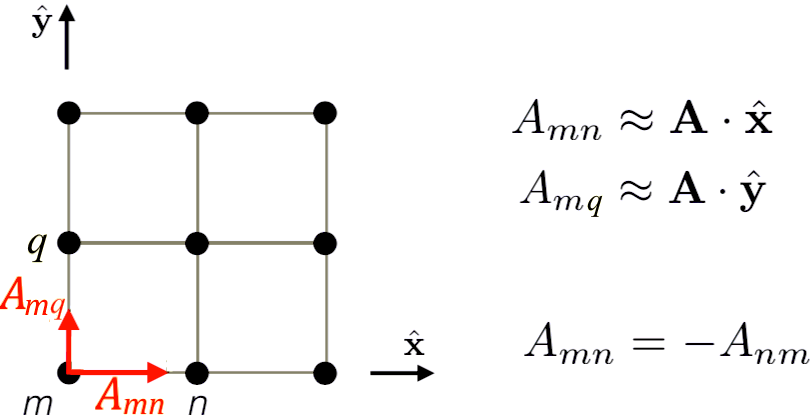}
\captionsetup{format=hang}
\caption[Lattice gauge theory with gauge fields on the links]{The Kogut Susskind prescription for gauge fields on the edges of a lattice. The gauge fields are labeled by the two lattice sites that make up the end points of the link they lie on. Adapted from \cite{oleg}.}
\label{fig22}
\end{figure}

\begin{figure}[H]
\centering
\includegraphics[scale=0.5]{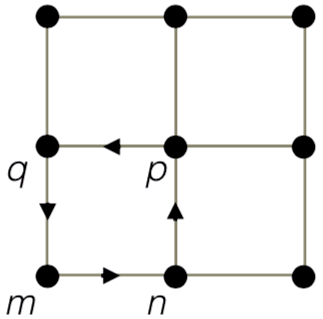}
\captionsetup{format=hang}
\caption[Orientation of edges in lattice gauge theory]{Orientation of edges on the lattice. As per equation \ref{rules}, we have that the right and up direction are positive, and the left and down direction are negative. This explains the minus signs in equation \ref{fluflu}. Image from \cite{oleg}.}
\label{orient}
\end{figure}

With these discrete gauge field variables on the lattice edges, the continuous Equations \ref{contE} and  \ref{contflux} can be made discrete (in the latter case just by summing the elements around a plaquette/square)

\begin{eqnarray}
\label{EA}
E_{mn} &=& - \dot{A}_{mn}, \\
\label{fluflu}
\Phi_{mnpq} &=& A_{mn} + A_{np} + A_{pq} + A_{qm} \nonumber \\
 &=& A_{mn} + A_{np} - A_{qp} - A_{mq},
\end{eqnarray}

where minus signs come from the directional rule of Figure \ref{fig22} above,

\begin{equation}
\label{rules}
    A_{mn}=-A_{nm}
\end{equation}

and ascribing an orientation to the edges as follows in Figure \ref{orient}.

We will elaborate on equations \ref{EA} and \ref{fluflu} below.

\pagebreak
\subsubsection{Kinetic energy, canonical quantization, and wavefunctions} 

A fruitful way to view the system is actually from a classical mechanics perspective \cite{oleg}. Notably, consider the gauge field A to be the canonical variable equivalent of position. Thus, an equivalent of velocity would be $\dot{A}$. From here we can derive the form of the kinetic energy by plugging our new canonical variables, $x \equiv A$ and $(v=\dot{x}) \equiv \dot{A}$, into the basic kinetic energy expression, $\frac{1}{2} m v^2$. For simplicity, we'll say the ``mass" (``moment of inertia") of the gauge field A is just $1$, in which case the kinetic energy of the field reads

\begin{eqnarray}
\label{T}
T &=& \frac{1}{2} m v^2 \nonumber \\
&=& \frac{1}{2} (m=1) (\dot{x})^2 \nonumber \\
&\equiv& \frac{1}{2} \dot{A}^2 \nonumber \\
&=& \frac{1}{2} \sum_{edges}  \dot{A}_{mn}^2,
\end{eqnarray}

In keeping to the classical mechanics perspective, the momentum (conjugate to the position) reads

\begin{eqnarray}
p_{mn} &=&  \frac{\partial T}{\partial \dot{x}}  \nonumber \\
&\equiv&  \frac{\partial T}{\partial \dot{A}}  \nonumber \\
&=& \dot{A}_{mn}  \nonumber \\
&=&  - E_{mn},
\end{eqnarray}

where \ref{EA} was used in the last equality to show that in this context, the electric field plays the role of conjugate momentum when the gauge field is the canonical variable. This perspective yields the energy density expected from electrodynamics as well (with the vacuum permittivity $\epsilon_0$ and mass both set to 1) 

\begin{eqnarray}
T &=& \frac{p^2}{2 m}  \nonumber \\
&\equiv& \frac{(-E)^2}{2 (m=1)} \nonumber \\
&=& \frac{1}{2} (\epsilon_0=1) E^2.
\end{eqnarray}

The relation between the position and momentum are then subjected to the same canonical quantization as standard quantum mechanics (with $\hbar=1$),

\begin{eqnarray}
\label{CCR}
[x,p] &=& i \nonumber \\
\left[A,-E\right] &\equiv& i \nonumber \\
- \left[A, E\right] &=& i \nonumber \\
 \left[E, A\right] &=& i \nonumber \\
 \left[E_a, A_b\right] &=& i \delta_{ab}
\end{eqnarray}

where $a$ and $b$ are potentially differing edges, accounting for the same property we saw earlier regarding operators commuting if they didn't share any edges in equations \ref{xzanti} and \ref{xzcom}.

Further, just like the quantum mechanical momentum operator, we now have (with $\hbar$ set to $1$)

\begin{eqnarray}
p &=& -i \frac{\partial}{\partial x} \nonumber \\
-E &\equiv& -i \frac{\partial}{\partial A} \nonumber \\
E &=&  i \frac{\partial}{\partial A}.
\end{eqnarray}

This derivative notation begs the question what is the operator acting on? Notably, the derivative representation of the quantum mechanical momentum operator, $p = -i \frac{\partial}{\partial x}$, arises from considering the action of a spatial derivative on the wavefunction solution of the Schrödinger equation. The form of our wavefunction is dictated beyond the generic plane wave because of the U(1) symmetry. If our canonical variable is, instead of an unbounded position $x$, a gauge field A which takes periodic values because of the U(1) symmetry

\begin{equation}
    A \in \left[0,2\pi\right),
\end{equation}

then so does our wavefunction. Thus, just like the solution to the Schrödinger equation for a particle on a 1-dimensional ring,

\begin{eqnarray}
\label{bloch}
\psi (A+2\pi) &=& \psi (A) \\
\psi_n(A) &=& \frac{1}{\sqrt{2\pi}}e^{- i n A}
\end{eqnarray}
for $n= {0,\pm 1, \pm 2, ...}$.

We can now act with this momentum operator (the electric field) on the wavefunction to tell us that the electric field conjugate variable is quantized

\begin{eqnarray}
E \psi_n &=&  i \frac{\partial}{\partial A} \bigg(\psi_n = \frac{1}{\sqrt{2\pi}}e^{- i n A}\bigg) \nonumber \\
&=&  n \psi_n.
\end{eqnarray}
\pagebreak
\subsubsection{Raising/lowering and translation operators}

Two additional operators to consider are raising/lowering operators and the translation operator. The basic properties of exponentials tell us that raising/lowering operators take the form $e^{\mp i A}$ \cite{oleg}

\begin{eqnarray}
e^{-iA} \psi_m &\propto& e^{-iA(1+m)} \propto \psi_{m+1}, \\
e^{iA} \psi_m &\propto& e^{iA(1-m)} \nonumber \\
&=& e^{-iA(-1+m)} \propto \psi_{m-1}.
\end{eqnarray}

And moreover, the electric field acts as a shift/translation operator on the gauge fields \cite{QSL}. This is to be expected, since we've already made the identification with the electric field playing the role of conjugate momentum in this system, and in quantum mechanics we know that the momentum operator is the generator (Lie algebra element) of translations (Lie group elements) \cite{townsend}.\footnote{Using the Lie algebra element (symmetry generator) — Lie group element (operator) relationship characteristic of geometric perspectives of gauge theory

\begin{equation*}
    G = e^{iA} \quad G = \text{group element} \quad A = \text{algebra element}
\end{equation*}} 
Notably, the translation operator T which shifts the state $\psi(x)$ by some distance $x=a$ reads (with $\hbar=1$)

\begin{equation}
\label{trans}
  \hat{T}(a) = e^{-i a \hat{p}}.
\end{equation}

And based on our analogy between position $x$ and our gauge field $A$ (which, recall, is valued in $\left[0,2\pi\right)$ and so we'll consider a translation of some \textit{angle} $A=\theta$), we have

\begin{eqnarray}
\label{naive}
\hat{T}(\theta) &=& e^{-i (a \sim \theta) (\hat{p} \sim -E)} \nonumber \\
&=& e^{i \theta E}.
\end{eqnarray}
\pagebreak

\subsubsection{Potential energy and pure gauge (gapless) Hamiltonian}
\label{sec:gapless}

While deriving a kinetic energy term from classical mechanics analogies was practically trivial, the same method is not so fruitful. Instead we will start from basic electromagnetism and import those insights to a lattice picture. 

To make things as concrete as possible, we will find the expression for the energy per unit volume stored in a magnetic field ($u=\frac{1}{2} B^2$, with the vacuum permeability $\mu_0$ set to 1) by using an inductor as an example before getting more general. Using Faraday's law to equate electromotive force and energy, one finds that the energy in an inductor to be \cite{ut}

\begin{equation}
U= \frac{1}{2} L I^2,
\end{equation}

and from Ampere's law one can find that the inductance $L$ and current $I$ of a solenoid which simplifies this further to (with the vacuum permeability $\mu_0$ set to 1)

\begin{eqnarray}
u &=& \frac{U}{V} \nonumber \\
&=& \frac{1}{V} \frac{1}{2} \bigg(\frac{\mu_0 N^2 A}{2 \ell}\bigg) \bigg( \frac{B \ell}{\mu_0 N}\bigg)^2  \nonumber \\
&=& \frac{\ell A}{V} \frac{1}{2}  \frac{B^2}{\mu_0}  \nonumber \\
&=& \frac{1}{2} B^2. 
\end{eqnarray}

There is nothing stopping us from writing this in terms of our gauge field A via $B=\nabla \times A$, our only task is to generalize the concept of the curl to a discrete lattice like in \ref{EA}. The standard determinant/component approach to the curl generalizes to the following calculus limit definition \cite{math}

\begin{equation}
\label{eq:curl}
    (\nabla \times A) \cdot n = \lim_{S\to0} \frac{1}{S} \oint_\ell A \cdot d\ell,
\end{equation}

where $n$ is normal to the area S which has a boundary $\ell$. This definition can be easily exported to a lattice description, as there is no shrinking area $S$ and taking the line integral is nothing more than summing the gauge fields $A$ around a loop/plaquette. Following the same orientation as in Figure \ref{orient}, 

\begin{eqnarray}
\label{curl}
 \nabla \times A &=& \oint_\ell A \cdot d\ell \nonumber \\
 &=& \sum_{edges} A_{ij} \nonumber \\
 &=& A_{mn} + A_{np} + A_{pq} + A_{qm} \nonumber \\
\Phi_{mnpq} &=& A_{mn} + A_{np} - A_{qp} - A_{mq}.
\end{eqnarray}

Thus the potential energy reads

\begin{eqnarray}
 U &=& \int \frac{1}{2} B^2 \nonumber \\
 &=& \frac{1}{2} \int (\nabla \times A)^2 \nonumber \\
 &=& \frac{1}{2} \sum_{plaquettes} \Phi_{mnpq}^2.
\end{eqnarray}

We have however missed a subtlety in this term — the fact that it depends on the gauge field $A$. By the same Bloch's theorem argument of Equation \ref{bloch}, since $A$ is periodic, the potential energy $U(A)$ must be periodic as well. We can remedy this in a slick way however to match the $U$ just derived about. We will give $U$ periodicity by unitizing a trig function

\begin{equation}
    U= - \sum_{plaquettes} cos (\Phi_{mnpq}).
\end{equation}

We have added the negative sign to reproduce the above definition of $U$, since in the limit of small flux $\Phi_{mnpq}$ we can use the small angle approximation 

\begin{eqnarray}
-cos(\Phi_{mnpq}) &\approx& - \bigg(1 -  \frac{1}{2} \Phi_{mnpq}^2) + ...\bigg) \nonumber \\
&\approx& - 1 +  \frac{1}{2} \Phi_{mnpq}^2.
\end{eqnarray}

Since adding a constant to a Hamiltonian does nothing to the dynamics of the system, we can safely ignore the $-1$ and we see that the small flux little of the periodic form of the potential energy is precisely what one would expect from electrodynamics. 

And so in total our pure gauge (which we call gapless and which will be expanded on in the next section) Hamiltonian reads

\begin{eqnarray}
H &=& T + U \nonumber \\
&=& \frac{1}{2} \sum_{edges}  \dot{A}_{mn}^2  - \sum_{plaquettes} cos (\Phi_{mnpq}) \nonumber \\
&=& \frac{1}{2} \sum_{edges} E_{mn}^2  - \sum_{plaquettes} cos (\Phi_{mnpq})  \\
&\approx& \frac{1}{2} \sum_{edges} E_{mn}^2 + \sum_{plaquettes} \frac{1}{2} \Phi_{mnpq}^2 \nonumber \\
&\equiv& \frac{1}{2} \int (E^2 +B^2). \nonumber
\end{eqnarray}

\pagebreak
\subsubsection{Localized charge as a constant of motion}

We have stated that the gauge field $A$ is our canonical variable, and that the electric field is the momentum conjugate to that via our kinematics argument.

If there exists some object $E$ representing the electric field, there is necessary some charge giving rise to it. Where is that charge?

In the same way that discretizing the curl enabled us to represent the $B$ field in the system, we can discretizing the divergence enables us to develop a concept of localized charge in the system via Gauss's law (with the permittivity of free space $\epsilon_0$ set to 1) and the divergence theorem

\begin{eqnarray}
\label{gausseq}
Q &=& \Phi_E \nonumber \\
&=& \oint E \cdot dA \nonumber \\
&=& \int (\nabla \cdot E) dV.
\end{eqnarray}

\begin{figure}[H]
\centering
\includegraphics[scale=0.5]{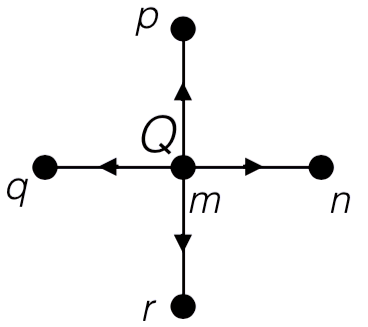}
\captionsetup{format=hang}
\caption[Orientation of lattice for quantifying localized charges at a vertex]{Orientation of the lattice for the purpose of quantifying the localized charge on a vertex. Image adapted from \cite{oleg}.}
\label{divfig}
\end{figure}

The standard determinant/component approach to the divergence generalizes to the following calculus limit definition \cite{math}

\begin{equation}
\label{eq:div}
    \nabla \cdot E |_m = \lim_{V\to0} \frac{1}{V} \oint_{\partial V \equiv S} E \cdot \hat{n} dS,
\end{equation}

where $m$ is the point we are calculating the divergence at and $\hat{n}$ is normal to the boundary $\partial V \equiv S $ of the volume $V$. Exactly as in equation \ref{eq:curl}, this definition can be easily exported to a lattice description, as there is no shrinking volume $V$ and taking the desired integral is nothing more than summing the electric fields on the surface plane that is formed around the particular vertex site $p$, i.e. those on the links attached to $p$. Using the orientation of the links shown in Figure \ref{divfig} as a visual guide, 

we use the discretization of the divergence in equation \ref{eq:div} to represent localized charge as per  Gauss's law in equation \ref{gausseq},

\begin{eqnarray}
\label{div}
Q_m &=& (\text{div} E)_m \nonumber \\
&=& \nabla \cdot E |_m \nonumber \\
 &=& \oint_{S} E \cdot dS \nonumber \\
 &=& \sum_{edges} E_{ij} \nonumber \\
 &=& E_{mn} + E_{mp} - E_{mq} - E_{mr}.
\end{eqnarray}

\pagebreak
\subsubsection{Full (gapped) Hamiltonian}

The Hamiltonian of Section \ref{sec:gapless} does not include sources (charges).\footnote{Here, we use the terms ``sources" as it relates to mass and charge respectively creating non-trivial dynamics in terms of the curvature of spacetime in general relativity, and in terms of the curvature in the topographical picture of the electric potential.} It represents a gapless photonic mode of the system because a photon has no charge/mass and is not a source. Thus the energy of that phase of the system is continuously connected to the vacuum, i.e. the energy spectrum is gapless. Thinking of a linear dispersion relation like for photons (or massless quasiparticles in Dirac matter for that matter) is helpful in picturing a gapless mode.

On the other hand, introducing a source like mass into the system, like in the Klein-Gordon equation where the dispersion is now (in natural units)

\begin{equation}
    E(k) = \sqrt{k^2 + m^2},
\end{equation}

which is certainly no longer linear because the source (mass in this case) has induced a gap.

\begin{figure}[H]
\centering
\includegraphics[height=6cm,width=8cm]{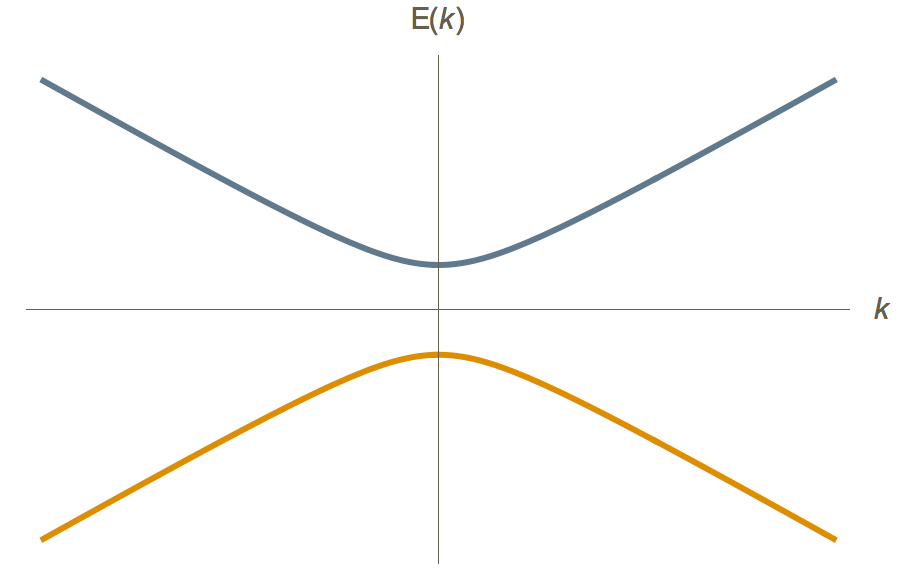}
\captionsetup{format=hang}
\caption[Gapped dispersion relation due to sources (charge or mass)]{Gapped dispersion relation, $E(k)=\sqrt{k^2 + m^2}$.}
\end{figure}

Charge can also include such a gap. And so to be as general as possible, we need to include such excitation in our full Hamiltonian.

Mimicking the electrostatic potential energy one finds from integrating a Coulomb force in a work-energy derivation 
\begin{equation}
    U= k_e \frac{qQ}{r}.
\end{equation}
But just like in Equations \ref{eq:curl} and \ref{eq:div}, the $\frac{1}{r}$ can be neglected on the lattice.
And so we will represent the term simply by 
\begin{eqnarray}
   U &=& K Q^2 \nonumber \\
   &=& K (\text{div} E)^2, 
\end{eqnarray}
where $K$ is now just a factor to give units of energy to the term in the full Hamiltonian. We'll adopt similar tuning parameters for the other ``kinetic" and ``potential" terms in the gauge Hamiltonian as well to consider possible limits in the future,
\begin{eqnarray}
H &=& \frac{1}{2} T \sum_{edges}  E_{mn}^2  - V \sum_{plaquettes} cos (\Phi_{mnpq}) + K \sum_{vertices} Q_m{}^2 \nonumber \\
&=& \frac{1}{2} T \sum_{edges}  E_{mn}^2  - V \sum_{plaquettes} cos (\Phi_{mnpq}) + K \sum_{vertices} (\text{div}E)_m{}^2.
\end{eqnarray}

\begin{center}
    \rule{4cm}{0.05cm}
\end{center}

This concludes our foray into lattice gauge theory, up to a point of familiarity with concepts of quantized electrodynamics on the lattice — since these are the concepts that will be generalized in the tensor gauge picture of fracton theory that we wish to cover.

We will however use some of these insights to motivate the toric code from a QED on the lattice perspective. 

\pagebreak

\subsection{QED on the lattice as a physical motivation for the toric code}
\label{return}

\subsubsection{Equivalence of operators and fields}

Recall the toric code Hamiltonian, but where we will rewrite $A_v$ as $S_v$ (for ``star" operator) and $B_p$ as $P_p$ (for ``plaquette" operator) to avoid getting the operators confused with the gauge fields $A_{mn}$ and magnetic field $B$ from the curl of $A$ in the Kogut Susskind lattice gauge theory (KS LGT) picture \cite{QSL}

\begin{equation}
    H = - \sum_v S_v - \sum_p P_p,
\end{equation}

where the operators are defined with the Pauli spins like before as 

\begin{eqnarray}
S_v &=& \prod_i \sigma_i{}^x, \\
P_p &=& \prod_i \sigma_i{}^z.
\end{eqnarray}

In what follows we will show how the charge $Q_m$ and the periodic curl of the gauge field $cos (\Phi_{mnpq})$ in KS LGT correspond to the star operator $S_v$ and plaquette operator $P_p$ in the toric code respectively.

Firstly, with an eye towards the toric code as a stabilizer code, not a kinetic theory, we will take the limit of the KG LGT theory being at rest. Since the $E_{mn}$ fields represented the conjugate momentum in the theory, we assume this is zero at rest, so the kinetic energy term in the Hamiltonian goes to zero and we have so far (also with the prefactors $T=V=K=1$)

\begin{quote}
    
This MUST be incorrect! Later on in this derivation you use E in the vertex term. If E=0, then div E = 0, and Q = 0.\footnote{Thanks is due to Professor Tzu-Chieh Wei for asking about the logic of this ``at-rest limit" we consider above, as this question and our erroneous response to it led to the realizations we discuss below \cite{weichat}.}

The answer comes, as promised in the footnote \ref{foothere}, from a key property of the toric code we need to reproduce, the commutation of the $A_v$ and $B_p$ operators.

Note that in the Kogut-Susskind Hamiltonian above,

\begin{equation}
    H = \frac{1}{2} T \sum_{edges}  E_{mn}^2  - V \sum_{plaquettes} cos (\Phi_{mnpq}) + K \sum_{vertices} (\text{div}E)_m{}^2
\end{equation}

there are 3 terms, not the 2 of the toric code? Which shall we dispense with, and why?\footnote{Thanks is due to Hiroki Sukeno for pointing out the following route to this derivation \cite{Hiroki}.}

Note the orientation of the curl and divergence from equations \ref{curl} and \ref{div} which gives those terms an equal number of positive and negative terms,

\begin{eqnarray}
\label{terms}
    \Phi_{mnpq} &=& A_{mn} + A_{np} - A_{qp} - A_{mq} \nonumber \\
    (\text{div} E)_m &=& E_{mn} + E_{mp} - E_{mq} - E_{mr} \equiv Q_m
\end{eqnarray}

Based on the fact that the commutator — of the objects $A_{mn}$ and $E_{mn}$ occupying the edge — is not equal to zero, but rather according to equation \ref{CCR}

\begin{equation}
\label{ccrnow}
    [E_a,A_b]=i \delta_{ab}
\end{equation}

we are going to make the ansatz that the $cos (\Phi_{mnpq})$ and $Q_m{}^2$ terms in the KG Hamiltonian commute, and thus THOSE are the 2 terms we will equate to the 2 toric code terms $A_v$ and $B_p$ since those commute as well.

Recall the small angle approximation of the periodic $cos (\Phi_{mnpq})$ is (up to a constant) $\Phi_{mnpq}^2$.

Looking at the non-trivial case for the operators as we did in on page \pageref{where}, let's consider the overlap of two edges between the $\Phi_{mnpq}^2$ and $Q_m{}^2$ terms.

Note that by the property of commutators, \cite{wikicomm}

\begin{equation}
\label{wikicomm}
    \ls AB,CD \rs = A \ls B,C \rs D + \ls A,C \rs BD + CA \ls B,D \rs + C \ls A,D \rs B,
\end{equation}

if $A=B$ and $C=D$ as in our proposed 

\begin{equation}
\label{obj}
    \ls Q_m{}^2, \Phi_{mnpq}^2 \rs
\end{equation}

then it suffices to prove that 

\begin{equation}
    \ls Q_m, \Phi_{mnpq} \rs =0
\end{equation} 

to prove equation \ref{obj} since that would lead to a zero in each term of equation \ref{wikicomm}. I.e.

\begin{equation}
    \ls A,B \rs = 0 \qquad \rightarrow \qquad \ls A^2, B^2 \rs =0
\end{equation}

Recall the orientation of up = positive, down = negative; and right = positive, left = negative from Figures \ref{orient} and \ref{divfig}. Using a similar convention as Figure \ref{conv}, we will look at the commutator between the terms adapted from equation \ref{terms} to allow for easier visualization,

\begin{eqnarray}
(\text{div} E)_m = Q_m \equiv Q_0 &=& E_1 + E_3 - E_4 - E_2 \nonumber \\
\Phi_{mnpq} \equiv \Phi_{5612} &=& A_5 + A_6 - A_1 - A_2,
\end{eqnarray}

where the $1$ through $6$ subscripts signify differing edges on the lattice in Figure \ref{figab} immediately below.

\begin{figure}[H]
\centering
\includegraphics[height=8cm,width=10cm]{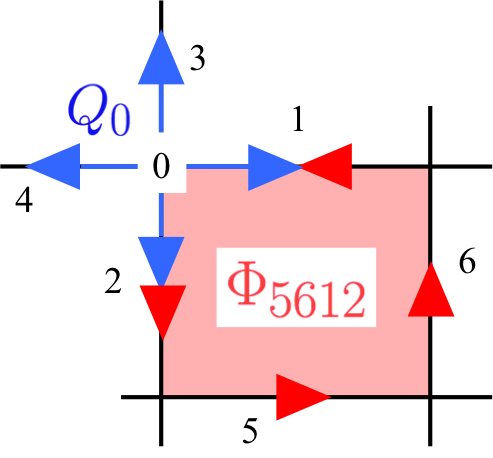}
\captionsetup{format=hang}
\caption[Non-trivial commutator of Kogut-Susskind divergence and curl terms]{Examining the non-trivial commutator of Kogut-Susskind divergence and curl terms using the orientation prescribed by Figures \ref{orient} and \ref{divfig}, we see that one of the edges, $2$, share parity (both as in the negative direction, down) and another edge, $1$, has opposite parity (one is in the positive direction and one is in the negative direction, right and left respectively).}
\label{figab}
\end{figure}

Using a 4-term generalization of the distributive property of commutators \cite{wikicomm}

\begin{equation}
    \ls A +B, C+D \rs = \ls A,C \rs + \ls A,D \rs + \ls B,C \rs + \ls B,D \rs,
\end{equation}

and the commutation of $E$ and $A$ on differing edges, 

\begin{equation}
    [E_a,A_b]=i \delta_{ab},
\end{equation}

we have 

\begin{eqnarray}
    \ls Q_0, \Phi_{5612} \rs &=& \ls E_1 + E_3 - E_4 - E_2, A_5 + A_6 - A_1 - A_2 \rs \nonumber \\
    &=& - \ls E_1, A_1 \rs + \ls E_2, A_2 \rs \nonumber \\
     &=& - i + i \nonumber \\
     &=& 0
\end{eqnarray}

Hence, the $\Phi_{mnpq}^2$ and $Q_m{}^2$ terms of the KG Hamiltonian commute, just like the $A_v$ and $B_p$ terms of the toric code.

This concludes our correction of the erroneous ``at-rest limit" discussed prior to this interlude.

\end{quote}

Now that we have proven loosely the equivalence of the $cos (\Phi_{mnpq})$ and $Q_m{}^2$ terms in the KG Hamiltonian and the toric code terms $A_v$ and $B_p$ (via showing that those KG terms commute just like the latter toric code terms), we now neglect the first term of the KG Hamiltonian, the kinetic $\frac{1}{2} T \sum_{edges}  E_{mn}^2 $. This is stated concisely, but without the above derivation, as the $T=0$ limit in \cite{QSL}. 

So we now have only

\begin{equation}
    H = -  \sum_{plaquettes} cos (\Phi_{mnpq}) +  \sum_{vertices} Q_m{}^2 \nonumber
\end{equation}

To show that the first term is equivalent to the plaquette operator $P_p$ consider the following identification

\begin{equation}
\label{id} 
    \sigma_i{}^z \equiv e^{i A_{mn}}.
\end{equation}

Making this identification allows us to see the how the plaquette operator is equivalent to the flux term $cos (\Phi_{mnpq})$ in the KS LGT Hamiltonian,

\begin{eqnarray}
P_p &=& \prod_i \sigma_i{}^z \nonumber \\
&=& \prod_{mn} e^{i A_{mn}} \nonumber \\
&=& e^{i \sum_{mn} A_{mn}} \nonumber \\
&=& e^{i (\nabla \times A)} \nonumber \\
&=& e^{i \Phi_{mnpq}} \nonumber \\
&\sim& cos (\Phi_{mnpq}),
\end{eqnarray}

where in the last line we have taken the real part of the exponential via Euler's formula.

Showing the equivalence between the charge $Q_m$ and the star operator $S_v$ requires slightly more tact, but only in terms on the underlying qubit structure of the toric code. 

We begin similarly by making the following identification

\begin{equation}
    \sigma_i{}^x \equiv e^{i \pi E_{mn}}.
\end{equation}

In doing so, we can write the star operator as follows,

\begin{eqnarray}
S_v &=& \prod_i \sigma_i{}^x \nonumber \\
&=& \prod_{mn} e^{i \pi E_{mn}} \nonumber \\
&=& e^{i \pi \sum_{mn}  E_{mn}} \nonumber \\
&=& e^{i \pi (\text{div} E)} \nonumber \\
&=& e^{i \pi Q_m}.
\end{eqnarray}

By taking into consideration that the toric code has a discrete $\mathbb{Z}_2$ symmetry as opposed to continuous $U(1)$, we can allow charge to either occupy a site or not, i.e. $Q_m$ is either $0$ or $1$. In this way, the above expression becomes

\begin{eqnarray}
S_v &=& e^{i \pi Q_m} \nonumber \\
&=& e^{i \pi (0/1)} \nonumber \\
&=& \pm 1,
\end{eqnarray}
which is precisely the convention the toric code uses for the ground state (+1) and excited state (-1) of a particular qubit at a vertex. And so we see what the charge in the KS LGT model is the star operator in the toric code.

Note that the same discretization occurs for the initially periodic $A_{mn}$ in the identification \ref{id}, it was just that we didn't need to make use of it to make the identification across theories. Because the qubit must have the $\pm 1$, we must restrict the $U(1)$ range of $A_{mn}$ from $\left[0,2\pi\right)$ to only $0$ and $\pi$.

\pagebreak

\section{Tensor gauge theory}
\label{TGT}

We will begin with summary of the previous chapter, but using the notation present in \cite{pretEM}.

A compact $U(1)$ gauge theory means that the gauge field is periodic

\begin{equation}
    A_i \equiv A_i +2\pi.
\end{equation}

The electric field $E_i$ (playing the role of an angular momentum operator) and gauge potential $A_i$ (playing the role of a position operator) are conjugates of one another. More rigorously,

\begin{equation}
    E=\frac{\partial \mathcal{L}}{\partial \dot{A}},
\end{equation}

and they obey the following commutation relation that is a generalization of equation \ref{CCR} \cite{kogut}

\begin{equation}
\label{CCR2}
    [E_i(x),A_j(y)] = i \delta_{ij} \delta(x-y).
\end{equation}

If $A_i$ follows the gauge transformation 

\begin{equation}
\label{gaugetrans}
    A_i \rightarrow A_i + \partial_i \alpha,
\end{equation}

then (as we will touch on in Section \ref{gauss}) the source-free Gauss's law that follows from this gauge invariance if we have 

\begin{equation}
    \partial_i E^i =0.
\end{equation}

The low-energy theory that obeys this gauge transformation is

\begin{equation}
    H= \frac{1}{2} g \sum_{edges} E^2 - \sum_{plaquettes} cos(B),
\end{equation}

where the magnetic field $B$ is constructed via

\begin{eqnarray}
B&=&\nabla \times A\nonumber \\
    B_i &=& \epsilon_{ijk} \partial^j A^k,
\end{eqnarray}

which is just the component by component picture of the curl operator \cite{curlwiki}.

When the coupling constant $g$ is small, the fluctuations of the cosine are small about its minimum and by the small angle approximation

\begin{equation}
    H= \frac{1}{2} \int d^3 x (gE^2 + B^2).
\end{equation}

\pagebreak

\subsection{Conservation of charge and conservation of dipole moment}
\label{charge}

Allowing for charges in the system, 

\begin{equation}
    \partial_i E^i = \rho, 
\end{equation}

enables an energy gap to form and the Hamiltonian is modified to include

\begin{equation}
    H= \frac{1}{2} g \sum_{edges} E^2 - \sum_{plaquettes} cos(B) + U \sum_{vertices} (\partial_i E^i)^2.
\end{equation}

Charge is conserved since it is a total derivative term 

\begin{equation}
    \int (\rho = \partial_i E^i) = 0.
\end{equation}

In a system where only the \textit{total} charge is conserved, it is physically viable to have charges appear from the vacuum, which have no movement restrictions so long as both a positive and negative charge of equal magnitude are created.

\begin{figure}[H]
\centering
\includegraphics[scale=0.5]{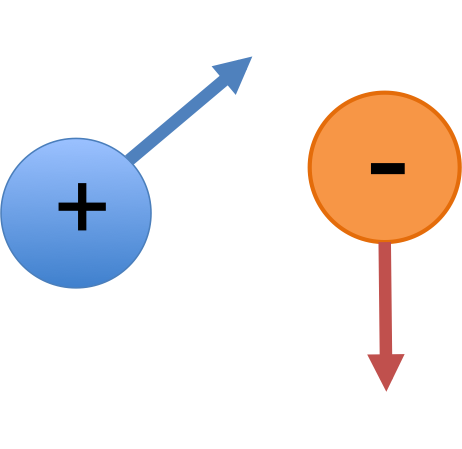}
\captionsetup{format=hang}
\caption[Charges moving freely with only total charge conservation]{Positive and negative pair of charges moving freely with total charge conserved. Here, the arrows denote the charges' movement in space. Figure adapted from \cite{fractonCS}.}
\label{free}
\end{figure}

The core tenet of tensor gauge theory is that there is no reason, a priori, to restrict the gauge field $A$ to be a vector $A_i$. Rather it could be a tensor $A_{ij}$.

In this case the gauge transformation(s) and the corresponding Gauss's law(s) comes in three flavors which are called the scalar, vector, and traceless theories respectively as written in Table \ref{table}.

\begin{table}[H]
\centering 
\renewcommand{\arraystretch}{1.3}
 \begin{tabular}{| l || l |} 
 \hline
Gauge invariance & Gauss's law  \\ 
 \hline\hline 
$A_{ij} = A_{ij} + \partial_i \partial_j \alpha$ & $\partial_i \partial_j E^{ij} = 0$   \\ 
 \hline  
$A_{ij} = A_{ij} + \frac{1}{2}(\partial_i\alpha_j + \partial_j \alpha_i)$ & $\partial_i E^{ij} = 0$  \\
 \hline
$A_{ij} = A_{ij} + \delta_{ij} \alpha$ & $E^{i}{}_i = 0$  \\
 \hline
\end{tabular}
\caption{Equivalence between gauge field transformations and Gauss's laws for various tensor gauge theories.}
\label{table}
\end{table}

Again, as mentioned below equation \ref{gaugetrans}, we will explain how one arrives at these particular Gauss's laws from a given transformation of the gauge field in Section \ref{gauss}.

As a side note, both $A_{ij}$ and $E_{ij}$ are symmetric tensors. As a result their commutation reads 

\begin{equation}
\label{CCR3}
    [E_{ij}(x),A_{kl}(y)] = i (\delta_{ik}\delta_{jl}+\delta_{il}\delta_{jk}) \delta(x-y).
\end{equation}

which is a generalization of equation \ref{CCR2} and \ref{CCR} from \cite{kogut} and \cite{oleg} respectively.\footnote{Note that this has a different sign than is written in \cite{primarysource}. The latter convention matches the more pedagogically clear kinematic analogy of \cite{oleg} and so we stick to this convention. To enable the reader to trust whichever source they like, two other sources that match \cite{primarysource} are \cite{QSL} and \cite{lgt2}.}

We begin with the scalar theory. Allowing for charges, we again have charge conservation

\begin{equation}
    \int (\rho = \partial_i\partial_j E^{ij}) = 0.
\end{equation}

Now however, we also have conservation of dipole moment, characteristic of fracton phenomenology as we will illustrate conceptually below.

\begin{eqnarray}
\int \vec{x} \rho &=& \int x^k \partial_i \partial_j E^{ij} \\ 
    &=& - \int \partial_i E^{ik} \\
  &=& 0
\end{eqnarray}

where in the last line we note the integral is over a total derivative, and in the second line we have integrated by parts in the following way

\begin{eqnarray}
\int \partial_i \partial_j (x^k E^{ij}) &=& \int \left[ \partial_i \partial_j (x^k) E^{ij} +  x^k \partial_i \partial_j  E^{kj} \right] \nonumber \\ 
0 &=& \int \left[  \partial_i (\partial_j x^k) E^{ij} +  x^k \partial_i \partial_j  E^{ij} \right] \nonumber\\ 
-\int \partial_i (\delta_j{}^k) E^{ij} &=&  \int x^k \partial_i \partial_j  E^{ij}  \nonumber \\ 
-\int \partial_i E^{ik} &=&  \int x^k \partial_i \partial_j  E^{ij}.
\end{eqnarray}

In contrast to the free movement of Figure \ref{free}, while charges can indeed still be created out of the vacuum (with total charge conserved still of course) they are not free to move as they please because of the dipole conservation. Instead they are allowed to move only in pairs, such that the dipole moment is conserved.

\begin{figure}[H]
\centering
\includegraphics[scale=0.5]{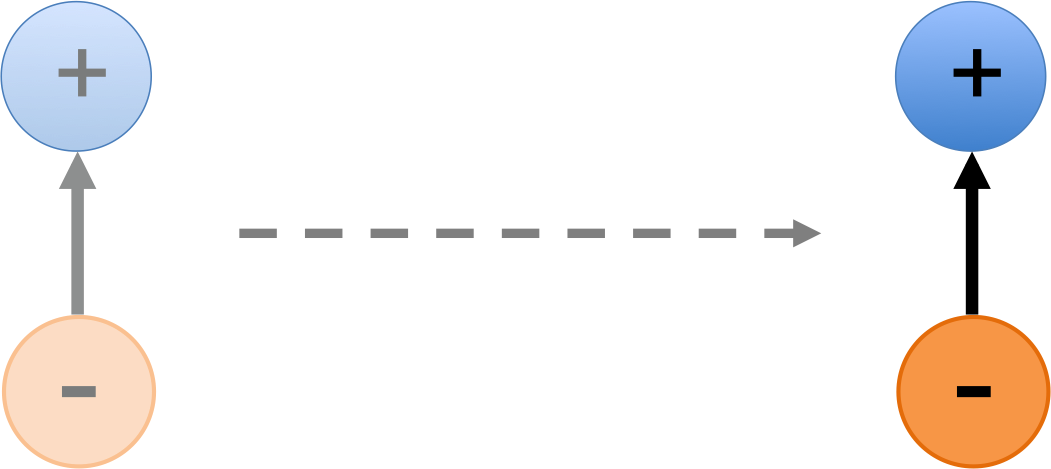}
\captionsetup{format=hang}
\caption[Lineon motion due to dipole moment conservation]{A pair of charges moving toward such that dipole moment is conserved. Here, the solid arrows denote the dipole moment of the charge configuration, and the dashed arrow denotes movement in space.}
\end{figure}

These are precisely the lineons of Figure \ref{lineons}.

By considering the creation of a zero total dipole moment configuration from the vacuum,

\begin{figure}[H]
\centering
\includegraphics[scale=0.5]{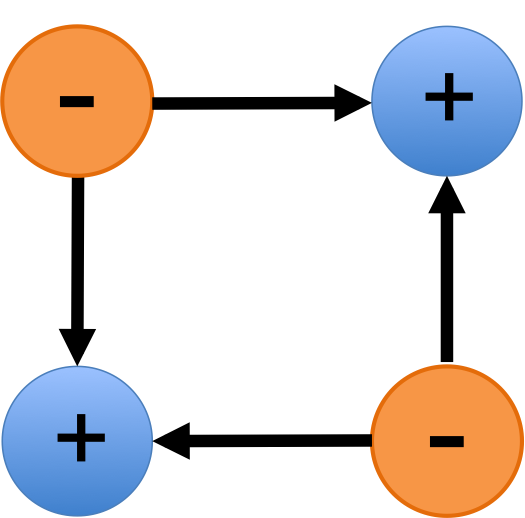}
\captionsetup{format=hang}
\caption[Quadruple of charges such that the total dipole moment of the system is zero]{A quadruple of charges with that dipole moment of zero. Here, the solid arrows denote the dipole moment of the charge configuration. Figure adapted from \cite{fractonCS}.}
\label{quad1}
\end{figure}

we can see the impossibility of moving a single charge, embodying the same fractal propagation that Figure \ref{single} illustrated.

\begin{figure}[H]
\centering
\includegraphics[scale=.7]{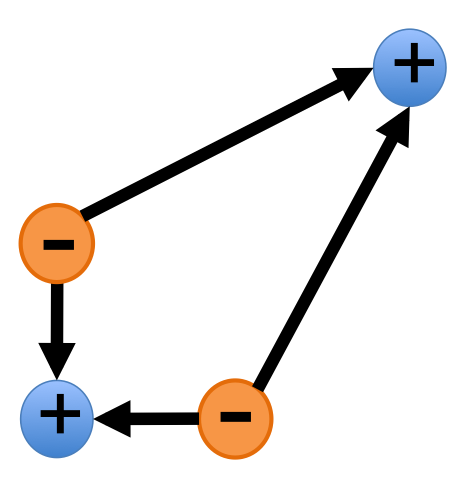}
\captionsetup{format=hang}
\caption[Violation of dipole moment conservation due to moving a singular charge]{Any attempt at moving a single charge results in a violation of dipole conservation. The initially $(0,0)$ dipole moment vector has become $(1,1)$.}
\label{28}
\end{figure}

Moving a charge while only respecting the total charge conservation leads to a change in dipole moment, notably from the vector $(0,0)$ to $(1,1)$ as illustrated in Figure \ref{28},

To respect the dipole conservation we need to add another charge to counter this new non-zero dipole moment, which then necessitates \textit{another} charge to keep the total charge constant.

\begin{figure}[H]
\centering
\includegraphics[scale=.55]{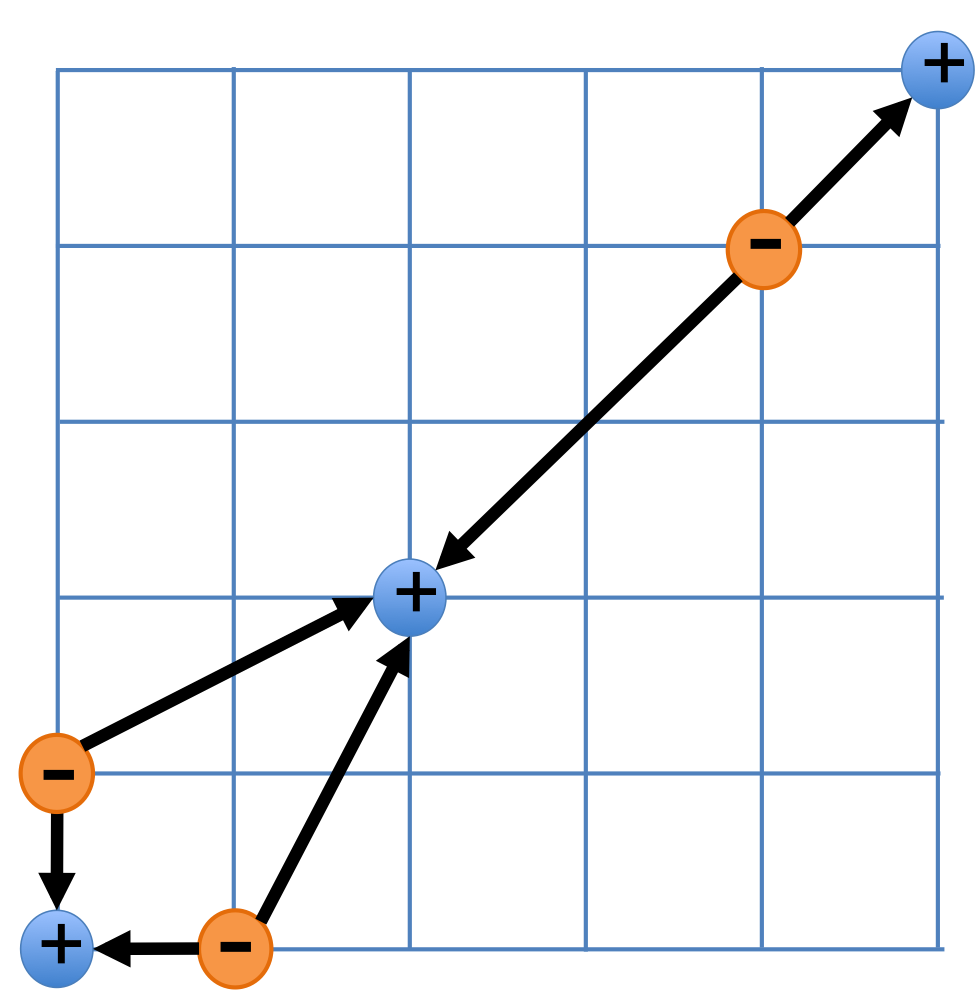}
\captionsetup{format=hang}
\caption[Rectifying altered dipole moment via fractalization of charges]{To address the resultant $(1,1)$ dipole moment, one could try to add a negative charge above and to the right of the moved positive charge such that it creates a $(-1,-1)$ dipole. This would leave 3 negative charges and 2 positive however, and so more subtle placement \textit{and} an additional positive charge is necessary.}
\label{quad3}
\end{figure}
\pagebreak
\subsection{Gauge invariance intro: a more precise momentum operator}
\label{gauss}

What IS gauge invariance? 

It is the invariance of the state $\psi(x)$ under some variation of $x$. Notably

\begin{equation}
\psi(x) = \psi (x+\Delta x).
\end{equation}

Note that in our theory, the canonical variable $x$ is instead the gauge field $A_i$. Writing the variation $\Delta x$ as $\partial_i \alpha$ (to respect the lattice description with the index $i$) we have that gauge invariance requires 

\begin{equation}
\psi(A_i) = \psi (A_i + \partial_i \alpha).
\end{equation}

How can we simplify the right hand side of this equation? Recall the definition of the translation operator from quantum mechanics in equation \ref{trans} (with $\hbar=1$)

\begin{equation}
\hat{T}(\Delta x) = e^{-i\Delta x \hat{p}},
\end{equation}

which translates the state by $\Delta x$

\begin{eqnarray}
\psi(x) &=& \psi (x+\Delta x) \nonumber \\
&=& \hat{T}(\Delta x) \psi(x).
\end{eqnarray}

Doing the same thing with the state $\psi(A_i)$ and a translation of $\partial_i \alpha$ we have 

\begin{eqnarray}
\psi(A_i) &=& \psi (A_i+\partial_i \alpha) \nonumber \\
&=& \hat{T}(\partial_i \alpha) \psi(A_i) \nonumber \\
&=& e^{i (\partial_i \alpha) E_i} \psi(A_i),
\end{eqnarray}

since $-E^i$ is the analogue of momentum in our theory. This is what we need to show to have gauge invariance.

We however have to be a bit more careful about the momentum and translation operator.

Recall the difference between the commutation operators of standard quantum mechanics and our lattice gauge theory from equations \ref{CCR3}, \ref{CCR2} and \ref{CCR}

\begin{eqnarray}
[x,p] &=& i \qquad \text{(with } \hbar=1) \nonumber \\
\left[A,-E\right] &=& i \nonumber \\
-\left[A,E\right] &=& i \nonumber \\
 \left[E, A\right] &=& i \nonumber \\
\left[E_{ij}(x),A_{kl}(y)\right] &=& i (\delta_{ik}\delta_{jl}+\delta_{il}\delta_{jk}) \delta(x-y).
\end{eqnarray}

This $\delta(x-y)$ difference carries over to the momentum and translation operator as well. Instead of the standard quantum mechanical 

\begin{equation}
\hat{T}(\Delta x) = e^{-i\Delta x \hat{p}},
\end{equation}

and even our naive version of the lattice gauge theory translation operator from equation \ref{naive}
 
\begin{equation}
\hat{T}(\theta) = e^{i\theta E},
\end{equation}

we need to account for the $\delta(x-y)$ as follows

\begin{eqnarray}
\left[ E^i(x), A_j(y) \right] &=& i \delta_{ij} \delta(x-y), \nonumber \\
\left[\int dx E^i(x),  A_j(y) \right] &=& i \delta_{ij} \int dx \left[ \delta(x-y)\right] \nonumber \\
&=& i \delta_{ij},
\end{eqnarray}

which tells us that the more precise momentum operator is 

\begin{equation}
\hat{p_i} = \int dx E^i(x).
\end{equation}

Thus, in the translation operator, we have 

\begin{equation}
\hat{T}(\partial_i \alpha) = e^{i \int dx \left[ (\partial_i\alpha) E^i(x)\right] }.
\end{equation}

\pagebreak

\subsection{Gauge invariance and Gauss's law}
\label{novel3}

In order to show that, indeed, if $A_i=A_i+\partial_i \alpha$ then (where we drop the $E(x)$ in favor of $E$ to clean up the notation)

\begin{eqnarray}
\psi(A_i) &=& \psi (A_i+\partial_i \alpha) \nonumber \\
&=& \hat{T}(\partial_i \alpha) \psi(A_i) \nonumber \\
&=& e^{i \int dx \left[ (\partial_i \alpha) E^i\right] } \psi(A_i),
\end{eqnarray}

and we will make use of the following integration by parts

\begin{eqnarray}
\partial_i(\alpha E^i) &=& (\partial_i \alpha) E^i + \alpha (\partial_i E^i), \nonumber \\
(\partial_i \alpha) E^i &=& \partial_i(\alpha E^i) - \alpha (\partial_i E^i),
\end{eqnarray}

to make the following argument

\begin{eqnarray}
\psi(A_i) &=& \psi (A_i+\partial_i \alpha) \nonumber \\
&=& \hat{T}(\partial_i \alpha) \psi(A_i) \nonumber \\
&=& e^{i \int  dx \left[ (\partial_i \alpha E^i)\right] } \psi(A_i) \nonumber \\
&=& e^{i \int  dx  \left[ (\partial_i(\alpha E^i) - \alpha (\partial_i E^i))\right] } \psi(A_i) \nonumber \\
&=& e^{-i \int  dx \left[ \alpha (\partial_i E^i)\right] } \psi(A_i),
\end{eqnarray}

where we neglected the boundary term in the integral.

Since $\alpha$ is arbitrary, the only way for  $e^{-i \int dA \alpha (\partial_i E^i)}$ to equal 1 is for 

\begin{equation}
\partial_i E^i = 0,
\end{equation}

such that 

\begin{eqnarray}
\psi(A_i) &=& \psi (A_i+\partial_i \alpha) \nonumber \\
&=& \hat{T}(\partial_i \alpha) \psi(A_i) \nonumber \\
&=& e^{i \int dx\left[  (\partial_i \alpha) E^i\right] } \psi(A_i) \nonumber \\
&=& e^{i \int dx\left[  (\partial_i(\alpha E^i) - \alpha (\partial_i E^i)) \right] } \psi(A_i) \nonumber \\
&=& e^{-i \int dx \left[ \alpha (\partial_i E^i)\right] } \psi(A_i) \nonumber \\
&=& e^{-i \int dx\left[  \alpha (0)\right] } \psi(A_i) \nonumber \\
&=& e^{0} \psi(A_i) \nonumber \\
&=& \psi(A_i). 
\end{eqnarray}

And so the gauge invariance $A_i = A_i + \partial_i \alpha$ necessitates the Gauss’s law $\partial_i E^i = 0$. 

\begin{table}[H]
\centering 
\renewcommand{\arraystretch}{1.3}
 \begin{tabular}{| l || l | l|} 
 \hline
Gauge invariance & Gauss's law & Name \\ 
 \hline\hline 
 $A_i = A_i + \partial_i \alpha$ & $\partial_i E^i = 0$ & standard rank-1 (vector) gauge theory \\
 \hline
$A_{ij} = A_{ij} + \partial_i \partial_j \alpha$ & $\partial_i \partial_j E^{ij} = 0$  & scalar tensor gauge theory \\ 
 \hline  
$A_{ij} = A_{ij} + \frac{1}{2}(\partial_i\alpha_j + \partial_j \alpha_i)$ & $\partial_i E^{ij} = 0$ &vector tensor gauge theory \\
 \hline
$A_{ij} = A_{ij} + \delta_{ij} \alpha$ & $E^{i}{}_i = 0$  & traceless scalar tensor gauge theory\\
 \hline
\end{tabular}
\end{table}

This is for the standard vector gauge theory. What about higher-rank gauge theory?

At least for the first brand of tensor gauge theory, the scalar type, where the gauge invariance reads

\begin{equation}
A_{ij} = A_{ij} + \partial_i\partial_j \alpha.
\end{equation}

The procedure to show this corresponds to the Gauss’s law of the form

\begin{equation}
\partial_i\partial_j E^{ij} = 0,
\end{equation}

is nearly identical to that of the standard vector gauge theory we just did above. 

Here, the shift parameter is $\partial_i\partial_j \alpha$ as opposed to $\partial_i \alpha$ so we have 

\begin{eqnarray}
\psi(A_{ij}) &=& \psi (A_{ij}+\partial_i \partial_j \alpha) \nonumber \\
&=& \hat{T}(\partial_i \partial_j \alpha) \psi(A_{ij}) \nonumber \\
&=& e^{i \int dx\left[  (\partial_i \partial_j \alpha) E^{ij}\right] } \psi(A_{ij}) \nonumber \\
&=& e^{i \int dx\left[  (\partial_i\partial_j(\alpha E^{ij}) - \alpha (\partial_i\partial_j E^{ij}))\right] } \psi(A_{ij}) \nonumber \\
&=& e^{-i \int dx\left[  \alpha (\partial_i\partial_j E^{ij})\right] } \psi(A_{ij}) \nonumber \\
&=& e^{-i \int dx\left[  \alpha (0) \right] } \psi(A_{ij})  \nonumber \\
&=& e^{0} \psi(A_{ij}) \nonumber \\
&=& \psi(A_{ij}),
\end{eqnarray}

where the same exact steps were followed as in the vector gauge theory case.

One thing that we will use that is different in the vector tensor gauge theory is that the tensor gauge fields $A_{ij}$ and the conjugates $E^{ij}$ are symmetric tensors

\begin{eqnarray}
A_{ij} &=& A_{ji}, \nonumber  \\
E^{ij} &=& E^{ji}.
\end{eqnarray}

Here the gauge invariance is of the form 

\begin{equation}
A_{ij} = A_{ij} + \frac{1}{2}(\partial_i\alpha_j + \partial_j \alpha_i),
\end{equation}  
and we aim to show that the corresponding Gauss’s law is

\begin{equation}
\partial_i E^{ij} =0, 
\end{equation}

where the unsaturated indices explain the label ``vector” tensor gauge theory since the object $\partial_i E^{ij} \equiv x^j$ is a vector. This is different from the ``scalar” tensor gauge theory were the $\partial_i \partial_j E^{ij} \equiv c$ is a scalar. 

Here, the shift parameter is $\frac{1}{2}(\partial_i\alpha_j + \partial_j \alpha_i)$ as opposed to $\partial_i \alpha$ so we have

\begin{eqnarray}
\psi(A_{ij}) &=& \psi (A_{ij}+\frac{1}{2}(\partial_i\alpha_j + \partial_j \alpha_i)) \nonumber \\
&=& \hat{T}(\frac{1}{2}(\partial_i\alpha_j + \partial_j \alpha_i)) \psi(A_{ij}) \nonumber \\
&=& e^{i \int dx\left[  (\frac{1}{2}(\partial_i\alpha_j + \partial_j \alpha_i)) E^{ij}\right] } \psi(A_{ij}) \nonumber \\
&=& e^{i \int dx\left[  (\frac{1}{2}\partial_i\alpha_j E^{ij} + \frac{1}{2}\partial_j \alpha_i E^{ij})\right] } \psi(A_{ij}).
\end{eqnarray}

We then use the symmetry of $E^{ij}$ to combine the two terms in the exponential

\begin{eqnarray}
\frac{1}{2}\partial_i\alpha_j E^{ij} + \frac{1}{2}\partial_j \alpha_i E^{ij} &=& \frac{1}{2}\partial_i\alpha_j E^{ij} + \frac{1}{2}\partial_i \alpha_j E^{ji} \nonumber \\
&=&\frac{1}{2}\partial_i\alpha_j E^{ij} + \frac{1}{2}\partial_i \alpha_j E^{ij} \nonumber \\
&=&\partial_i\alpha_j E^{ij},
\end{eqnarray}

where in the first line we simply relabeled the indices (Shakespeare’s theorem). Now we just continue as usual with integration by parts

\begin{eqnarray}
\psi(A_{ij}) &=& \psi (A_{ij}+\frac{1}{2}(\partial_i\alpha_j + \partial_j \alpha_i)) \nonumber \\
&=& \hat{T}(\frac{1}{2}(\partial_i\alpha_j + \partial_j \alpha_i)) \psi(A_{ij}) \nonumber \\
&=& e^{i \int dx\left[  (\frac{1}{2}(\partial_i\alpha_j + \partial_j \alpha_i)) E^{ij}\right] } \psi(A_{ij}) \nonumber \\
&=& e^{i \int dx\left[  (\frac{1}{2}\partial_i\alpha_j E^{ij} + \frac{1}{2}\partial_j \alpha_i E^{ij}) \right]} \psi(A_{ij})   \nonumber \\
&=& e^{i \int dx\left[  (\partial_i\alpha_j E^{ij}) \right] } \psi(A_{ij})  \nonumber \\
&=& e^{-i \int dx\left[  \alpha_j (\partial_i E^{ij}) \right] } \psi(A_{ij})  \nonumber \\
&=& e^{-i \int dx\left[  \alpha_j (0) \right] } \psi(A_{ij}) \nonumber \\
&=& e^{0} \psi(A_{ij}) \nonumber \\
&=& \psi(A_{ij}).
\end{eqnarray}

\pagebreak

\section{Fracton field theory}
\label{fft}

Note that while these higher rank theories, with their higher-order charge conservation, do represent fractons (as we saw in Section \ref{charge}) we do not yet have a concise description like in the Hamiltonian pictures of lattice gauge theory as in Section \ref{LGT}. That is our current objective, to craft a theory\footnote{To follow the literature chronologically, a ``theory" here means a Lagrangian \cite{pretko}. There is however work on the Hamiltonian side \cite{banaj} as well.} describing the dynamics of fractons in terms of degrees of freedom while respecting the conservation laws we outlined above.

\subsection{Noether currents and continuity equations}
\label{noether}
\iffalse
\begin{quote}
2. Suppose you have a Lagrangian $L$ which is a functional of the complex scalar field $\Phi$, its complex conjugate $\Phi^*$, their time derivatives, and any number of spatial derivatives acting on these fields. Write down the Euler-Lagrange equation of motion for such a theory. Since you haven't given an explicit form of $L$, this equation will involve all manner of functional derivatives of $L$ with respect to all the things on which it can depend. By construction, this will take the form of a forced continuity equation: $\partial_t \rho + \partial_i j^i = f$, and you can identify $\rho$, $j^i$, and $f$ in terms of variations of the Lagrangian. For example, I believe $\rho = \Phi ( \partial L / \partial ( \partial_t \Phi ) )$.\footnote{This route forward was due to conservation with the author of \cite{kev} and \cite{newkev}, Kevin Grosvenor  \cite{kevcomm}.}
\end{quote}
\fi

To begin, we will include a field theory refresher following Tong \cite{tong} and Banerjee \cite{banerjee}.

%The generic variation of a Lagrangian reads:

%\begin{equation}
 %   \delta \LL = \frac{\delta \LL}{\partial \phi_a} + \frac{\delta \LL}{\partial (\partial_\mu \phi_a)} \partial_\mu (\delta \phi_a)
%\end{equation}

Consider the symmetry transformations $\delta \phi_a$ of equation 2.20 in \cite{thesis}

\begin{equation}
\label{u1sym}
    \phi \rightarrow e^{i\alpha} \phi \Longrightarrow \delta \phi = i \alpha \phi,
\end{equation}

where $\alpha$ is small. This is known as a global U(1) symmetry, since the symmetry parameter $\alpha$ alters the phase of the fields by the same amount regardless of the position, unlike if we had instead $\alpha(x)$, which would lead us to proper (local) gauge theory like in particle physics \cite{thesis}.

A Lagrangian which has this symmetry is the Schrödinger Lagrangian,

\begin{equation}
\label{schro}
    \LL = \frac{i}{2} (\phi^* \dot{\phi} - \dot{\phi}^* \phi) - \frac{1}{2m} \partial_i \phi^* \partial_i \phi.
\end{equation}

Using the Euler-Lagrange equations (which follow from setting the variation of the action, $S = \int \LL$, to zero, $\delta S =0$, see Tong equation 1.6 \cite{tong})

\begin{equation}
    \partial_\mu \bigg[ \frac{\partial \LL}{\partial(\partial_\mu \phi_a)} \bigg] - \frac{\partial \LL}{\partial \phi_a} = 0,
\end{equation}

on the Schrödinger Lagrangian \ref{schro}, appropriately yields the (albeit un-quantized, see Tong's section 2.8) Schrödinger equation,

\begin{eqnarray}
    \partial_\mu \bigg[ \frac{\partial \LL}{\partial(\partial_\mu \phi_a)} \bigg] - \frac{\partial \LL}{\partial \phi_a} &=& 0 \nonumber \\
    \partial_t \bigg[ \frac{\partial \LL}{\partial(\dot{\phi})} \bigg] - \partial_i \bigg[ \frac{\partial \LL}{\partial(\partial_i(\phi))} \bigg] - \frac{\partial \LL}{\partial \phi} &=& 0 \nonumber \\
    \partial_t \bigg[ \frac{i \phi^*}{2} \bigg] - \partial_i \bigg[ - \frac{\partial_i \phi^*}{2m} \bigg] - \frac{-i \dot{\phi}^*}{2} &=& 0 \nonumber \\
    \frac{i \dot{\phi}^*}{2}   + \frac{ \partial_i \partial_i \phi^*}{2m} + \frac{i \dot{\phi}^*}{2} &=& 0 \nonumber \\
    i \dot{\phi}^*   + \frac{ \nabla^2 \phi^*}{2m}  &=& 0, 
\end{eqnarray}

which we can conjugate to get the other field's equation

\begin{eqnarray}
\bigg(   i \dot{\phi}^*   + \frac{ \nabla^2 \phi^*}{2m} \bigg)^* &=& 0 \nonumber \\
    - i \dot{\phi}   + \frac{ \nabla^2 \phi}{2m}  &=& 0. 
\end{eqnarray}

As stated above, the theory has the following symmetry 
\begin{eqnarray}
\delta \phi &=& i \alpha \phi, \nonumber \\
\delta \phi^* &=& -i \alpha \phi^*, 
\end{eqnarray}

By Noether's theorem, we have that the continuous symmetry leads to a conservation law — more formally, leads to a conserved current $j^\mu$,

\begin{equation}
\label{current}
    j^\mu = \frac{\partial \LL}{\partial(\partial_\mu \phi_a)}\delta \phi_a - F^\mu.
\end{equation}

The formalization of the conservation of this current is known at the continuity equation — decomposed into components with a Minkowski metric $(+,-,-,-)$

\begin{eqnarray}
\label{contref}
    \partial_\mu j^\mu &=& 0 \nonumber \\
    \partial_0 j^0 - \partial_i j^i &=& 0 \nonumber \\
    \frac{\partial \rho}{\partial t} - \nabla \cdot j &=& 0,
\end{eqnarray}

where the identification $\rho = j^0$ comes from $Q = \int dV \rho$ in electrodynamics and leads to the concept of a Noether charge (a conserved quantity associated with the flowing current)

\begin{equation}
\label{chargeref}
Q = \int dV j^0,
\end{equation}

which we will look at more deeply in Section \ref{dipcharge}.

Note that the $F^\mu$ in equation \ref{current} are arbitrary functions of $\phi$, and simply allows for flexibility in the obtaining $\delta S=0$ since varying the Lagrangian like a total derivative $\delta \LL = \partial_\mu F^\mu$ would end up as an integral of a total derivative in the variation of the action since $S=\int \LL$, and we would still have the continuity equation $\partial_\mu j^\mu = 0$.

We can ignore $F^\mu$ in this warm up with the Schrödinger Lagrangian (and more generally ignore it for the rest of this work, as the dipole symmetry we cover later will not require any $F^\mu$ bookkeeping either). In this case, we can be more specific about the action principle, $\delta S =0$, and require 

\begin{equation}
    \delta \LL = 0.
\end{equation}

We can use the definition of a conserved current, equation \ref{current}, to prove that the continuity equation above is satisfied on-shell (when the Euler Lagrange equations of motion hold). The key thing to keep in mind, which we will circle back to in concluding this section, is that \textit{both} the continuity equation and the Euler Lagrange equations follow from the requirement that the Lagrangian not vary.

Processing to find the currents, as per equation \ref{current}, for the Schrödinger Lagrangian and plugging then into the continuity equation, we have

\begin{eqnarray}
\partial_0 j^0 &=& \partial_t \bigg[ \frac{\partial \LL}{\partial(\partial_t \phi_a)}\delta \phi_a \bigg] \nonumber \\
&=& \partial_t \bigg[ \frac{\partial \LL}{\partial(\dot{\phi})} i \alpha \phi + \frac{\partial \LL}{\partial(\dot{\phi}^*)} (-i \alpha \phi^*) \bigg] \nonumber \\
&=& \partial_t \bigg[ \frac{i \phi^*}{2} i \alpha \phi + \bigg(-\frac{i \phi}{2}\bigg) (-i \alpha \phi^*) \bigg] \nonumber \\
&=& \partial_t \bigg[ - \alpha \phi^* \phi \bigg] \nonumber \\
&=& - \alpha (\dot{\phi}^* \phi+\phi^*\dot{\phi}),
\end{eqnarray}

and

\begin{eqnarray}
\partial_i j^i &=& \partial_i \bigg[ \frac{\partial \LL}{\partial(\partial_i \phi_a)}\delta \phi_a \bigg] \nonumber \\
&=& \partial_i \bigg[ \frac{\partial \LL}{\partial(\partial_i\phi)} i \alpha \phi + \frac{\partial \LL}{\partial(\partial_i \phi^*)} (-i \alpha \phi^*) \bigg] \nonumber \\
&=& \frac{i \alpha}{2m} \partial_i \bigg[ (-\partial_i \phi^*) \phi -  (-\partial_i \phi) \phi^* \bigg] \nonumber \\
&=& \frac{i \alpha}{2m} \partial_i \bigg[ \phi (-\partial_i \phi^*)  +  \phi^* (\partial_i \phi)  \bigg] \nonumber \\
&=& \frac{i \alpha}{2m} ( -\partial_i \phi \partial_i \phi^* - \phi \nabla^2 \phi^*   +  \partial_i \phi^*  \partial_i \phi + \phi^* \nabla^2 \phi) \nonumber \\
&=& \frac{i \alpha}{2m} ( - \phi \nabla^2 \phi^*   + \phi^* \nabla^2 \phi).
\end{eqnarray}

Note that the final form of the current reads

\begin{eqnarray}
\label{anchor}
    j^0 &=&  - \phi^* \phi \nonumber \\
    j^i &=& \frac{i}{2m} (  \phi^* \partial_i \phi - \phi \partial_i \phi^*)  
\end{eqnarray}

where the symmetry parameter $\alpha$ is ignored as per standard procedure in field theory (see equation 2.22 in \cite{thesis}). In the next section we will examine $ j^0 =  - \phi^* \phi $ from a quantum mechanical perspective.

One quick step before showing that: with these definitions of $j^0$ and $j^i$ along with the Euler Lagrange equations of motion, that the continuity equation $\partial_0 j^0 - \partial_i j^i = 0$ holds, let's shift the $i$'s in the equations of motion to match what we'll encounter when writing out $\partial_0 j^0 - \partial_i j^i = 0$,

\begin{eqnarray}
i \dot{\phi}^*   + \frac{ \nabla^2 \phi^*}{2m}  &=& 0, \nonumber \\
- \dot{\phi}^*   + \frac{ i \nabla^2 \phi^*}{2m}  &=& 0, 
\end{eqnarray}
similarly,
\begin{eqnarray}
    - i \dot{\phi}   + \frac{ \nabla^2 \phi}{2m}  &=& 0, \nonumber \\
     \dot{\phi}   + \frac{i \nabla^2 \phi}{2m}  &=& 0. 
\end{eqnarray}

Now we have 

\begin{eqnarray}
\partial_0 j^0 - \partial_i j^i &=& - \alpha (\dot{\phi}^* \phi+\phi^*\dot{\phi}) - \frac{i \alpha}{2m} ( - \phi \nabla^2 \phi^*   + \phi^* \nabla^2 \phi) \nonumber \\
&=&- \dot{\phi}^* \phi-\phi^*\dot{\phi} + \frac{i }{2m}  \phi \nabla^2 \phi^*   - \frac{i }{2m}\phi^* \nabla^2 \phi) \nonumber \\
&=& \phi(- \dot{\phi}^* + \frac{i }{2m}  \nabla^2 \phi^* ) -   \phi^* (\dot{\phi}    + \frac{i }{2m} \nabla^2 \phi) \nonumber \\
&=& \phi(0) -   \phi^* (0) \nonumber  \\
&=& 0.
\end{eqnarray}

Thus, when the Euler-Lagrange equations are satisfied (which, as mentioned before, follow from the modified action principle $\delta \LL = 0$) we have the continuity equation

\begin{equation}
    \delta \LL = 0 \qquad \Longleftrightarrow \qquad \partial_0 j^0 - \partial_i j^i =0.
\end{equation}

Equation 22.7 from Srednicki's text on quantum field theory\footnote{Thanks is due to Hiroki Sukeno for pointing out this equation \cite{Hiroki}.} summarizes this relationship between the variation of the action/Lagrangian, Euler-Lagrange equation, and continuity equation very succinctly \cite{Srednicki}

\begin{equation}
\partial_\mu j^\mu = \delta \LL - \frac{\delta S}{\delta \phi_a} \delta \phi_a,
\end{equation}

where we have $\frac{\delta S}{\delta \phi_a}=0$ when the Euler-Lagrange equations hold such that 

\begin{equation}
\partial_\mu j^\mu = \delta \LL.
\end{equation}

Now either the symmetry doesn't lead to any variation in the Lagrangian and $\delta \LL =0$ (as will be the case for this work) or as mentioned earlier, $\delta \LL = \partial_\mu F^\mu$, and the $F^\mu$ is absorbed into the definition of $j^\mu$. Either way, we end up with the continuity equation $\partial_\mu j^\mu=0$.

If we work through $\delta \LL = 0$ but for a more general Lagrangian (notably higher derivatives), we will end up with another version of the continuity equation, notably one that gives us conservation of dipole moment as opposed to only charge conservation above.

\pagebreak

\subsection{Quantum view of the U(1) Noether charge}

Note that all that we have discussed above is with respect to the field $\phi$. What can we learn from looking at the states? We have from quantum mechanics \cite{sakurai} \cite{tong},

\begin{eqnarray}
|k_i\rangle &=& \hat{a_i}^\dagger |0\rangle \nonumber \\
|n_i\rangle &=& (\hat{a_i}^\dagger)^n |0\rangle \nonumber \\
0 &=& a_i |0\rangle \nonumber \\
\delta_{ij}|0\rangle &=& a_i |k_j\rangle
\end{eqnarray}

where $|0\rangle$ is the vacuum, $|k_i\rangle$ is a state with eigenvalue $k_i$, $|n_i\rangle$ is the state with $n$ particles of eigenvalue $k_i$, $\hat{a}^\dagger$ is the creation operator, and $\hat{a}$ the annihilation operator.

One can construct a ``number operator" $N$ from the creation and annihilation operators

\begin{equation}
\label{nnn}
    N = \int \frac{d^3 k_i}{2 \pi^3} a_i{}^\dagger a_i
\end{equation}

where the $\int \frac{d^3 k_i}{2 \pi^3}$ arises due to Fourier transforming between the creation and annihilation operator picture and the field picture. As seen in Tong's equation 2.18 \cite{tong}

\begin{equation}
    \phi \approx \int \frac{d^3 k_i}{2 \pi^3} (a_i + a_i{}^\dagger )
\end{equation}

where we've neglected all the prefactors and wave solution aspects of the equation to focus on the decomposition of the field in terms of creation and annihilation operators. Bringing equation \ref{nnn} into position space would read (see equations 2.74 and 2.75 of Tong for a similar calculation, and the discussion above 2.104 for the charge \cite{tong})

\begin{eqnarray}
N &=& \int \frac{d^3 k_i}{2 \pi^3} a_i{}^\dagger a_i \nonumber \\
Q &=& - \int d^3x \phi^* \phi
\end{eqnarray}

where we can recognize $-\phi^* \phi$ as the temporal part of the Noether current $j^0$ from equation \ref{anchor} such that $Q = \int dV j^0$ as expected. This similarity between number operator and charge operator makes sense from the point of view that: if our particles are charged particles, and we assume unit charge for each particle, then the number of particles tells us the charge.

\iffalse
Thus, if the matter field transforms as above, the creation operator does as well, because of it’s having built up the field from the vacuum

\begin{equation}
\hat{a}^\dagger \rightarrow  e^{i \alpha} \hat{a}^\dagger.
\end{equation}

By conjugating, we have 

\begin{equation}
\hat{a} \rightarrow  e^{-i \alpha} \hat{a}.
\end{equation}

Consider the operator in quantum mechanics for the number of particles (where, if we are considering charged particles, can also be regarded as an operator for total charge)

\begin{equation}
\hat{N} =  \hat{a}^\dagger \hat{a}.
\end{equation}

By both or the creation and annihilation operator transformation above, 

\begin{eqnarray*}
\hat{N} &=& \hat{a}^\dagger \hat{a} \\
&\rightarrow& e^{i \alpha} \hat{a}^\dagger e^{-i \alpha} \hat{a} \\
&=& e^{i \alpha}  e^{-i \alpha} \hat{a}^\dagger \hat{a} \\
&=& e^{0} \hat{a}^\dagger \hat{a} \\
&=& \hat{a}^\dagger \hat{a} \\
&=& \hat{N}.
\end{eqnarray*}

Thus, a U(1) symmetry in a system tells us that the particle number/total charge is conserved.

In the field point of view 

\begin{eqnarray}
\label{posit}
\hat{N} |0\rangle &=& \hat{a}^\dagger \hat{a} |0\rangle \nonumber \\
&=& |\phi^* \phi\rangle
\end{eqnarray}

This $\phi^* \phi$ is the conserved charge $Q$ from Noether's theorem (as we will cover momentarily) as we discuss conserved currents and continuity equations following Tong's section 1.3 \cite{tong}.
\fi
\pagebreak

\subsection{Tensor gauge theory and dipole moment}
\label{conscon}
Total charge being conserved isn’t all we need for fractonic systems! As we learned earlier, we need conservation of dipole moment too. How can we accommodate that conservation law in terms of a invariance in the fields?

Recall the main advancement of TENSOR gauge theory. 

As opposed to a shift by a vector, like $\partial_i \alpha$ in the transformation $A_i = A_i + \partial_i \alpha$, we upgraded to shifting by a tensor, like $\partial_i\alpha_j$ in the transformation $A_{ij} = A_{ij} + \frac{1}{2}(\partial_i\alpha_j + \partial_j \alpha_i)$.

So maybe instead of $\phi \rightarrow e^{i \alpha} \phi$ from equation \ref{u1sym}, we ought to upgrade the rank of the symmetry parameter $\alpha$ 

\begin{equation}
\label{sym1}
\phi \rightarrow e^{i \vec{\alpha}} \phi.
\end{equation}

Why isn’t this allowed? Well, mathematically there is a vector in the exponential. Recall that in the translation operator

\begin{equation}
\hat{T}(\Delta x) = e^{-i\Delta x \hat{p}} \rightarrow \hat{T}(\partial_i \alpha) e^{-i (\partial_i \alpha) E^i}.
\end{equation}

BOTH $\hat{p}$ and $E^i$ were vectors. Thus instead of $\phi \rightarrow e^{i \vec{\alpha}} \phi$ we need some other vector up in the exponential to dot with the $\vec{\alpha}$ to yield an overall scalar.

Pretko states this invariance takes the form  

\begin{equation}
\label{sym2}
\phi \rightarrow e^{i \vec{\alpha}\cdot \vec{x}} \phi = e^{i \alpha_i x^i} \phi
\end{equation}

Where does this symmetry come from?\footnote{Significant inspiration in answering this via Noether currents is due to conservation with Hirkoki Sukeno and Li Yabo \cite{phds}.} We will proceed to answer this as follows: 

\begin{enumerate}

   \item firstly in Section \ref{kevhelp} we will prove the existence of a dipole moment version of the continuity equation and Noether charge. This will lead to a conserved dipole moment, which is precisely what we hope to obtain because we saw in Section \ref{charge} that conservation of dipole moment accounts for the fractal phenomenology of fractons.

   \item for the remainder of the Chapter, we will show that (as opposed to the $U(1)$ symmetry $\delta \phi = i \alpha \phi$ discussed in the past two sections that led to a conservation of total charge) the symmetry of equation \ref{sym2} is the necessary symmetry to arrive at a conserved charge (in this case, the charge will be the dipole moment, so $Q^i$ as opposed to $Q$) due to Noether's theorem. We will find out along the way that the Lagrangian must depend on higher derivatives (corresponding to a higher rank object — just like tensors of the previous chapter on tensor gauge theory) in order for the symmetry to lead to a higher rank Noether charge, $Q^i$.
 
\end{enumerate}

%\vspace{7cm}
\pagebreak

\subsection{Dipole moment continuity equation} 
\label{dipcharge}
\label{kevhelp}

\subsubsection{Scalar Noether charges}

Before we delve into the dipole moment (vector) Noether charge, let's review the standard (scalar) version we promised beneath equation \ref{chargeref}. We said that the Noether charge $Q$ is a conserved quantity, i.e. we want to show that \cite{tong}

\begin{equation}
\frac{d}{d t} Q = 0,
\end{equation}

which reads as follows when we utilize the definition of the $Q$ from \ref{chargeref} (noting that the order in which we take the time derivative and do the spatial integration is irrelevant)

\begin{eqnarray}
\frac{d}{d t} \int dV j^0 &=& 0 \nonumber \\
\int dV \bigg(\frac{d}{d t} j^0\bigg) &=& 0.
\end{eqnarray}

Using the continuity equation \ref{contref}, we can replace that $\frac{d}{d t} j^0$ with $-\nabla \cdot j$

\begin{eqnarray}
\int dV \bigg(\frac{d}{d t} j^0\bigg) &=& 0 \nonumber \\
- \int dV (\nabla \cdot j) &=& 0.
\end{eqnarray}

Since this is a volume integral of a divergence, we can use the divergence theorem \cite{wiki}

\begin{eqnarray}
\label{finalcharge}
\int dV (\partial_i j^i) &=& 0 \nonumber \\
\int dS (n_i j^i) &=& 0 \nonumber \\
\int dS_i  j^i &=& 0.
\end{eqnarray}

The key question to ask now is, when is equation \ref{finalcharge} true? It's true when the volume dV (equivalently, the surface area dS) is large enough. For example, consider the case where the volume (surface area) we are integrating over is infinitely large (infinitely far away). What does this mean? Firstly, what is this quantity $\int dS_i  j^i$? It is a flux quantifying the amount of current flowing out of an area. As an analogy: consider the current, electric field, and electric flux due to a single charge $Q$. A point charge $Q$ will cause an electric field $E$, but this does not extend infinitely far away, but rather falls off like $1/r^2$. Thus, if the volume $V$ (surface area $S$) enclosing the charge $Q$ is large enough (far enough away), the flux approaches $0$ since the $E$ does as well as we travel farther from $Q$.

So we can say equation \ref{finalcharge} holds exactly in the limit of integrating at spatial infinity,

\begin{equation}
\label{infty}
\frac{d}{d t} Q = \int_\infty dS_i  j^i = 0.
\end{equation}

Otherwise, we simply have the local conservation,

\begin{equation}
\frac{d}{d t} Q = \int dS_i  j^i,
\end{equation}

which is to be interpreted as the statement that: if there is any charge $Q$ leaving a volume, it must be accompanied by a current $j^i$ flowing out of said volume \cite{tong}.

\subsubsection{Vector Noether charges}

Now we will move onto the dipole moment Noether charge. Our objective here is to show that if the charge density $\rho$ satisfies a continuity equation of the sort $\partial_t \rho + \partial_i \partial_j J^{ij} = 0$, then the dipole moment is conserved. That is, $\partial_t Q^i = 0$, where $Q^i = \int d^d x \rho x^i$, in d spatial dimensions.\footnote{This route forward was due to conservation with the author of \cite{kev} and \cite{newkev}, Kevin Grosvenor  \cite{kevcomm}.}

The dipole moment is defined as 

\begin{equation}
    Q^i = \int dV x^i \rho (t,x),
\end{equation}

just as in equation 3.98 of \cite{griffiths}, the generalized version of a physical dipole (equation 3.101)

\begin{equation}
    \vec{p} = q \vec{d}.
\end{equation}

The continuity equation reads
\begin{eqnarray}
\label{ccc}
    \nabla \cdot J + \frac{\partial \rho}{\partial t} &=&  0, \nonumber \\
    \partial_i \partial_j J^{ij} + \frac{\partial \rho}{\partial t} &=&  0.
\end{eqnarray}

Looking at $\frac{d Q^i}{dt}$

\begin{eqnarray}
    \frac{d Q^i}{dt} &=& \frac{d}{dt} \int dV x^i \rho (t,x) \nonumber \\
    &=& \int dV x^i \frac{d \rho}{dt} \nonumber \\
    &=& -\int dV x^i (\partial_j\partial_k J^{jk}),
\end{eqnarray}

and then integrating by parts,

\begin{eqnarray}
    \partial_j \partial_k (x^i J^{jk}) &=& \partial_j \partial_k (x^i) J^{jk} + x^i \partial_j \partial_k ( J^{jk}) \nonumber \\
   \partial_j \partial_k (x^i J^{jk}) &=& \partial_j \delta_k{}^i J^{jk} + x^i \partial_j \partial_k ( J^{jk}) \nonumber \\
   \partial_j \partial_k (x^i J^{jk}) &=& \partial_j J^{ji} + x^i \partial_j \partial_k ( J^{jk}) \nonumber \\
\int dV  \left[   \partial_j \partial_k (x^i J^{jk})\right] &=& \int dV  \left[ \partial_j J^{ji} + x^i \partial_j \partial_k ( J^{jk}) \right]\nonumber \\
0 &=& \int dV  \left[ \partial_j J^{ji} + x^i \partial_j \partial_k ( J^{jk}) \right]\nonumber \\
 \int dV   \partial_j J^{ji} &=&  -\int dV x^i \partial_j \partial_k ( J^{jk}), 
\end{eqnarray}

so that we have 

\begin{eqnarray}
    \frac{d Q^i}{dt} &=& \frac{d}{dt} \int dV x^i \rho (t,x) \nonumber \\
    &=& \int dV x^i \frac{d \rho}{dt} \nonumber \\
    &=& -\int dV x^i (\partial_j\partial_k J^{jk}) \nonumber \\
    &=&  \int dV \partial_j J^{ji}.
\end{eqnarray}

Note that $\partial_j$ is still just a divergence, the quantity $\partial_j J^{ji}$ is just the divergence of a tensor as opposed to a vector \cite{wiki}. This enables us to use the divergence theorem just like we did in equation \ref{finalcharge} to state

\begin{eqnarray}
    \frac{d Q^i}{dt}    &=&  \int dV (\partial_j J^{ji}) \nonumber \\
    &=& \int dS (n_j J^{ji}) \nonumber \\
     &=& \int dS_j J^{ji}
\end{eqnarray}

And just like in equation \ref{infty} we can either enforce a sufficiently large volume (surface area) and integrate at infinity, 

\begin{equation}
    \frac{d Q^i}{dt} = \int_\infty dS_j J^{ji} = 0,
\end{equation}

or settle with the local conservation

\begin{equation}
    \frac{d Q^i}{dt} = \int dS_j J^{ji}, 
\end{equation}

which is the be interpreted exactly as in the scalar case: any change in dipole moment $Q^i$ in a volume must be accounted for by the flow of current $J^{ji}$ into or out of said volume.

\iffalse
which is identical to the statement of conservation of charge $Q = \int dV \rho$, and reads
\begin{equation}
      \frac{d Q}{dt}  = \int dS J^i,
\end{equation}
implying that any charge leaving a volume $V$ must be accompanied by a flow of current $J^i$ out of the volume \cite{tong}.

%\edit{\textbf{HOW DO I SAY AN EQUIVALENT THING IN THIS DIPOLE CASE?}} 
Attempting to verbally replicate such a statement, we can say:
\begin{quote}
    any change in dipole moment $Q^i$ within a volume must be accounted for by the flow of a current $J^i$, which is a component of a tensor, $J^i= \partial_j J^{ji}$, out of the volume.
\end{quote}
\fi
Now we need to start studying the Noether currents/charges corresponding to symmetries like those encountered in equation \ref{sym2} to show that such a symmetry indeed leads to the continuity equations we saw in this section to link the conserved dipole moment with a particular Lagrangian and symmetry.

\pagebreak

\subsection{Necessity of higher derivatives/tensors }
\label{need}

We start by upgrading the constant symmetry of equation \ref{u1sym}, to a shift  — notably one of a linear polynomial shift symmetry (according to the original work of Gromov and Grosvenor \cite{gromov} \cite{kev}, but here we follow Banerjee \cite{banerjee}),

\begin{eqnarray}
 \delta \phi = i \alpha \phi \qquad \longrightarrow \qquad \delta \phi &=& i \alpha(x) \phi \nonumber \\
 &=& i (\alpha_0 + \alpha_i x^i) \phi \nonumber \\
 &=& i \alpha_0 \phi + i \alpha_i x^i \phi,
\end{eqnarray}
   
where $\alpha_0$ and $\alpha_i$ are arbitrary constants.

Following Tong's equation 1.36, we can write out the invariance under a variation of a generic Lagrangian $\LL(\phi, \partial_\mu \phi)$ using the symmetry (sans the imaginary $i$ to clean things up) written above \cite{tong} \footnote{All variations of Lagrangians for the rest of this chapter contain implicitly variations with respect to the complex conjugate of the field as well, just like in Section \ref{noether}. We neglect them here for brevity, but will reintroduce them in Section \ref{gaugetheory} once the theory has been solidly grounded.}

\begin{eqnarray}
\label{full}
0&=& \delta \LL = \frac{\partial \LL}{\partial \phi_a} \delta \phi_a + \frac{\partial \LL}{\partial (\partial_\mu \phi_a)} \partial_\mu \delta \phi_a \nonumber \\
&=& \frac{\partial \LL}{\partial \phi} \delta \phi + \frac{\partial \LL}{\partial (\partial_t \phi)} \partial_t \delta \phi + \frac{\partial \LL}{\partial (\partial_i \phi)} \partial_i \delta \phi \nonumber \\
&=& \frac{\partial \LL}{\partial \phi} (\alpha_0 + \alpha_i x^i) \phi + \frac{\partial \LL}{\partial \dot{\phi}} \partial_t \left[(\alpha_0 + \alpha_i x^i) \phi\right] + \frac{\partial \LL}{\partial (\partial_i \phi)} \partial_i \left[(\alpha_0 + \alpha_i x^i) \phi\right] \nonumber \\
&=& \frac{\partial \LL}{\partial \phi} (\alpha_0 + \alpha_i x^i) \phi + \frac{\partial \LL}{\partial \dot{\phi}} (\alpha_0 + \alpha_i x^i) \dot{\phi}  + \frac{\partial \LL}{\partial (\partial_i \phi)}  \alpha_i \partial_i(x^i) \phi +\frac{\partial \LL}{\partial (\partial_i \phi)} (\alpha_0 + \alpha_i x^i) \partial_i \phi \nonumber \\
&=& \frac{\partial \LL}{\partial \phi} (\alpha_0 + \alpha_i x^i) \phi + \frac{\partial \LL}{\partial \dot{\phi}} (\alpha_0 + \alpha_i x^i) \dot{\phi}  + \frac{\partial \LL}{\partial (\partial_i \phi)}  \alpha_i (1) \phi +\frac{\partial \LL}{\partial (\partial_i \phi)} (\alpha_0 + \alpha_i x^i) \partial_i \phi \nonumber \\
 &=& \left[ \alpha_0 + \alpha_i x^i \right] \bigg[\frac{\partial \LL}{\partial \phi} ( \phi)  +  \frac{\partial \LL}{\partial \dot{\phi}}   \dot{\phi} + \frac{\partial \LL}{\partial (\partial_i \phi)} ( \partial_i \phi)\bigg]  + \frac{\partial \LL}{\partial (\partial_i \phi)}  \alpha_i \phi, 
\end{eqnarray}

which is identical to Banerjee's equation 8, except that the last term is split off and written on its own like in Banerjee's two equations:

\begin{eqnarray}
0&=& \left[ \alpha_0 + \alpha_i x^i \right] \bigg[\frac{\partial \LL}{\partial \phi}  \phi  +  \frac{\partial \LL}{\partial \dot{\phi}}   \dot{\phi} + \frac{\partial \LL}{\partial (\partial_i \phi)} ( \partial_i \phi) \bigg] , \nonumber \\
0&=& \frac{\partial \LL}{\partial (\partial_i \phi)}  \alpha_i \phi, 
\end{eqnarray}

and these are further simplified to exclude the symmetry constants $\alpha$ 

\begin{eqnarray}
\label{eqq1}
0&=& \frac{\partial \LL}{\partial \phi}  \phi  +  \frac{\partial \LL}{\partial \dot{\phi}}   \dot{\phi} + \frac{\partial \LL}{\partial (\partial_i \phi)} ( \partial_i \phi),   \nonumber \\
0&=& \frac{\partial \LL}{\partial (\partial_i \phi)}   \phi. 
\end{eqnarray}

Why are these TWO equations?\footnote{Much thanks is due to Li Yabo for pointing out this route to solve this issue \cite{yabo}.}

Consider the following representation of our equation above to help visualize the logic behind this step,

\begin{eqnarray}
0 &=& \left[ \alpha_0 + \alpha_i x^i \right] \bigg[\frac{\partial \LL}{\partial \phi} ( \phi)  +  \frac{\partial \LL}{\partial \dot{\phi}}   \dot{\phi} + \frac{\partial \LL}{\partial (\partial_i \phi)} ( \partial_i \phi)\bigg]  + \frac{\partial \LL}{\partial (\partial_i \phi)}  \alpha_i \phi \nonumber \\
&=& (A+B) (C+D+E) + BF.
\end{eqnarray}

Consider the fact that $\alpha_0$ and $\alpha_i$ are arbitrary constants. I.e. our equation above must hold for any and all choice of $\alpha_0$ and $\alpha_i$. Notably, what if $B=0$ while $A$ continues to be arbitrary (can take on any value)?

\begin{eqnarray}
0 &=& (A+B) (C+D+E) + BF \nonumber \\
&=& A (C+D+E).
\end{eqnarray}

But if $A$ is arbitrary, we can't set it to zero, thus we must have $C+D+E=0$ (which is the first of Banerjee's equations above). THEN we are left with 

\begin{eqnarray}
0 &=& (A+B) (C+D+E) + BF \nonumber \\
&=& BF \nonumber \\
&=& \alpha_i \frac{\partial \LL}{\partial (\partial_i \phi)}\phi \nonumber \\
&=& \frac{\partial \LL}{\partial (\partial_i \phi)}\phi,
\end{eqnarray}
where, again, $B=\alpha_i$ is arbitrary so we can't set it to zero. Likewise, setting $\phi$ to zero is a trivial solution, so we must have 

\begin{equation}
\label{eqq2}
    \frac{\partial \LL}{\partial (\partial_i \phi)} = 0.
\end{equation}

This tells us that $\LL$ cannot depend on $\partial_i \phi$, only on the field $\phi$ and it's time-derivative $\dot{\phi}$.

What's the problem here?\footnote{Much thanks is due to Hiroki Sukeno for explaining the problem with such a theory \cite{Hiroki}.}

Recall that the whole of quantum field theory was born out of the progression of wave mechanics to quantum wave mechanics á la Louis de Broglie, Erwin Schrödinger, Max Born, and company. Our mention of waves is deliberate, since without any dependence on $\partial_i \phi$, we have no way to view wave behavior in the field $\phi$.

Notice there is spatial derivatives ($\nabla^2$) in: the wave equation, the Schrödinger equation, and the Klein-Gordon equation,

\begin{eqnarray}
\bigg(\frac{1}{c^2}\frac{\partial^2}{\partial t^2} - \nabla^2\bigg) f(t,\vec{x}) &=& 0, \nonumber \\
\bigg(i \hbar \frac{\partial }{\partial t} - \frac{\hbar^2}{2m}\nabla^2\bigg) |\psi(t)\rangle &=& 0, \nonumber \\
\bigg(\frac{1}{c^2}\frac{\partial^2}{\partial t^2} - \nabla^2\bigg) \psi(t,\vec{x}) &=& 0.
\end{eqnarray}

And so, if we don't have spatial dependence in our theory, we don't have any wave behavior and thus there is no way for information to travel from one point to another since wave propagation is the mechanism of transferring information in both quantum mechanics and field theory. So either we have a very boring theory, or we have one that violates the causality of special relativity. Neither of these options are preferable, so we look at ways to include spatial dependence in the theory — notably, by including higher spatial dependence in the Lagrangian.
\pagebreak

\subsection{Varying a higher order Lagrangian}
\label{higher}

As Tong mentions in the paragraph below equation 1.4, without the restriction of trying to keep Lorentz invariance, there is no reason not to include higher spatial derivative terms. It makes sense to not include higher temporal derivatives in our case\footnote{Tong's general classical mechanics comment below his equation 1.4 also justifies this: ``Recall that in particle mechanics $L$ depends on $q$ and $\dot{q}$ but not $\ddot{q}$. In field theory we similarly restrict to Lagrangians $\LL$ depending on $\phi$ and $\dot{\phi}$ but not $\ddot{\phi}$."}  since the problem term we encounters was with regards to spatial dependence of $\LL$ missing \cite{tong}.

Thus instead of a full set of 2nd derivatives $\partial_\mu \partial_\nu$ we will only consider 2 spatial derivatives $\partial_i \partial_j$.

Proceeding in varying $\LL$ as we did before with our given symmetry,

\begin{eqnarray}
\label{var}
0= \delta \LL &=& \frac{\partial \LL}{\partial \phi} \delta \phi + \frac{\partial \LL}{\partial (\partial_\mu \phi)} \partial_\mu \delta \phi + \frac{\partial \LL}{\partial (\partial_\mu\partial_\nu \phi)} \partial_\mu\partial_\nu \delta \phi \nonumber \\
&=& \frac{\partial \LL}{\partial \phi} \delta \phi + \frac{\partial \LL}{\partial \dot{\phi}} \partial_t(\delta \phi) + \frac{\partial \LL}{\partial (\partial_i \phi)} \partial_i (\delta \phi)  +\frac{\partial \LL}{\partial (\partial_i\partial_j \phi)} \partial_i\partial_j (\delta \phi) 
%&=& \frac{\partial \LL}{\partial \phi} \delta \phi + \frac{\partial \LL}{\partial \dot{\phi}} \partial_t(\delta \phi) + \frac{\partial \LL}{\partial (\partial_i \phi)} \partial_i (\delta \phi) + \frac{\partial \LL}{\partial (\partial_t\partial_i \phi)} \partial_i \partial_t (\delta \phi) +\frac{\partial \LL}{\partial (\partial_i\partial_j \phi)} \partial_i\partial_j (\delta \phi) \nonumber \\
%&=& \frac{\partial \LL}{\partial \phi} (\alpha_0 + \alpha_i x^i) \phi + \frac{\partial \LL}{\partial \dot{\phi}} (\alpha_0 + \alpha_i x^i) \dot{\phi}  + \frac{\partial \LL}{\partial (\partial_i \phi)}  \alpha_i  \phi +\frac{\partial \LL}{\partial (\partial_i \phi)} (\alpha_0 + \alpha_i x^i) \partial_i \phi \nonumber \\
%&+& \frac{\partial \LL}{\partial (\partial_t\partial_i \phi)} \partial_i \left[ (\alpha_0+\alpha_i x^i) \dot{\phi} \right] +\frac{\partial \LL}{\partial (\partial_i\partial_j \phi)} \partial_i\partial_j \left[ (\alpha_0+\alpha_i x^i) \phi \right] 
\end{eqnarray}

The next step is aimed at mimicking the procedure of Section \ref{noether}, which can be summarized as\footnote{Thanks is due to Li Yabo for pointing out this route to the derivation \cite{yabo}.}

\begin{quote}
    We used the formulas for conserved currents to prove that the continuity equation above is satisfied on-shell (when the Euler Lagrange equations of motion hold).
\end{quote}

In this case however, we will be rearranging the above variation eq. \ref{var}, throwing out terms that all together equal zero on-shell due to the Euler-Lagrange equations of motion (which are all those proportional to $\delta \phi$, see Tong's equation 1.5 \cite{tong}), and leaving terms that will ultimately take on the roles of the currents in the continuity equation. 

Note that varying the Lagrangian is done according to the principle of least action, $\delta S =0$ where the action $S$ is the integral of the Lagrangian, $S= \int \LL$. This is notable because we are then allowed to rearrange the terms of the variation eq. \ref{var} above using an integration by parts procedure.

Notably,

\begin{equation}
\label{eq1}
\frac{\partial \LL}{\partial \dot{\phi}} \partial_t(\delta \phi) = \partial_t\bigg(\frac{\partial \LL}{\partial \dot{\phi}} \delta \phi \bigg) - \partial_t \bigg( \frac{\partial \LL}{\partial \dot{\phi}} \bigg) \delta \phi,
\end{equation}

where the first term is the one we already have in eq. \ref{var}, and the last term is the proportional to $\delta \phi$ term for the Euler-Lagrange equation of motion, and we can anticipate that the middle term will be part of the current.

Continuing with the next term of eq. \ref{var}

\begin{equation}
\label{eq2}
\frac{\partial \LL}{\partial (\partial_i \phi)} \partial_i (\delta \phi) = \partial_i \bigg( \frac{\partial \LL}{\partial (\partial_i \phi)} \delta \phi \bigg) - \partial_i \bigg( \frac{\partial \LL}{\partial (\partial_i \phi)} \bigg) \delta \phi,
\end{equation}
which follows the same: ``term we have — current — equation of motion" ordering. 
The $\partial_i$ clues us in to the fact that we need to be more crafty \footnote{I.e. via brute algebraic rearranging with good foresight, motivated by equation 2.7 of \cite{jelle}, as opposed to using an integration by parts technique.} getting the double $\partial_i \partial_j$ term to fit into the continuity equation $\partial_t j^0 + \partial_i j^i =0.$

%\begin{eqnarray}
%\frac{\partial \LL}{\partial(\partial_i \partial_j \phi)} \partial_i \partial_j (\delta \phi) &=& \frac{\partial \LL}{\partial(\partial_i \partial_j \phi)} \partial_i \partial_j (\delta \phi) + 0 + 0 \nonumber \\
%&=& \frac{\partial \LL}{\partial(\partial_i \partial_j \phi)} \partial_i \partial_j (\delta \phi) + \bigg[  \partial_i \partial_j \bigg( \frac{\partial \LL}{\partial(\partial_i \partial_j \phi)}\bigg) \delta \phi - \partial_i \partial_j \bigg( \frac{\partial \LL}{\partial(\partial_i \partial_j \phi)}\bigg) \delta \phi \bigg] + \bigg[ \partial_i \bigg( \frac{\partial \LL}{\partial(\partial_i \partial_j \phi)}\bigg)\partial_j \delta \phi - \partial_i \bigg( \frac{\partial \LL}{\partial(\partial_i \partial_j \phi)}\bigg)\partial_j \delta \phi \bigg] 
%\end{eqnarray}

\begin{align}
\label{eq3}
    \frac{\partial \LL}{\partial(\partial_i \partial_j \phi)} \partial_i \partial_j (\delta \phi) ={}& \frac{\partial \LL}{\partial(\partial_i \partial_j \phi)} \partial_i \partial_j (\delta \phi) + 0 + 0  \\
\begin{split}
     ={}& \frac{\partial \LL}{\partial(\partial_i \partial_j \phi)} \partial_i \partial_j (\delta \phi) + \bigg[  \partial_i \partial_j \bigg( \frac{\partial \LL}{\partial(\partial_i \partial_j \phi)}\bigg) \delta \phi - \partial_i \partial_j \bigg( \frac{\partial \LL}{\partial(\partial_i \partial_j \phi)}\bigg) \delta \phi \bigg] \\
         & + \bigg[ \partial_i \bigg( \frac{\partial \LL}{\partial(\partial_i \partial_j \phi)}\bigg)\partial_j \delta \phi - \partial_i \bigg( \frac{\partial \LL}{\partial(\partial_i \partial_j \phi)}\bigg)\partial_j \delta \phi \bigg]  
\end{split} \nonumber\\
\begin{split}
={}& \partial_i \partial_j \bigg( \frac{\partial \LL}{\partial(\partial_i \partial_j \phi)}\bigg) \delta \phi + \partial_i \bigg( \frac{\partial \LL}{\partial(\partial_i \partial_j \phi)}\bigg)\partial_j \delta \phi + \frac{\partial \LL}{\partial(\partial_i \partial_j \phi)} \partial_i \partial_j (\delta \phi) \\
& - \partial_i \partial_j \bigg( \frac{\partial \LL}{\partial(\partial_i \partial_j \phi)}\bigg) \delta \phi - \partial_i \bigg( \frac{\partial \LL}{\partial(\partial_i \partial_j \phi)}\bigg)\partial_j \delta \phi
\end{split} \nonumber\\
\begin{split}
={}& \partial_i \partial_j \bigg( \frac{\partial \LL}{\partial(\partial_i \partial_j \phi)}\bigg) \delta \phi + \partial_i \bigg( \frac{\partial \LL}{\partial(\partial_i \partial_j \phi)}\bigg)\partial_j \delta \phi + \frac{\partial \LL}{\partial(\partial_i \partial_j \phi)} \partial_i \partial_j (\delta \phi) \\
& - \partial_i \partial_j \bigg( \frac{\partial \LL}{\partial(\partial_i \partial_j \phi)}\bigg) \delta \phi - \partial_j \bigg( \frac{\partial \LL}{\partial(\partial_j \partial_i \phi)}\bigg)\partial_i \delta \phi
\end{split} \nonumber\\
\begin{split}
={}& \partial_i \partial_j \bigg( \frac{\partial \LL}{\partial(\partial_i \partial_j \phi)}\bigg) \delta \phi + \partial_i \bigg( \frac{\partial \LL}{\partial(\partial_i \partial_j \phi)}\bigg)\partial_j \delta \phi + \frac{\partial \LL}{\partial(\partial_i \partial_j \phi)} \partial_i \partial_j (\delta \phi) \\
& - \partial_i \partial_j \bigg( \frac{\partial \LL}{\partial(\partial_i \partial_j \phi)}\bigg) \delta \phi - \partial_j \bigg( \frac{\partial \LL}{\partial(\partial_i \partial_j \phi)}\bigg)\partial_i \delta \phi
\end{split} \nonumber\\
={}& \partial_i \partial_j \bigg( \frac{\partial \LL}{\partial(\partial_i \partial_j \phi)}\bigg) \delta \phi + \partial_i \bigg( \frac{\partial \LL}{\partial(\partial_i \partial_j \phi)} \partial_j (\delta \phi) \bigg) - \partial_i \bigg( \partial_j \bigg( \frac{\partial \LL}{\partial(\partial_i \partial_j \phi)}\bigg) \delta \phi \bigg). \nonumber
\end{align}

Making all the substitutions dictated by equations \ref{eq1}, \ref{eq3} into the original variation \ref{var}, we end up with

\vspace{5cm}

\begin{align}
0= \delta \LL ={}& \frac{\partial \LL}{\partial \phi} \delta \phi + \frac{\partial \LL}{\partial \dot{\phi}} \partial_t(\delta \phi) + \frac{\partial \LL}{\partial (\partial_i \phi)} \partial_i (\delta \phi)  +\frac{\partial \LL}{\partial (\partial_i\partial_j \phi)} \partial_i\partial_j (\delta \phi)  \nonumber \\
\begin{split}
    ={}& \frac{\partial \LL}{\partial \phi} \delta \phi + 
    \bigg[\partial_t\bigg(\frac{\partial \LL}{\partial \dot{\phi}} \delta \phi \bigg) - \partial_t \bigg( \frac{\partial \LL}{\partial \dot{\phi}} \bigg) \delta \phi \bigg] \\ 
    &+ \bigg[ \partial_i \bigg( \frac{\partial \LL}{\partial (\partial_i \phi)} \delta \phi \bigg) - \partial_i \bigg( \frac{\partial \LL}{\partial (\partial_i \phi)} \bigg) \delta \phi \bigg] \\
    &+ \bigg[ \partial_i \partial_j \bigg( \frac{\partial \LL}{\partial(\partial_i \partial_j \phi)}\bigg) \delta \phi + \partial_i \bigg( \frac{\partial \LL}{\partial(\partial_i \partial_j \phi)} \partial_j (\delta \phi) \bigg) - \partial_i \bigg( \partial_j \bigg( \frac{\partial \LL}{\partial(\partial_i \partial_j \phi)}\bigg) \delta \phi \bigg) \bigg]. \nonumber \\
\end{split} 
\end{align}

Now collecting terms according to our objective of 

\begin{quote}
    ... rearranging the above variation eq. \ref{var}, throwing out terms that all together equal zero on-shell due to the Euler-Lagrange equations of motion (which are all those proportional to $\delta \phi$), and leaving terms that will ultimately take on the roles of the currents in the continuity equation. 
\end{quote}

\begin{align}
0= \delta \LL ={}& \frac{\partial \LL}{\partial \phi} \delta \phi + \frac{\partial \LL}{\partial \dot{\phi}} \partial_t(\delta \phi) + \frac{\partial \LL}{\partial (\partial_i \phi)} \partial_i (\delta \phi)  +\frac{\partial \LL}{\partial (\partial_i\partial_j \phi)} \partial_i\partial_j (\delta \phi)  \nonumber \\
\begin{split}
    ={}& \frac{\partial \LL}{\partial \phi} \delta \phi + 
    \bigg[\partial_t\bigg(\frac{\partial \LL}{\partial \dot{\phi}} \delta \phi \bigg) - \partial_t \bigg( \frac{\partial \LL}{\partial \dot{\phi}} \bigg) \delta \phi \bigg] \\ 
    &+ \bigg[ \partial_i \bigg( \frac{\partial \LL}{\partial (\partial_i \phi)} \delta \phi \bigg) - \partial_i \bigg( \frac{\partial \LL}{\partial (\partial_i \phi)} \bigg) \delta \phi \bigg] \\
    &+ \bigg[ \partial_i \partial_j \bigg( \frac{\partial \LL}{\partial(\partial_i \partial_j \phi)}\bigg) \delta \phi + \partial_i \bigg( \frac{\partial \LL}{\partial(\partial_i \partial_j \phi)} \partial_j (\delta \phi) \bigg) - \partial_i \bigg( \partial_j \bigg( \frac{\partial \LL}{\partial(\partial_i \partial_j \phi)}\bigg) \delta \phi \bigg) \bigg] \nonumber \\
\end{split}  \\
\vspace{2cm} \nonumber \\
\begin{split}
    ={}& \bigg[\frac{\partial \LL}{\partial \phi} \delta \phi -  \partial_t \bigg( \frac{\partial \LL}{\partial \dot{\phi}} \bigg) \delta \phi - \partial_i \bigg( \frac{\partial \LL}{\partial (\partial_i \phi)} \bigg) \delta \phi+ \partial_i \partial_j \bigg( \frac{\partial \LL}{\partial(\partial_i \partial_j \phi)}\bigg) \delta \phi \bigg] \\ 
    &+ \partial_t\bigg(\frac{\partial \LL}{\partial \dot{\phi}} \delta \phi \bigg) \\
    &+ \bigg[ \partial_i \bigg( \frac{\partial \LL}{\partial (\partial_i \phi)} \delta \phi \bigg)  + \partial_i \bigg( \frac{\partial \LL}{\partial(\partial_i \partial_j \phi)} \partial_j (\delta \phi) \bigg) - \partial_i \bigg( \partial_j \bigg( \frac{\partial \LL}{\partial(\partial_i \partial_j \phi)}\bigg) \delta \phi \bigg) \bigg] \nonumber \\
\end{split} \\
\begin{split}
    ={}& \bigg[\frac{\partial \LL}{\partial \phi}  -  \partial_t \bigg( \frac{\partial \LL}{\partial \dot{\phi}} \bigg)  - \partial_i \bigg( \frac{\partial \LL}{\partial (\partial_i \phi)} \bigg) + \partial_i \partial_j \bigg( \frac{\partial \LL}{\partial(\partial_i \partial_j \phi)}\bigg)  \bigg] \delta \phi \\ 
    &+ \partial_t\bigg(\frac{\partial \LL}{\partial \dot{\phi}} \delta \phi \bigg) \\
    &+ \partial_i \bigg[  \bigg( \frac{\partial \LL}{\partial (\partial_i \phi)} \delta \phi \bigg)  +  \bigg( \frac{\partial \LL}{\partial(\partial_i \partial_j \phi)} \partial_j (\delta \phi) \bigg) -  \bigg( \partial_j \bigg( \frac{\partial \LL}{\partial(\partial_i \partial_j \phi)}\bigg) \delta \phi \bigg) \bigg]. \nonumber \\
\end{split}
\end{align}

Thus, on shell the terms in the parenthesis proportional to $\delta \phi$ goes to zero, and we are left with a continuity equation

\begin{align}
\label{final}
\begin{split}
   0= \delta \LL ={}& \bigg[\frac{\partial \LL}{\partial \phi}  -  \partial_t \bigg( \frac{\partial \LL}{\partial \dot{\phi}} \bigg)  - \partial_i \bigg( \frac{\partial \LL}{\partial (\partial_i \phi)} \bigg) + \partial_i \partial_j \bigg( \frac{\partial \LL}{\partial(\partial_i \partial_j \phi)}\bigg)  \bigg] \delta \phi \\ 
    &+ \partial_t\bigg(\frac{\partial \LL}{\partial \dot{\phi}} \delta \phi \bigg) \\
    &+ \partial_i \bigg[  \bigg( \frac{\partial \LL}{\partial (\partial_i \phi)} \delta \phi \bigg)  +  \bigg( \frac{\partial \LL}{\partial(\partial_i \partial_j \phi)} \partial_j (\delta \phi) \bigg) -  \bigg( \partial_j \bigg( \frac{\partial \LL}{\partial(\partial_i \partial_j \phi)}\bigg) \delta \phi \bigg) \bigg]  
\end{split} \\
   ={}& 0 + \partial_t\bigg(\frac{\partial \LL}{\partial \dot{\phi}} \delta \phi \bigg) + \partial_i \bigg[  \bigg( \frac{\partial \LL}{\partial (\partial_i \phi)} \delta \phi \bigg)  +  \bigg( \frac{\partial \LL}{\partial(\partial_i \partial_j \phi)} \partial_j (\delta \phi) \bigg) -  \bigg( \partial_j \bigg( \frac{\partial \LL}{\partial(\partial_i \partial_j \phi)}\bigg) \delta \phi \bigg) \bigg] \nonumber \\
   ={}& \partial_t j^0 + \partial_i j^i, \nonumber
\end{align}

so that 

\begin{eqnarray}
\label{conteq}
j^0 &=& \frac{\partial \LL}{\partial \dot{\phi}} \delta \phi \nonumber \\
j^i &=& \bigg( \frac{\partial \LL}{\partial (\partial_i \phi)} \delta \phi \bigg)  +  \bigg( \frac{\partial \LL}{\partial(\partial_i \partial_j \phi)} \partial_j (\delta \phi) \bigg) -  \bigg( \partial_j \bigg( \frac{\partial \LL}{\partial(\partial_i \partial_j \phi)}\bigg) \delta \phi \bigg).
\end{eqnarray}

And thus we have the conserved total charge

\begin{equation}
\label{belowthis}
    Q = \int dV j^0
\end{equation}

Note that the charge $Q$ is a rank-0 scalar, and its corresponding $j^i$ is a rank-1 vector. This dimensional relationship following from the integration definition, and in particular as Seiberg points out, we need a rank-2 tensor current $j^{ij}$  if we want a rank-1 vector charge like the dipole moment $Q^i$ \cite{seiberg}.

\iffalse 
His going from 1.7 

$$
\partial_\mu J^{[\mu\nu]}=0
$$

to 1.8 (where the first equation is a continuity equation, and the second equation is an additional constraint on the system — which can be relaxed as we'll discuss in a moment)

\begin{eqnarray*}
\partial_0 J_{0}^j - \partial_i J^{[ij]} &=&0 \\
\partial_i J_{0}^i  &=&0 
\end{eqnarray*}

goes as follows:
 
\begin{eqnarray*}
0 &=& \partial_\mu J^{[\mu\nu]} \\
&=& \frac{1}{2} \partial_\mu(J^{\mu\nu}-J^{\nu\mu})\\
&=& \frac{1}{2} (\partial_\mu J^{\mu\nu}-\partial_\mu J^{\nu\mu}) \\
&=& \frac{1}{2} [(\partial_0 J^{0\nu}-\partial_i J^{i\nu})-(\partial_0 J^{\nu 0}-\partial_i J^{\nu i})]
\end{eqnarray*}

plugging in $\nu=0$ and $\nu=i$ yield both equations of 1.8 in \cite{seiberg}.

\fi

\pagebreak
\subsection{Lemma for dipole conservation continuity equation}

In our penultimate step we will use both of the previous sections' work: the full symmetry of section \ref{need}, as well as the higher order variation of section {higher}. Starting with equation \ref{final} but without\footnote{Thanks is due to Li Yabo for pointing out this alternative approach to the problem \cite{yabo}.} imposing the on-shell condition, we have (via expanding terms and using equation \ref{full})

\begin{align}
\begin{split}
   0= \delta \LL ={}& \bigg[\frac{\partial \LL}{\partial \phi}  -  \partial_t \bigg( \frac{\partial \LL}{\partial \dot{\phi}} \bigg)  - \partial_i \bigg( \frac{\partial \LL}{\partial (\partial_i \phi)} \bigg) + \partial_i \partial_j \bigg( \frac{\partial \LL}{\partial(\partial_i \partial_j \phi)}\bigg)  \bigg] \delta \phi \\ 
    &+ \partial_t\bigg(\frac{\partial \LL}{\partial \dot{\phi}} \delta \phi \bigg) \\
    &+ \partial_i \bigg[  \bigg( \frac{\partial \LL}{\partial (\partial_i \phi)} \delta \phi \bigg)  +  \bigg( \frac{\partial \LL}{\partial(\partial_i \partial_j \phi)} \partial_j (\delta \phi) \bigg) -  \bigg( \partial_j \bigg( \frac{\partial \LL}{\partial(\partial_i \partial_j \phi)}\bigg) \delta \phi \bigg) \bigg],  
\end{split} \nonumber \\
\begin{split}
   0= \delta \LL ={}& \bigg[\frac{\partial \LL}{\partial \phi}  -  \partial_t \bigg( \frac{\partial \LL}{\partial \dot{\phi}} \bigg)  - \partial_i \bigg( \frac{\partial \LL}{\partial (\partial_i \phi)} \bigg) + \partial_i \partial_j \bigg( \frac{\partial \LL}{\partial(\partial_i \partial_j \phi)}\bigg)  \bigg] \delta \phi \\ 
    &+ \partial_t\bigg(\frac{\partial \LL}{\partial \dot{\phi}}\bigg) \delta \phi + \frac{\partial \LL}{\partial \dot{\phi}} \dot{\delta \phi} \\
    &+ \partial_i \bigg( \frac{\partial \LL}{\partial (\partial_i \phi)}\bigg) \delta \phi +  \frac{\partial \LL}{\partial (\partial_i \phi)} \alpha_i \phi +  \frac{\partial \LL}{\partial (\partial_i \phi)} (\alpha_0 +\alpha_i x^i) \partial_i \phi \\
    &+  \partial_i \bigg[ \bigg( \frac{\partial \LL}{\partial(\partial_i \partial_j \phi)} \partial_j (\delta \phi) \bigg) -  \bigg( \partial_j \bigg( \frac{\partial \LL}{\partial(\partial_i \partial_j \phi)}\bigg) \delta \phi \bigg) \bigg],  
\end{split} \nonumber \\
\begin{split}
   0= \delta \LL ={}& \bigg[\frac{\partial \LL}{\partial \phi}  + \partial_i \partial_j \bigg( \frac{\partial \LL}{\partial(\partial_i \partial_j \phi)}\bigg)  \bigg] \delta \phi \\ 
    & + \frac{\partial \LL}{\partial \dot{\phi}} \dot{\delta \phi} \\
    & +  \frac{\partial \LL}{\partial (\partial_i \phi)} \alpha_i \phi +  \frac{\partial \LL}{\partial (\partial_i \phi)} (\alpha_0 +\alpha_i x^i) \partial_i \phi \\
    &+  \partial_i \bigg[ \bigg( \frac{\partial \LL}{\partial(\partial_i \partial_j \phi)} \partial_j (\delta \phi) \bigg) -  \bigg( \partial_j \bigg( \frac{\partial \LL}{\partial(\partial_i \partial_j \phi)}\bigg) \delta \phi \bigg) \bigg],  
\end{split} \nonumber \\
\begin{split}
   0= \delta \LL ={}& \bigg[\frac{\partial \LL}{\partial \phi}  + \frac{\partial \LL}{\partial \dot{\phi}} \dot{\phi} +\frac{\partial \LL}{\partial (\partial_i \phi)} \partial_i \phi + \partial_i \partial_j \bigg( \frac{\partial \LL}{\partial(\partial_i \partial_j \phi)}\bigg)  \bigg] \delta \phi \\ 
    & +  \frac{\partial \LL}{\partial (\partial_i \phi)} \alpha_i \phi  \\
    &+  \partial_i \bigg[ \bigg( \frac{\partial \LL}{\partial(\partial_i \partial_j \phi)} \partial_j (\delta \phi) \bigg) -  \bigg( \partial_j \bigg( \frac{\partial \LL}{\partial(\partial_i \partial_j \phi)}\bigg) \delta \phi \bigg) \bigg].  
\end{split}  
\end{align}

Continuing with the final two terms after a page break, with $\alpha(x) = \alpha_0 + \alpha_i x^i$

\begin{align}
\begin{split}
   0= \delta \LL ={}& \bigg[\frac{\partial \LL}{\partial \phi}  + \frac{\partial \LL}{\partial \dot{\phi}} \dot{\phi} +\frac{\partial \LL}{\partial (\partial_i \phi)} \partial_i \phi + \partial_i \partial_j \bigg( \frac{\partial \LL}{\partial(\partial_i \partial_j \phi)}\bigg)  \bigg] \delta \phi +  \frac{\partial \LL}{\partial (\partial_i \phi)} \alpha_i \phi  \\
    &+  \partial_i \bigg( \frac{\partial \LL}{\partial(\partial_i \partial_j \phi)}\bigg) \partial_j (\delta \phi) +  \bigg( \frac{\partial \LL}{\partial(\partial_i \partial_j \phi)}\bigg) \partial_i \partial_j (\delta \phi) -  \partial_i \partial_j \bigg( \frac{\partial \LL}{\partial(\partial_i \partial_j \phi)}\bigg) \delta \phi  - \partial_j \bigg( \frac{\partial \LL}{\partial(\partial_i \partial_j \phi)}\bigg) \partial_i(\delta \phi),
\end{split} \nonumber \\
\begin{split}
   0= \delta \LL ={}& \alpha(x) \bigg[\frac{\partial \LL}{\partial \phi} \phi  + \frac{\partial \LL}{\partial \dot{\phi}} \dot{\phi} +\frac{\partial \LL}{\partial (\partial_i \phi)} \partial_i \phi \bigg]  +  \frac{\partial \LL}{\partial (\partial_i \phi)} \alpha_i \phi  \\
    &+  \bigg[ \partial_i \bigg( \frac{\partial \LL}{\partial(\partial_i \partial_j \phi)}\bigg) \partial_j (\delta \phi)\bigg] + \bigg[ \frac{\partial \LL}{\partial(\partial_i \partial_j \phi)} \partial_i \partial_j  (\delta \phi)\bigg]  - \bigg[\partial_j \bigg( \frac{\partial \LL}{\partial(\partial_i \partial_j \phi)}\bigg) \partial_i(\delta \phi) \bigg],
\end{split} \nonumber \\
\begin{split}
   0= \delta \LL ={}& \alpha(x) \bigg[\frac{\partial \LL}{\partial \phi} \phi  + \frac{\partial \LL}{\partial \dot{\phi}} \dot{\phi} +\frac{\partial \LL}{\partial (\partial_i \phi)} \partial_i \phi \bigg] +  \frac{\partial \LL}{\partial (\partial_i \phi)} \alpha_i \phi  \\
    &+  \bigg[ \partial_i \bigg( \frac{\partial \LL}{\partial(\partial_i \partial_j \phi)}\bigg) \alpha(x) \partial_j \phi \bigg] + \bigg[ \frac{\partial \LL}{\partial(\partial_i \partial_j \phi)} \alpha_i \partial_j  \phi + \frac{\partial \LL}{\partial(\partial_i \partial_j \phi)} \alpha(x) \partial_i \partial_j \phi  \bigg]  \\
    &- \bigg[\partial_j \bigg( \frac{\partial \LL}{\partial(\partial_i \partial_j \phi)}\bigg) \alpha_i \phi + \partial_j \bigg( \frac{\partial \LL}{\partial(\partial_i \partial_j \phi)}\bigg) \alpha(x) \partial_i \phi \bigg].
\end{split} 
\end{align}

Noting the symmetry of second partial derivatives — which allows us to cancel terms 

\begin{align}
\begin{split}
   0= \delta \LL ={}& \alpha(x) \bigg[\frac{\partial \LL}{\partial \phi} \phi  + \frac{\partial \LL}{\partial \dot{\phi}} \dot{\phi} +\frac{\partial \LL}{\partial (\partial_i \phi)} \partial_i \phi + \frac{\partial \LL}{\partial(\partial_i \partial_j \phi)}  \partial_i \partial_j \phi\bigg] \\ 
   & +  \frac{\partial \LL}{\partial (\partial_i \phi)} \alpha_i \phi + \frac{\partial \LL}{\partial(\partial_i \partial_j \phi)} \alpha_i \partial_j  \phi  - \partial_j \bigg( \frac{\partial \LL}{\partial(\partial_i \partial_j \phi)}\bigg) \alpha_i \phi,  
\end{split} 
\end{align}

and then integrating by parts on last term, we have

\begin{align}
\begin{split}
   0= \delta \LL ={}& (\alpha_0+\alpha_i x^i) \bigg[\frac{\partial \LL}{\partial \phi} \phi  + \frac{\partial \LL}{\partial \dot{\phi}} \dot{\phi} +\frac{\partial \LL}{\partial (\partial_i \phi)} \partial_i \phi + \frac{\partial \LL}{\partial(\partial_i \partial_j \phi)}  \partial_i \partial_j \phi\bigg] \\ 
   & +  \alpha_i \bigg[ \frac{\partial \LL}{\partial (\partial_i \phi)}  \phi + 2 \frac{\partial \LL}{\partial(\partial_i \partial_j \phi)} \partial_j  \phi \bigg].
\end{split} \nonumber 
\end{align}

Following the exact same logic we used in going from equation \ref{eqq1} to equation \ref{eqq2}, we know have

\begin{equation}
\label{lemma}
    \frac{\partial \LL}{\partial (\partial_i \phi)}  \phi + 2 \frac{\partial \LL}{\partial(\partial_i \partial_j \phi)} \partial_j  \phi =0. 
\end{equation}

How does this help us obtain a continuity equation of the form $\partial_t \rho + \partial_i \partial_j J^{ij} = 0$? 

We return to equation \ref{conteq} but plugging in the dipole symmetry $\delta \phi = \beta_j x^j \phi$ (where $\beta$ is used as opposed to $\alpha$ for the reader who wishes to verify this by hand, in which case partial derivatives $\delta$ and $\alpha$ are too similar and cause some typos) since the condition we just found above tells us what we need to have in order for both (charge and dipole) symmetries to be respected simultaneously.

\pagebreak 

\subsection{Dipole conservation continuity equation}

If we plug $\delta \phi = \beta_j x^j \phi$ into equations \ref{conteq} we have

\begin{eqnarray}
j^0 &=& \frac{\partial \LL}{\partial \dot{\phi}} \delta \phi \nonumber \\
&=& \frac{\partial \LL}{\partial \dot{\phi}} (\beta_j x^j \phi) \nonumber \\
&\equiv& \beta_j (x^j J^0)  \\
j^i &=&  \frac{\partial \LL}{\partial (\partial_i \phi)} (\beta_j x^j \phi)   +   \frac{\partial \LL}{\partial(\partial_i \partial_j \phi)} \partial_j (\beta_j x^j \phi)  -   \partial_j \bigg( \frac{\partial \LL}{\partial(\partial_i \partial_j \phi)}\bigg) (\beta_j x^j \phi) \nonumber \\
&=& \frac{\partial \LL}{\partial (\partial_i \phi)} \beta_j x^j \phi   +  \frac{\partial \LL}{\partial(\partial_i \partial_j \phi)}  \beta_j  \phi + \frac{\partial \LL}{\partial(\partial_i \partial_j \phi)} \beta_j x^j \partial_j \phi -   \partial_j  \bigg(\frac{\partial \LL}{\partial(\partial_i \partial_j \phi)}\bigg) \beta_j x^j \phi  \nonumber \\
&\equiv& \beta_j \bigg( \frac{\partial \LL}{\partial(\partial_i \partial_j \phi)} \phi + x^j J^i\bigg), \label{56}
\end{eqnarray}

where $J^0$ and $J^i$ are just equations \ref{conteq} with $\delta \phi \rightarrow \phi$.

We will define the quantities in parenthesis of the above equations to be 

\begin{equation}
    \rho_{\text{dipole}} = x^j J^0,
\end{equation}

\begin{equation}
    J^{ij} = \frac{\partial \LL}{\partial(\partial_i \partial_j \phi)} \phi + x^j J^i,
\end{equation}

for reasons that will become clear momentarily.

The continuity equation \ref{final} reads then 

\begin{eqnarray}
0&=& \partial_t j^0 +\partial_i j^i \nonumber \\
&=& \partial_t (\beta_j (x^j J^0)) +\partial_i \bigg[\beta_j \bigg( \frac{\partial \LL}{\partial(\partial_i \partial_j \phi)} \phi + x^j J^i\bigg)\bigg] \nonumber \\
&=& \beta_j (\partial_t \rho_{\text{dipole}} + \partial_i J^{ij}) \nonumber \\
&=& \partial_t \rho_{\text{dipole}} + \partial_i J^{ij}.
\end{eqnarray}

Look familiar? This is precisely the condition we need to ensure dipole conservation from equation \ref{ccc}.

To prove this is true, we will use the lemma \ref{lemma}.

\begin{eqnarray}
0&=& \partial_t j^0 +\partial_i j^i \nonumber \\
&=& \partial_t \rho_{\text{dipole}} + \partial_i J^{ij} \label{notethis}\\
&=& \partial_t (x^j J^0) +\partial_i  \bigg( \frac{\partial \LL}{\partial(\partial_i \partial_j \phi)} \phi + x^j J^i\bigg) \nonumber \\
&=& \partial_t (x^j J^0) +\partial_i  \bigg( \frac{\partial \LL}{\partial(\partial_i \partial_j \phi)}\bigg) \phi + \frac{\partial \LL}{\partial(\partial_i \partial_j \phi)} \partial_i \phi + \partial_i (x^j J^i) \nonumber \\
&=& \partial_t (x^j J^0) +\partial_i  \bigg( \frac{\partial \LL}{\partial(\partial_i \partial_j \phi)}\bigg) \phi + \frac{\partial \LL}{\partial(\partial_i \partial_j \phi)} \partial_i \phi + J^i +x^j \partial_i J^i \nonumber \\
&=& x^j (\partial_t  J^0 +  \partial_i J^i) + \partial_i  \bigg( \frac{\partial \LL}{\partial(\partial_i \partial_j \phi)}\bigg) \phi + \frac{\partial \LL}{\partial(\partial_i \partial_j \phi)} \partial_i \phi + \frac{\partial \LL}{\partial (\partial_i \phi)} \phi  + \frac{\partial \LL}{\partial(\partial_i \partial_j \phi)} \partial_j \phi -   \partial_j  \bigg(\frac{\partial \LL}{\partial(\partial_i \partial_j \phi)}\bigg) \phi  \nonumber \\
&=& x^j (\partial_t  J^0 +  \partial_i J^i)  + \bigg[\frac{\partial \LL}{\partial (\partial_i \phi)} \phi  + 2\frac{\partial \LL}{\partial(\partial_i \partial_j \phi)} \partial_j \phi\bigg]   \nonumber \\
&=& x^j (\partial_t  J^0 +  \partial_i J^i)  + 0   \nonumber \\
&=& \partial_t  J^0 +  \partial_i J^i.
\end{eqnarray}

This is nothing but the continuity equation for the conservation of total charge which we derived in equation
\ref{final}.

Note that \ref{notethis} fits the dimensional relationship that we pointed out below equation \ref{belowthis}.

All told, the symmetry of the fields 

\begin{eqnarray}
    \delta \phi &=& \alpha(x) \phi \nonumber \\
    &=& (\alpha_0 + \alpha_i x^i )\phi,
\end{eqnarray}

gives the system not only conservation of total charge, but also conservation of dipole moment — the key features of fracton models.

As a reminder, this detour into Noether currents and continuity equations became in an attempt to understand the invariance of the fields that Pretko posited in equation \ref{sym2}.

%\frac{\partial \LL}{\partial (\partial_i\partial_j \phi)} \partial_i\partial_j (\delta \phi)  = \partial_i \partial_j \bigg( \frac{\partial \LL}{\partial (\partial_i\partial_j \phi)} \delta \phi \bigg) - \partial_i \partial_j \bigg( \frac{\partial \LL}{\partial (\partial_i\partial_j \phi)} \bigg) \delta \phi

\pagebreak
\pagebreak

\section{Further plans}

\subsection{Fracton gauge field theory}
\label{gaugetheory}

We have in the last section justified the symmetry that Pretko proposed was necessary for a field-theoretic description of fractons \cite{pretko}

\begin{eqnarray}
    \phi &\rightarrow& e^{i \alpha_0} \phi, \nonumber \\
\phi &\rightarrow& e^{i \vec{\alpha}\cdot \vec{x}} \phi \nonumber \\
&=& e^{i \alpha_i x^i} \phi,
\end{eqnarray}

where $\alpha_0$ and $\vec{\alpha}$ are a constant scalar and constant vector respectively.

We can combine these invariance conditions using using the $\alpha(x) = \alpha_0 + \alpha_i x^i$ we've used previously

\begin{equation}
\label{symsym}
    \phi \rightarrow e^{i \alpha(x)}\phi.
\end{equation}

As we uncovered in Section \ref{higher}, we need higher order spatial derivatives to satisfy dipole moment conservation. So, with a $(+,-,-,-)$ Minkowski metric, a generic Lagrangian for the theory looks like 

\begin{equation}
    \LL = A |\partial_t \phi|^2 - B |\partial_i \phi|^2 - C|\partial_i \partial_j \phi|^2 - D |\phi|^2
\end{equation}

where we can replace $A$ with 1 and $D$ with $m^2$, as a mass term.

As Pretko works out however, the B and C terms above are not covariant, i.e. they do not transform like the field itself under the symmetry \ref{symsym}. In fact the only way to maintain covariance in these derivative terms is to introduce two copies of the field for the derivatives to act on. If the B and C terms are replaced with 

\begin{equation}
\label{covar}
    \phi \partial_i \partial_j \phi - \partial_i\phi \partial_j \phi,
\end{equation}

in which case the Lagrangian reads

\begin{equation}
    \LL =  |\partial_t \phi|^2 - B |\phi \partial_i \partial_j \phi- \partial_i\phi \partial_j \phi|^2 - C|\phi \partial_i \partial_i \phi- \partial_i\phi \partial_i \phi|^2 - m^2 |\phi|^2.
\end{equation}

These quartic terms that follow from the covariant \ref{covar} can be justified conceptually as well from our experience in Figures \ref{quad1} - \ref{quad3}, when we analyzed the consequences of moving a charge on a system with dipole moment conservation and how that epitomized fracton dynamics.

What remains unclear to us, at present, is the modification to the C term the Pretko has made in his equation 11 \cite{pretko}

\begin{equation}
    \LL =  |\partial_t \phi|^2 - B |\phi \partial_i \partial_j \phi- \partial_i\phi \partial_j \phi|^2 - C (\phi^*)^2 (\phi \partial_i \partial_i \phi- \partial_i\phi \partial_i \phi) - m^2 |\phi|^2.
\end{equation}

This is will be our next step in studying the field theory description of fractons. 

The derivation of this ungauged Lagrangian is followed by Pretko's gauging of the theory by: including covariant derivatives with gauge fields that transform like those we introduced in Section \ref{charge} and introducing field strengths/curvatures accounting for the dynamics of the gauge fields themselves.

Which such a gauge theory Lagrangian in hand, formal quantum field theory work began \cite{qft}. As mentioned in the introduction \pageref{here3}, much work in this area has been in formally grounding this quantum field theories \cite{seibergshao}.

\pagebreak
\subsection{Fractons and gauging algebras}
\label{inspriation}

The gauging of symmetry algebras is well known to be fruitful in high-energy theory, see the success of Yang-Mills theory in the Standard Model, but also in gravity. For example, general relativity and Newton-Cartan gravity can be viewed as gauge theories of the Poincaré and Bargmann algebras respectively \cite{bergshoeff} \cite{Roelthesis} \cite{thesis}.

In the following work \cite{gromov}, a symmetry algebra (taking the place of the Poincaré or Bargmann for example in comparison to the above mentioned gravity constructions) coined ``the multipole algebra" is constructed from the polynomial shift symmetries of \cite{kev} \cite{newkev}.

The so-called multipole algebra of \cite{gromov} is then gauged to construct an effective field theory that the author uses to study fractons. Studying how this gauging procedure  concerning symmetry algebra $\rightarrow$ effective field theory in condensed matter language compares to the gauging procedure concerning symmetry algebra $\rightarrow$ gravitational theory in high energy/gravity language could be fruitful for each community.

And so in the future, we personally plan to study the polynomial formalism of fracton models such as is covered in  \cite{haahthesis} \cite{williamson} \cite{zijianthesis}. In particular, the objective will be to learn about the polynomial shift symmetry algebra of \cite{gromov} and \cite{newkev} before learning how to gauge it — taking note along the way how the procedure differs from that of \cite{bergshoeff}.

\pagebreak

\subsection{Other quantum systems and fractons}
\subsubsection{Simulating gapped fracton models with ultracold Rydberg atoms}
\label{rydberg}

Recall the progress made using ultracold atomic physics in Feynman's dream of quantum simulation referenced on page \pageref{here}. 

Advances in Rydberg atom research have enabled 3-dimensional arrays \cite{atoms} — mimicking the spatial dimensionality of complex condensed matter spin systems that can now be studying via these atomic experiments \cite{quantummanybody}.

Notably, one of the spin system that can be studying in this framework is precisely the X-cube model we discussed in Section \ref{spin2} as these authors points out \cite{expfractons1}. More generally however, a more recent work on the same topic of realizing fractons experimentally with Rydberg atoms \cite{expfractons2}, aims to set the stage for implementation of long-range entangled quantum states \cite{chenwen} \cite{wenboulder}, an ingredient for further advancements in quantum information physics.

As discussed in the introduction on page \pageref{here2} regarding the inception of fractons, Haah had practical aims. Now that much of the theory behind (at least gapped) fracton models has been established, transitioning to putting these systems to practical use could be a fitting next step for the fracton community to focus on.

\subsubsection{Many-body localization (MBL), random unitary circuits, eigenstate thermalization hypothesis (ETH), and fractons}

Another way in which Rydberg atoms and other quantum systems can be related to the phenomenology of fractons is via the study of localization and thermalization.\footnote{The Google research group mentioned on page \pageref{here3} is working on these topics with their quantum computer as well \cite{goognew}.} The study of and thermalization has not only been benefited from advances in ultracold atomic physics \cite{quantummanybody}, but also from fracton physics. 

The eigenstate thermalization hypothesis (ETH) dictates the conditions under which quantum systems thermalize (reach equilibrium), while many-body localization (MBL) is the failure of ETH — when a system doesn't thermalize and retains memory of its initial conditions \cite{Nandkishore} \cite{MBL}. Studies on the time evolution of random unitary circuits have revealed that the fractalization (exponential growth of the Hilbert space) of fracton models violates the ETH \cite{circuit2}, and that the systems maintain a memory of their initial state, as is key to MBL \cite{circuit1}. Moreover, the dipole nature of fractons has been connected to these consequences for the ETH and MBL \cite{dipole1} \cite{dipole2} \cite{dipole3} \cite{dipole4}.

Given that MBL and the ETH are among those topics than can now be studied using ultracold physics \cite{quantummanybody}, and fractons have been found to be such a rich testing ground for these concepts, simulating fractons with ultracold physics as discussed above in Section \ref{rydberg} could very well lead to a radically new understanding of quantum and statistical mechanics.

\pagebreak

\end{document}